\documentclass{article}

\usepackage{arxiv}

\usepackage[utf8]{inputenc}
\usepackage[T1]{fontenc}
\usepackage{hyperref}
\usepackage{url}
\usepackage{booktabs}
\usepackage{amsfonts}
\usepackage{nicefrac}
\usepackage{microtype}
\usepackage{graphicx}
\usepackage{natbib}
\usepackage{doi}
\usepackage[USenglish]{babel}
\usepackage[intlimits]{amsmath}
\usepackage{bm}
\hypersetup{colorlinks=true, citecolor=blue}
\usepackage{amssymb}
\usepackage{fontawesome}
\usepackage{mathtools}
\usepackage{ulem}
\usepackage{xcolor}
\usepackage{lmodern}
\usepackage{siunitx}
\usepackage{booktabs}
\usepackage{etoolbox}

\renewrobustcmd{\bfseries}{\fontseries{b}\selectfont}
\renewrobustcmd{\boldmath}{}
\newrobustcmd{\B}{\bfseries}

\newcommand{\Mtiny}{\scriptscriptstyle}

\title{Another look at synthetic-type control charts}

\date{November 05, 2021}

\author{%
	\href{https://orcid.org/0000-0002-9666-5554}{\includegraphics[scale=0.06]{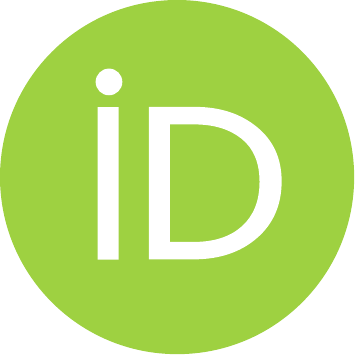}\hspace{1mm}Sven Knoth} \\
	Dep. of Mathematics \& Statistics\\
	Helmut Schmidt University\\
	Hamburg, Germany\\
	\texttt{knoth@hsu-hh.de}
}
	
\renewcommand{\shorttitle}{Synthetic Charts}

\begin{document}
	
\maketitle
	
\begin{abstract}
During the last two decades, in statistical process monitoring plentiful new
methods appeared with synthetic-type control charts being a prominent constituent.
These charts became popular designs for several reasons.
The two most important ones are simplicity and proclaimed excellent change point detection
performance. Whereas there is no doubt about the former, we deal here with the latter.
We will demonstrate that their performance is questionable. Expanding on some previous
skeptical articles we want to critically reflect upon recently developed variants of synthetic-type charts in order to emphasize
that there is little reason to apply and to push this special class of control charts.
\end{abstract}


\section{Introduction} \label{sec:intro}

From the statistical tools, we know as control charts, the majority was created in the 20$^\text{th}$ century.
During the last years, however, numerous new concepts were introduced. The synthetic chart, proposed
by \cite{Wu:Sped:2000a, Wu:Sped:2000b}, is a special example. On the one hand, it fascinates with
its simple design and its explicit solutions of the Average Run Length (ARL) equation and
related measures. The ARL is the expected number of samples or individual observations until
the control chart declares that a change happened alias lack of control was detected \citep{Shew:1925}.
It comes in various types, where the most popular ones are the zero-state and steady-state ARL \citep{Cros:1986}.
And on the other hand, in \cite{Wu:Sped:2000a} the synthetic chart
was proclaimed as superior in terms of the zero-state ARL, (unintentionally) concealing that it was equipped
with a solid head-start. So \cite{Davi:Wood:2002} criticized this pattern and suggested two
key elements: Enforce the steady-state ARL as performance measure which captures potentially misleading side effects
of introducing head-starts. Second, endow the older
runs rule chart, which differentiates between the change directions (called side-sensitive), as well with a head-start. 
This rather cautious critique did not block the further development of synthetic charts.
Instead, these charts became really popular. The more recent \cite{Knot:2016a} was already much more
explicit in its criticism. Nevertheless, synthetic charts remained highly attractive.
In particular, \cite{Raki:EtAl:2019} claimed that \cite{Knot:2016a} did consider only the original
synthetic chart of \cite{Wu:Sped:2000a}. This is partially correct, but the general message
would be the same anyway: Synthetic charts and all their derivatives (published so far) are
clearly dominated by older control charts. 
For example, two of the four synthetic-type charts in \cite{Chak:Raki:2021}, namely both standalone ones, were
analyzed in \cite{Knot:2016a}.
Here, we will utilize Exponentially Weighted Moving Average (EWMA)
charts, which will be compared to all four plain synthetic-type charts. For the sake of
a concise presentation, we touch only briefly combinations of synthetic with Shewhart-type charts,
which were called improved synthetic charts in \cite{Raki:EtAl:2019}. Their ``natural'' counterpart
is a Shewhart-EWMA combo \citep{Luca:Sacc:1990a, Capi:Masa:2010a}.
Thus, we provide a thorough ARL (zero- and steady-state) analysis
of 8 \citep[including charts without head-start like][]{Shon:Grah:2018} different synthetic-type charts and EWMA charts.

In Section~\ref{sec:class} we describe all considered control charts in more detail.
Later, in Section~\ref{sec:ststARL} we elaborate upon the steady-state ARL concept, where
some confusions have to be clarified. Our main results appear in Section~\ref{sec:cedStudy},
where we compare all the charts by looking at the zero-state ARL, conditional expected delay (CED)
and the steady-state ARL. Finally, we assemble our conclusions in Section~\ref{sec:Concl}.
In the Appendix some side results are given.

\section{Classification scheme of synthetic-type charts} \label{sec:class}

\begin{figure}[hbt]
\centering	
\includegraphics[width=.4\textwidth]{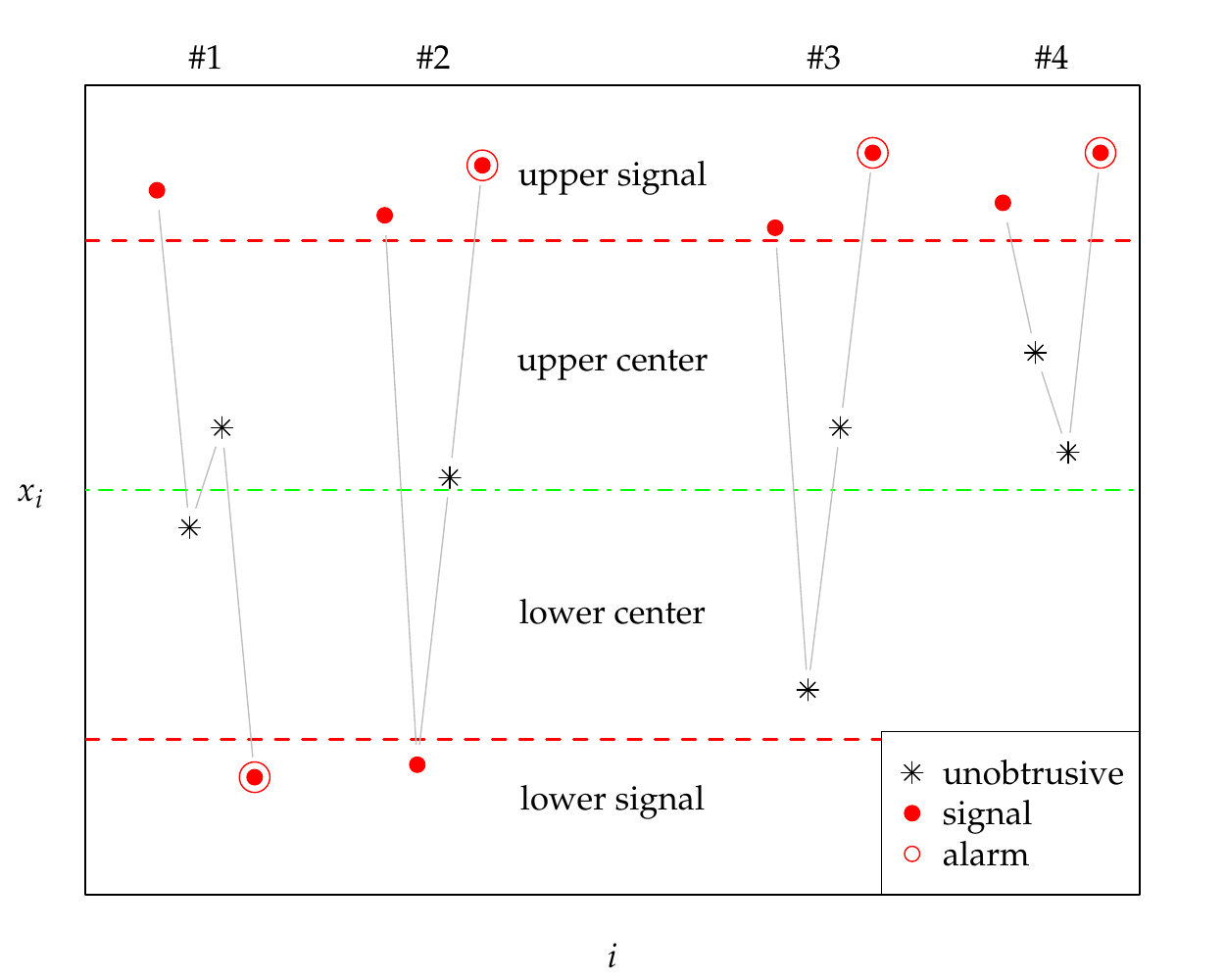}
\caption{Four observations that form an alarm pattern for four different synthetic-type charts with $H = 3$.}\label{fig:pattern}
\end{figure}

As \cite{Raki:EtAl:2019} and others mentioned, the constitutive element of a synthetic chart is
that two warnings or signals are needed to trigger the actual alarm. And these two signals should
not be too \textit{``far away from each other''}. Thus, synthetic charts are special runs (or scan) rules
charts, because they could be expressed as 2-of-$H+1$ runs rules, with $H \in \{1, 2, \ldots\}$,
cf. to \cite{Davi:Wood:2002, Bers:Kout:Raki:2020}. Differently to \cite{Raki:EtAl:2019},
we consider not only the head-start versions. Instead, following \cite{Shon:Grah:2018, Shon:Grah:2019} and \cite{Bers:Kout:Raki:2020},
we investigate common and head-start synthetic-type charts.
In the here following Table~\ref{tab:SGtab1}, we list the 8 synthetic-type charts with their initial reference.
\begin{table}[hbt]
\centering	
\caption{Simplified version of Table~1 in \cite{Shon:Grah:2018}, i.\,e. only 2-of-$H+1$ designs.} \label{tab:SGtab1}
\begin{tabular}{ccccllll} \toprule
	\# && label && \multicolumn{2}{c}{w/o head-start} & \multicolumn{2}{c}{w/ head-start} \\ \midrule
  1 && ``true'' synthetic
  && $R_1$ & \cite{Derm:Ross:1997}\footnotemark[1] & $S_1$ & \cite{Wu:Sped:2000a} \\
  2 && side-sensitive
  && $R_2$ & \cite{Klei:2000a} & $S_2$ & \cite{Davi:Wood:2002} \\
  3 && revised
  && $R_3$ & \cite{Mach:Cost:2014b}\footnotemark[2] & $S_3$ & \cite{Shon:Grah:2018} \\
  4 && modified
  && $R_4$ & \cite{Antz:Raki:2008a} & $S_4$ & \cite{Shon:Grah:2018} \\ \bottomrule
\end{tabular}	 
\end{table}
\footnotetext[1]{\cite{Derm:Ross:1997}:%
\textit{``Probably the easiest way to construct a control chart that considers each subgroup average in relation to those around
it is to define a chart that declares a process out of control if \uwave{two successive averages differ from $\mu$
by more than $c\sigma$} for some value $c$.''}}
\footnotetext[2]{\cite{Mach:Cost:2014b}:%
\textit{``The transient states describe the position of the last $L$ sample points; `\underline{1}' means that the
sample point fell below the $LCL$, \uwave{`0' means that the sample point fell in the central region},
and `1' means that the sample point fell above the $UCL$.''}}
Besides Table~\ref{tab:SGtab1}, which is a simplified and reduced version of Table~1 in \cite{Shon:Grah:2018},
we want to provide some more constructional details.
For simplicity, we assume individual normally distributed observations
with mean $\mu$ and standard deviation $\sigma$ (more details in the next section).
We set exemplary $H = 3$. Between two signals alias two observations beyond the limits, there must be
at most two ``unobtrusive'' observations to trigger an alarm. For ``true'' synthetic charts (in the narrower sense), it is not important whether
the two signals are raised on the same side of the chart, whereas the remaining three designs require the same side.
In Figure~\ref{fig:pattern}, we plotted the center line at the in-control mean $\mu_0$ and two limits at $\mu_0 \pm k \sigma_0$.
Here, $\sigma_0$ denotes the in-control standard deviation which is assumed to be constant and known. The design parameter $k$ controls
the detection behavior and is typically chosen to achieve a pre-defined in-control ARL.
Pattern \#1 in Figure~\ref{fig:pattern} would trigger an alarm only for chart \#1.
The next pattern raises an alarm for the common 2-of-4 runs rule, where the lower signal has no impact to the final alarm.
Chart \#3 requests that the observations enclosed by the two upper signals reside between the limits.
Later we will see that there are no big performance differences between these two chart designs.
The most involved design is \#4, where the in-between observations have to be on the same side like the signaling points.
The patterns are chosen so that pattern \#4 flags an alarm for all four charts, whereas pattern \#3 does it only for \#1, \#2 and \#3, etc.
Of course, these different patterns demand distinct values of $k$, namely $2.2087 > 2.0760 > 2.0723 > 1.9642$
for \#1, ..., \#4, respectively, for $H = 3$ and in-control ARL 500. Note that the values
for charts with head-start are slightly larger. After explaining the differences between the rows in Table~\ref{tab:SGtab1},
we want to describe the disparity between the columns. Thus, it is about head-start and no head-start.
The former presumes that the last data point, just before monitoring was started, would trigger a signal.
Hence, we need only one further signal to raise an alarm. Except for \#1, however, we have to know whether the signal was above
the upper or below the lower limit. This problem is dealt with pragmatically, that is, given the first observed signal, we just
imply that the hidden signal was on the same side, providing kind of a
wildcard head-start\footnote[3]{\cite{Davi:Wood:2002}: \textit{``The initial state is $0\pm$; that is,	the most recent observation at the onset of monitoring is considered to be beyond control limits on both sides of the center line.''}}.
There are some side effects to the Markov chain modeling, except for \#1, of course. Specifically,
we have to introduce further states of the chain that are related to this particular starting behavior. 
In Table~\ref{tab:classStates},
we indicate the resulting number of transient states we obtain for the
underlying Markov chain model, cf. to \cite{Shon:Grah:2018}.
We added as well the chart labels used in \cite{Shon:Grah:2018} and \cite{Bers:Kout:Raki:2020}.
%
%
\begin{table}[hbt]
\centering	
\caption{Number of transient states, 2-of-$H+1$ rules, see Table~\ref{tab:SGtab1}.} \label{tab:classStates}
\begin{tabular}{ccllll} \toprule
  \# && \multicolumn{2}{c}{w/o head-start} & \multicolumn{2}{c}{w/ head-start} \\ \midrule
  1 && DR:  & $H+1$     & WS:  & $H+1$ \\
  2 && KL:  & $H^2+H+1$ & DW:  & $H^2+2H+1$ \\
  3 && MC1: & $2H+1$    & MC2: & $3H+1$ \\
  4 && AR:  & $2H+1$    & MSS: & $4H$ \\ \bottomrule
\end{tabular}	 
\end{table}
The smallest value, $H+1$, is known from \cite{Davi:Wood:2002}. The latter reported as well the largest value
in Table~\ref{tab:classStates}, observed for the DW chart, namely $(H+1)^2$. From the latter
size, one can straightforwardly derive the number for the general KL chart, that is, $H^2 + H + 1$
\citep[][dealt with three synthetic-type charts: WS, KL and DW]{Knot:2016a}.
The dimension $2H+1$ was given in \cite{Mach:Cost:2014b}. The remaining numbers could be found
in \cite{Shon:Grah:2018}.

Before continuing with the competitor EWMA, we want to mention that the very recent \cite{Chak:Raki:2021} labeled the
synthetic-type charts differently. Their $S_1$ and $S_3$ correspond to WS (our $S_1$) and DW (our $S_2$), respectively.
The remaining two in \cite{Chak:Raki:2021}, $S_2$ and $S_4$, are just the latter combined with a Shewhart alarm rule.

As already mentioned, we utilize the common EWMA \citep{Robe:1959} chart with varying limits
as the main competitor to all the synthetic-type charts.
Picking an appropriate value for the smoothing constant $0 < \lambda \le 1$ (we favor here 0.25 and 0.1),
we create the following sequence of EWMA statistics \citep{Luca:Sacc:1990a, Mont:2019}:
\begin{align}
	Z_0 & = \mu_0\quad,\;
	Z_i = (1-\lambda) Z_{i-1} + \lambda X_i \quad,\; i = 1, 2, \ldots \,, \label{eq:ewmaseries} \\
	L_E & = \min \left\{ i\ge 1\!: |Z_i - \mu_0| > c_E \sqrt{ \big( 1-(1-\lambda)^{2i} \big) \frac{\lambda}{2-\lambda} } \sigma_0 \,\right\} \,. \label{eq:ewmarule}
\end{align}
Besides the series $\{Z_i\}$ we get the run-length alias stopping time $L_E$ which simply counts
the number of observations until the first alarm.
Fortunately, there are numerical routines \citep{Crow:1987a, Knot:2003, Knot:2005a}
for calculating all the measures we deploy in this contribution (see next section).
Of course, the results are only approximations (differently to the synthetic-type charts,
where the corresponding Markov chains are exact models), the accuracy of
the said numerical procedures is sufficiently high.
Eventually we want to note that we apply their implementations in
the \textsf{R} package \texttt{spc} \citep{K:2021}.

\section{Steady-state ARL and other measures} \label{sec:ststARL}

As told in the previous section, we consider an independent series $X_1, X_2, \ldots$ following a normal distribution with mean $\mu$ and standard deviation $\sigma$.
To incorporate a potential change, we apply the change point ($\tau$) model
\begin{equation}
	\mu = \begin{cases} \mu_0 = 0 & ,\; t < \tau \\ \mu_1 = \delta & ,\; t \ge \tau \end{cases} \quad .
	\label{eq:tau}
\end{equation}	
Regarding the standard deviation (variance) we make the common assumption that it is
known, $\sigma = \sigma_0 = 1$ (otherwise normalize the $X_t$), and it remains constant.

With $L$ we denote the run length (stopping time),
which is the number of observed $X_i$ values until an alarm is raised.
The expected values of $L$ for the two situations $\tau = 1 $ and $\tau = \infty$ 
constitute the well-known zero-state Average Run Length (ARL), cf. to \cite{Page:1954c, Cros:1986}.
Mostly, the control charts are setup to yield a pre-defined
in-control ARL, i.\,e. $E_\infty(L) = A$ for some suitably large number $A$ (here we set $A = 500$).
For a given control chart design, it is a common task to determine
out-of-control ARL values, $E_1(L)$, for specified values of $\delta$.
The resulting ARL profiles are typically used to judge the detection performance over a range
of changes $\delta$ and to compare charts to each other.

Besides the simple case $\tau = 1$ in \eqref{eq:tau}, we determine the series of conditional expected delays (CED)
\begin{align*}
	D_\tau & = E_\tau\big(L-\tau+1\mid L\ge \tau \big) \quad,\; \tau = 1, 2, \ldots
	\intertext{and its limit, the conditional steady-state ARL}
	\mathcal{D}_1 & = \lim_{\tau\to\infty} D_\tau \,.
\end{align*}
Both $\{D_\tau\}$ and $\mathcal{D}_1$ are functions of $\delta$.
For all charts considered here (EWMA and synthetic-type), the series $\{D_\tau\}$ converges quickly to $\mathcal{D}_1$.
Besides $\mathcal{D}_1$, one can utilize the cyclical steady-state ARL $\mathcal{D}_2$,
which incorporates re-starts after getting a false alarm. See \cite{Tayl:1968}, \cite{Cros:1986} and the
recent \cite{Knot:2021c} for more details. It is defined as follows:
\begin{align*}
 \mathcal{D}_2 & := \lim\limits_{\tau\to\infty} E_\tau \big(L_\star-\tau+1) \\ 
   \text{with } \quad L_\star & =  L_1 + L_2 + \ldots + L_{I_\tau-1} + L_{I_\tau}
  \quad \text{ and } \quad I_\tau = \min\left\{i\ge 1: \sum_{j=1}^i L_j \ge \tau\right\} \,. \nonumber
\end{align*}
Thus, after some number of false alarms ($L_1$, $L_2$, \ldots, $L_{I_\tau-1}$ as number of
observations to the next false alarm) the first true alarm appears at
observation $L_\star \ge \tau$. The term $L_\star-\tau+1$ denotes the resulting detection delay.
Of course, the restarting pattern (for EWMA typically at $\mu_0$, whereas for the synthetic-type charts
various ideas were investigated) influences the actual value of $\mathcal{D}_2$.

By denoting $\mathbb{Q}$ the transition matrix of transient states, $\mathbb{I}$ the identity matrix and
$\bm{1}$ a vector of ones, we start with the classical ARL (vector $\bm{\ell}$) result of \cite{Broo:Evan:1972}
\begin{equation*}
  \bm{\ell} = (\mathbb{I} - \mathbb{Q})^{-1} \bm{1} \,,
\end{equation*}
and continue with some prerequisites for the steady-state vectors \citep{Knot:2021c}:
\begin{align*}
  \varrho \bm{\psi}_1 & = \mathbb{Q}^\prime \bm{\psi}_1 \qquad \text{--- left eigenvector of the dominant eigenvalue $\varrho$} \,, \\
  \bm{\psi}_2 & = \big(\mathbb{I} - \mathbb{Q}^\prime\big)^{-1} \bm{e}_1 \qquad \text{--- $\bm{e}_1$ consists of zeros except for the restart state,
  where a 1 is set} \,.
\end{align*}
The equation for $\bm{\psi}_1$ was given in \cite{Broo:Evan:1972}, whereas the $\bm{\psi}_2$ equation was included in \cite{Darr:Sene:1965}.
Both vectors will be normalized (i.\,e. $\bm{1}^\prime \bm{\psi}_i=1$, $i=1,2$). Then the two
steady-state ARLs are calculated via $\mathcal{D}_i = \bm{\psi}_i^\prime \bm{\ell}$, $i=1,2$ (exact for synthetic-type, approximation
for EWMA). For the true synthetic chart \citep{Wu:Sped:2000a}, the following explicit solutions were
derived \citep{Knot:2016a}, $\Phi()$ denoting the cdf of the standard normal distribution:
\begin{align}
	p & = p(k; \delta) = 1 - \big[ \Phi(k-\delta) - \Phi(-k-\delta) \big] \quad,\;
	q = 1 - p \quad,\; r  = p ( 1 - q^H ) \,, \label{eq:kpqr} \\
	\bm{\ell}^\prime & = \bordermatrix{%
		& \Mtiny 0 & \Mtiny 1 & \Mtiny \ldots & \Mtiny H-1 & \Mtiny H \cr
		& \frac{1}{r} & \frac{1+q^H(q^{-1}-1)}{r} & \ldots  & \frac{1+q^H(q^{-(H-1)}-1)}{r} & \frac{1}{r} + \frac{1}{p} \cr
	} \,, \nonumber \\
	\bm{\psi}_1^\prime & = \bordermatrix{%
		& \Mtiny 0 & \Mtiny 1 & \Mtiny \ldots & \Mtiny H-1 & \Mtiny H \cr
		& s & \frac{q}{\varrho} s & \ldots & \left(\frac{q}{\varrho}\right)^{H-1} s  & \frac{\varrho}{p} s \cr
	} \quad,\; s = 1-q/\varrho \,, \nonumber \\
	\bm{\psi}_2^\prime & = \bordermatrix{%
		& \Mtiny 0 & \Mtiny 1 & \Mtiny \ldots & \Mtiny H-1 & \Mtiny H \cr
		& p & pq & \ldots & p q^{H-1} & q^H \cr
	} \,. \nonumber
\end{align}
Recall that the restart for $\bm{\psi}_2$ happens at state 0, which refers to the solid head-start situation.
Note that \cite{Wu:Ou:Cast:Khoo:2010} utilized the same restart state.
However, \cite{Shon:Grah:2019} considered a different restart state, namely $H$ that corresponds to the no head-start case.
The resulting vector is
\begin{equation*}
  \bm{\psi}_3^\prime = \frac{1}{2-q^H} \bordermatrix{%
  & \Mtiny 0 & \Mtiny 1 & \Mtiny \ldots & \Mtiny H-1 & \Mtiny H \cr
		& p & p q & \ldots & p q^{H-1} & 1 \cr
	} \,.
\end{equation*}
The vectors $\bm{\psi}_2$ and $\bm{\psi}_3$ differ in the entry for state $H$, namely $q^H$ and 1, respectively, and
in the normalizing constant, 1 and $1/(2-q^H)$, respectively. The impact to the resulting $\mathcal{D}_2$ is not
substantial. \cite{Shon:Grah:2019} did not explain why they used a different head-start. Moreover,
it remains as well unclear, why they proposed two ways of calculating the cyclical steady-state vector.
First, there are more than two approaches. Second, all these different procedures provide equal solutions
(except for the scaling constant). For an elaborated discussion refer to \cite{Knot:2021c}.
A more important problem, however, is the wrong result of both
\citet[p.192 and 195]{Shon:Grah:2019} and already \citet[p.2899]{Mach:Cost:2014b}
for $\mathcal{D}_1$ (conditional), in particular for
its steady-state vector ($\bm{\psi}_1$). By following the erroneous path in \cite{Cros:1986},
they obtained:
\begin{equation*}
  \bm{\psi}_4^\prime = \frac{1}{1 + H p} \bordermatrix{%
	& \Mtiny 0 & \Mtiny 1 & \Mtiny \ldots & \Mtiny H-1 & \Mtiny H \cr
	& p & p & \ldots & p & 1 \cr
} \;.
\end{equation*}
First, recall that
\cite{Cros:1986} introduced the terms conditional and cyclical, while he also provided Markov chain algorithms
to calculate these steady-state ARLs. His procedure for the cyclical steady-state ARL is correct, despite it is
not the one indicated in \citet[p.191]{Shon:Grah:2019}.
However, the approach to get the conditional steady-state
ARL by following \textit{``the matrix $\mathbb{R}$ ... can be	scaled up so that each row of the matrix sums to 1''}
\citep[][p.193]{Cros:1986} is wrong.
For more details we refer to \cite{Knot:2021c}. The surprisingly simple $\bm{\psi}_4$ is the output
of this wrong algorithm applied to the synthetic (in the narrower sense) chart. We wonder why none of the above authors questioned this nearly
uniform distribution. The good news are that the numerical differences when using $\bm{\psi}_1$, ..., $\bm{\psi}_4$
are not large, see Appendix \ref{app:steadyARL}.
Therefore it is not too restrictive to apply the conditional steady-state ARL $\mathcal{D}_1$
relying on $\bm{\psi}_1$ and the related CED $D_\tau$ for the rest of the paper. Note that neither in \citet[p.5]{Raki:EtAl:2019}
nor in \citet[p.13]{Chak:Raki:2021} the discussion of the steady-state ARL did touch these subtle complications.

\section{Comparison study} \label{sec:cedStudy}

\begin{figure}[hbt]
\centering
\begin{tabular}{cc}
  \footnotesize $S_1$ & \footnotesize $S_2$ \\[-1ex]
  \includegraphics[width=.48\textwidth]{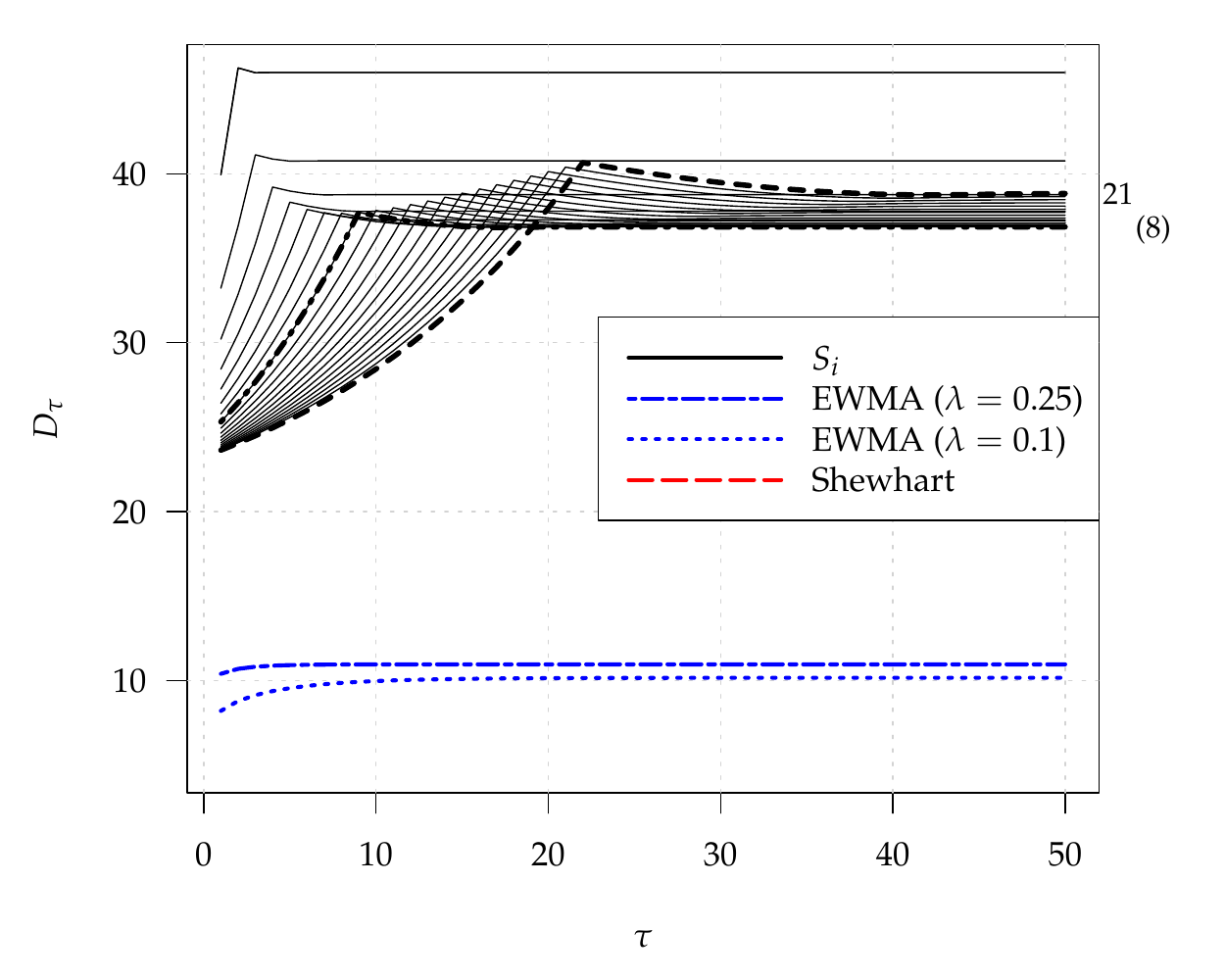} &	
  \includegraphics[width=.48\textwidth]{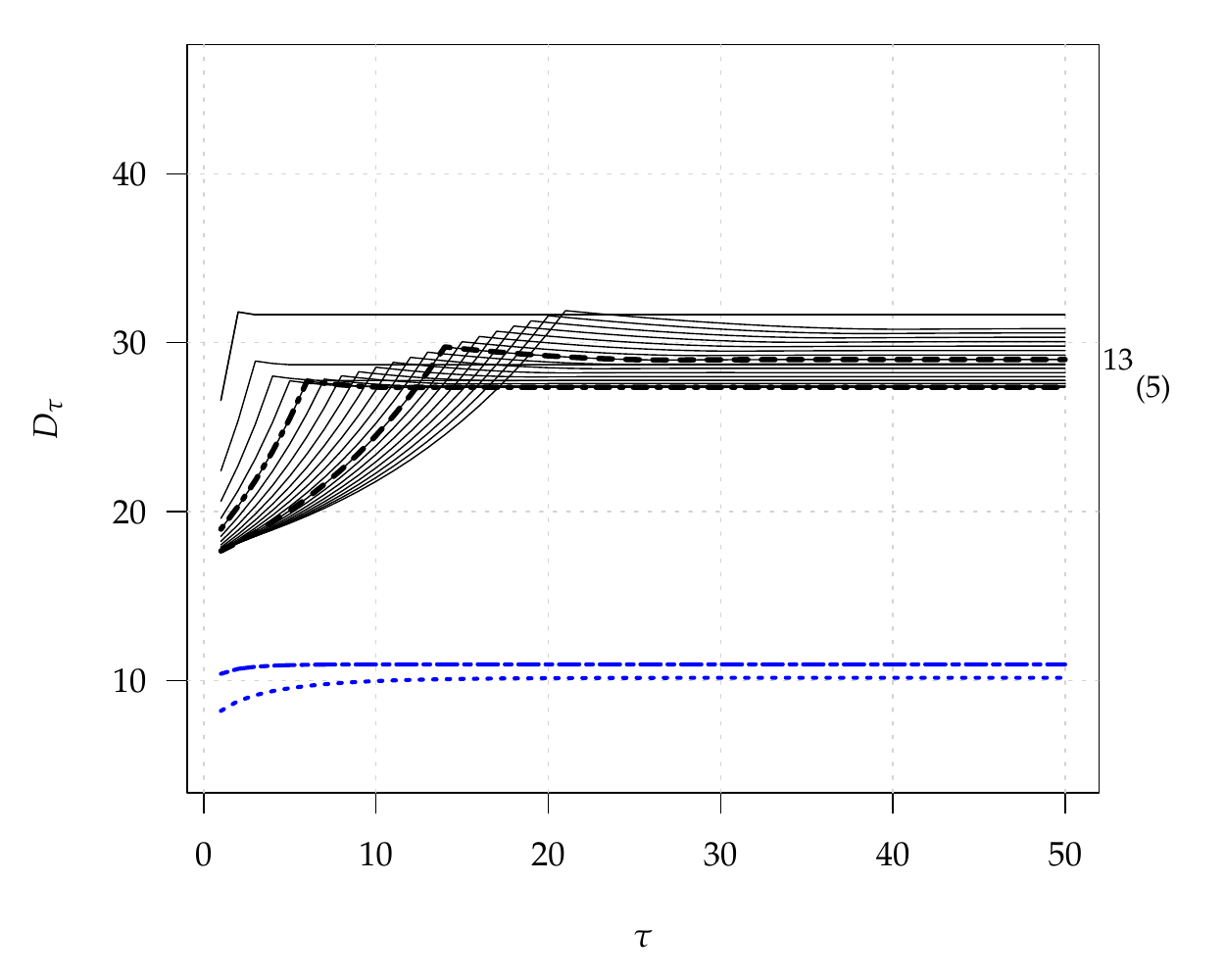} \\[1ex]
  \footnotesize $S_3$ & \footnotesize $S_4$ \\[-1ex]
  \includegraphics[width=.48\textwidth]{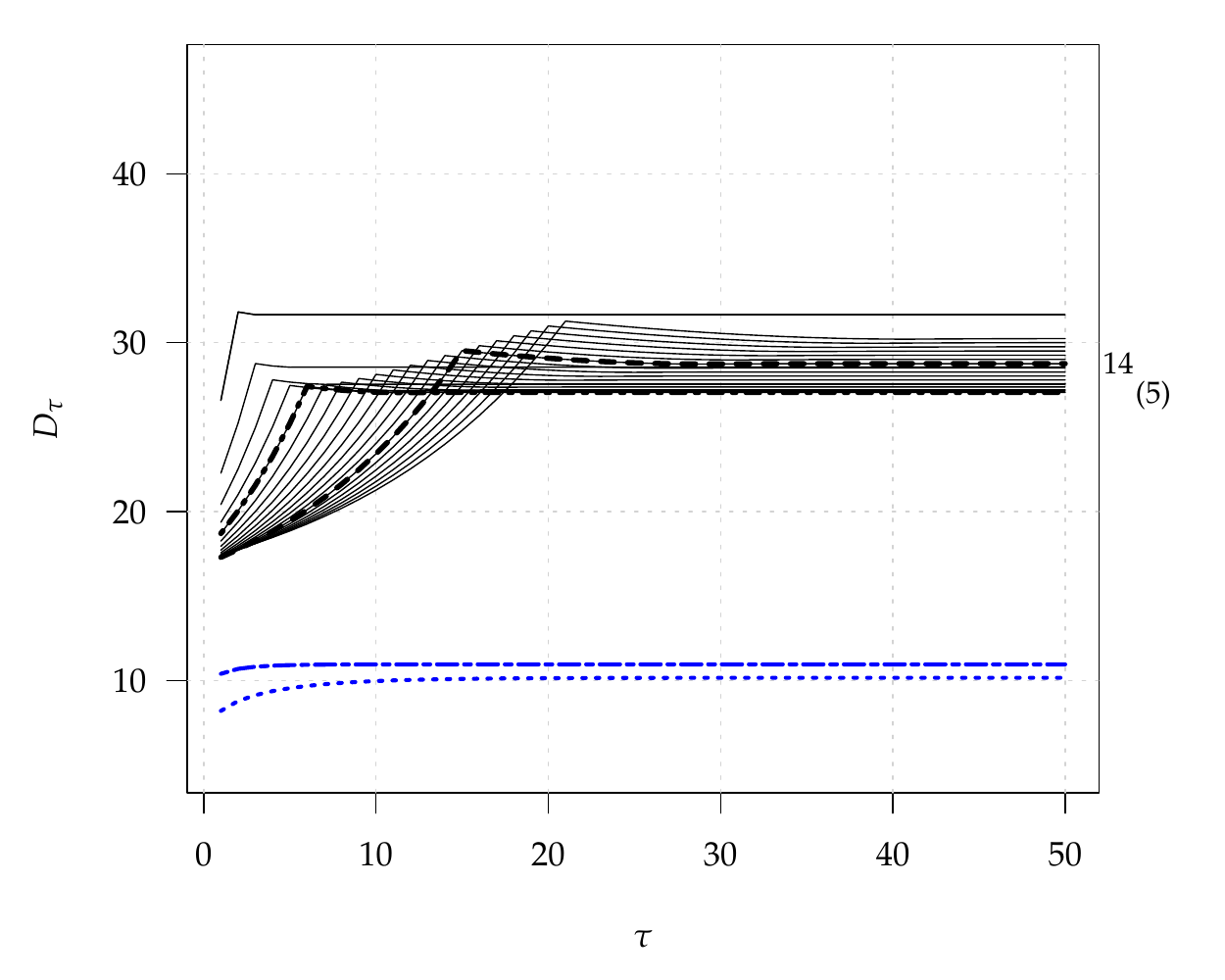} &	
  \includegraphics[width=.48\textwidth]{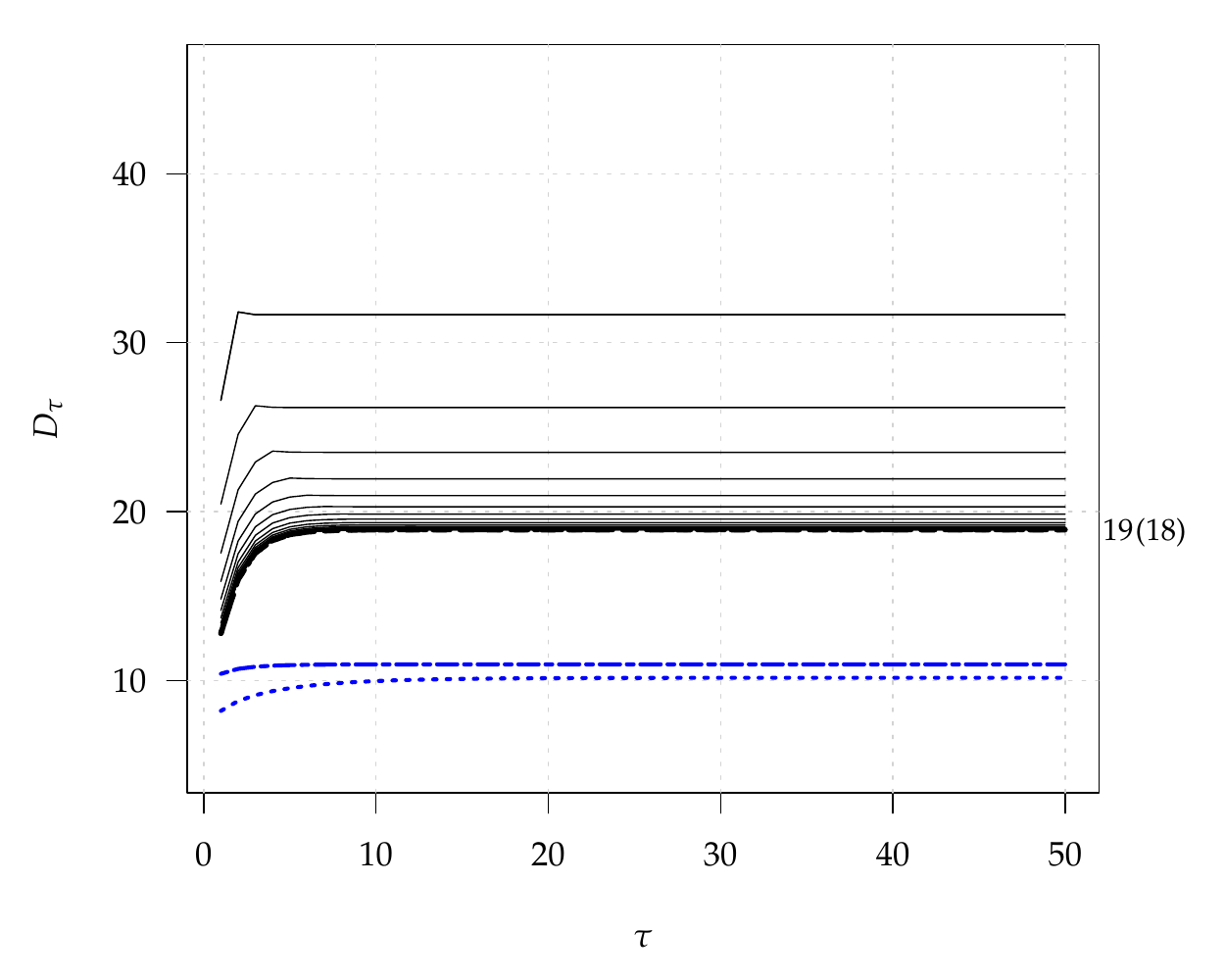}
\end{tabular}
\caption{$D_\tau$ profiles for four synthetic-type charts with head-start,
$H = 1, 2, \ldots, 25$, best scheme (zero-state and steady-state) bold (dashed and dash-dotted) lines,
shift $\delta = 1$, two EWMA charts; in-control ARL 500.} \label{fig:ced1}
\end{figure}

We start with a CED analysis of the four synthetic-type charts with head-start.
All considered control charts are designed to have an in-control ARL, $E_\infty(L)$, of 500.
For the aforementioned charts, labeled as $S_1$, \ldots, $S_4$, we determine
the CED $D_\tau$ for $\tau = 1, 2, \ldots, 50$. Moreover, we plot the CED profiles
for $H = 1, 2, \ldots, 25$. In Figure~\ref{fig:ced1}
we display besides the 25 mentioned profiles two EWMA ($\lambda=0.25$ and $=0.1$) CED profiles.

\begin{figure}[hbt]
\centering
\begin{tabular}{cc}
  \footnotesize $S_1$ & \footnotesize $S_2$ \\[-1ex]
  \includegraphics[width=.45\textwidth]{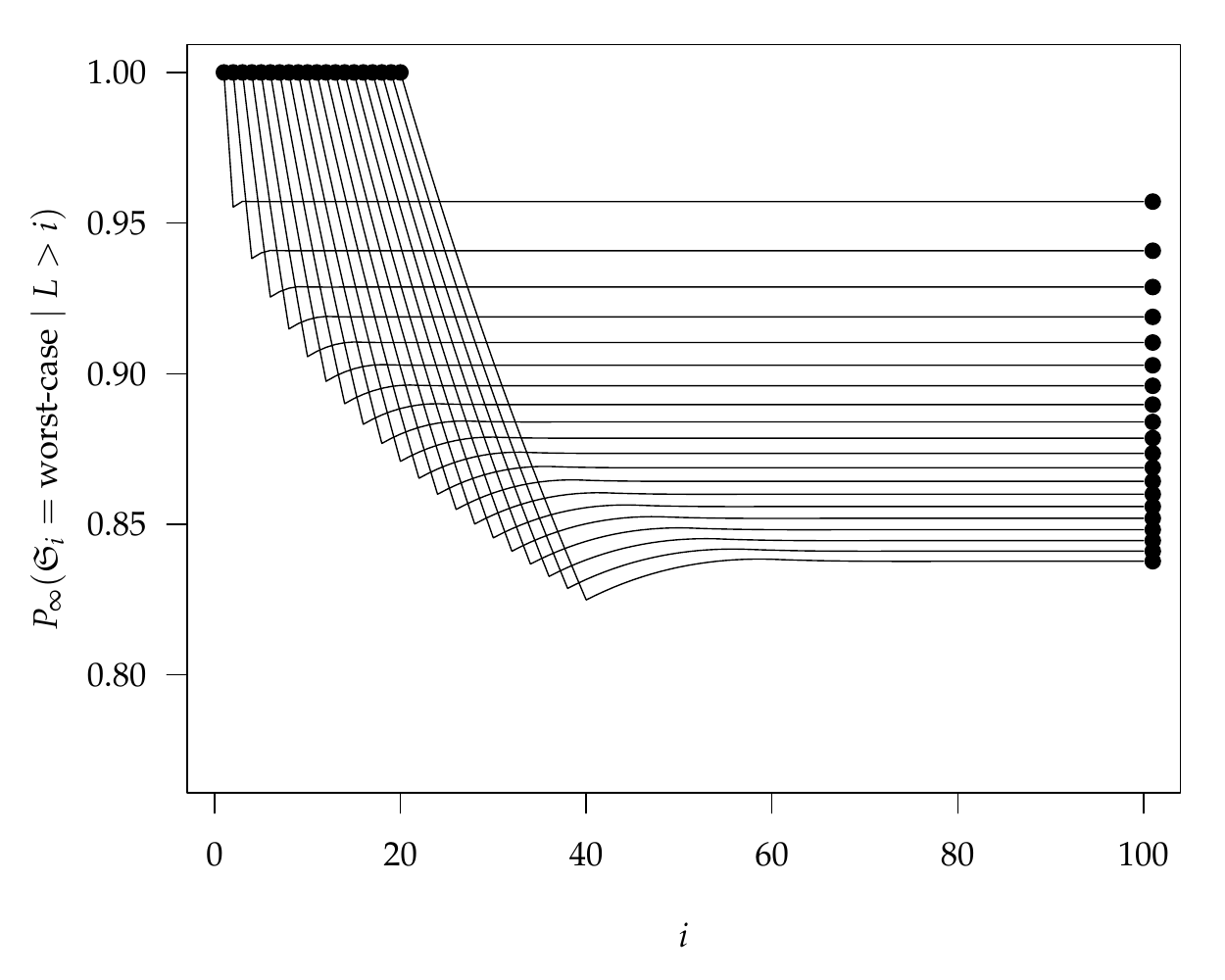} &	
  \includegraphics[width=.45\textwidth]{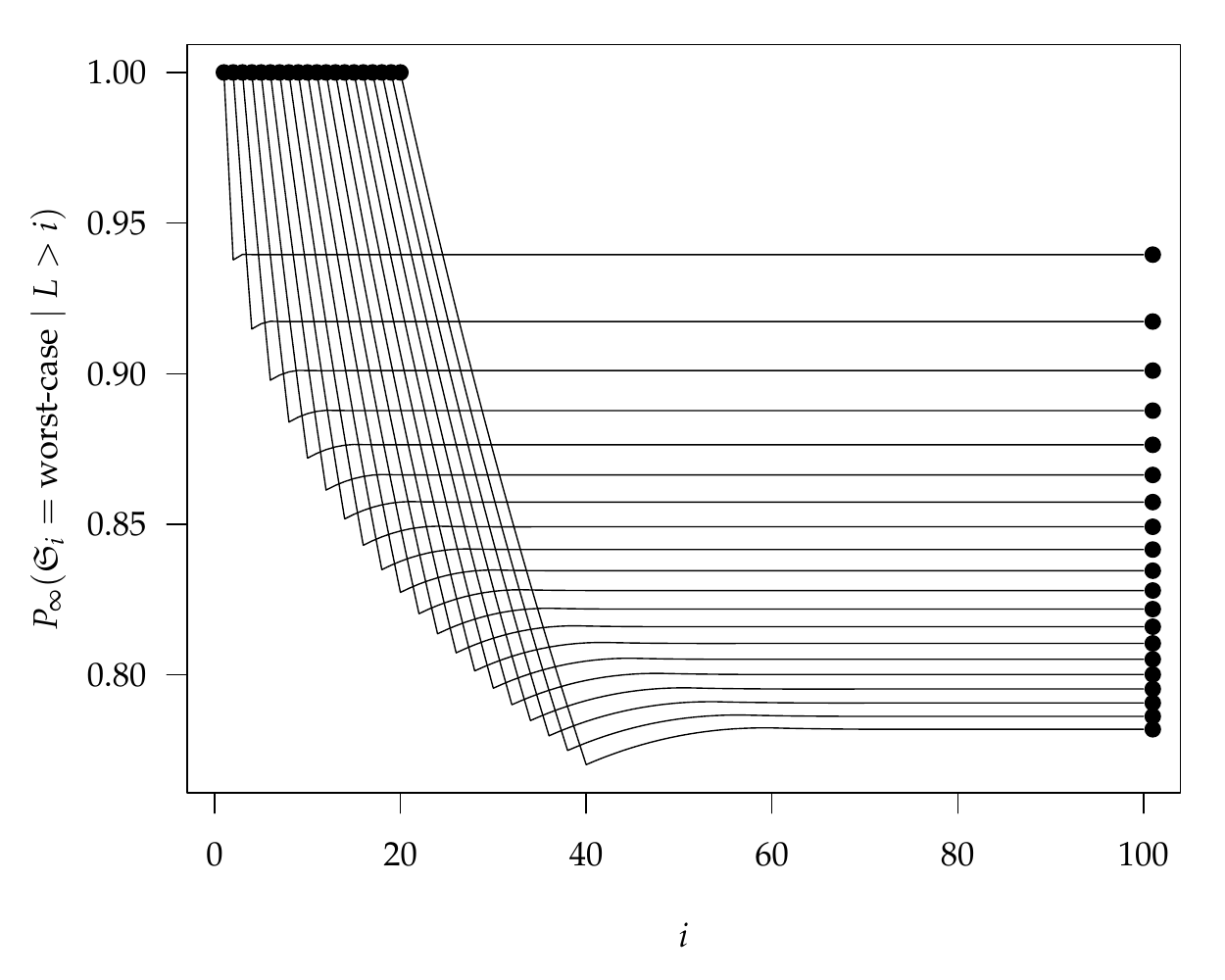} \\[1ex]
  \footnotesize $S_3$ & \footnotesize $S_4$ \\[-1ex]
  \includegraphics[width=.45\textwidth]{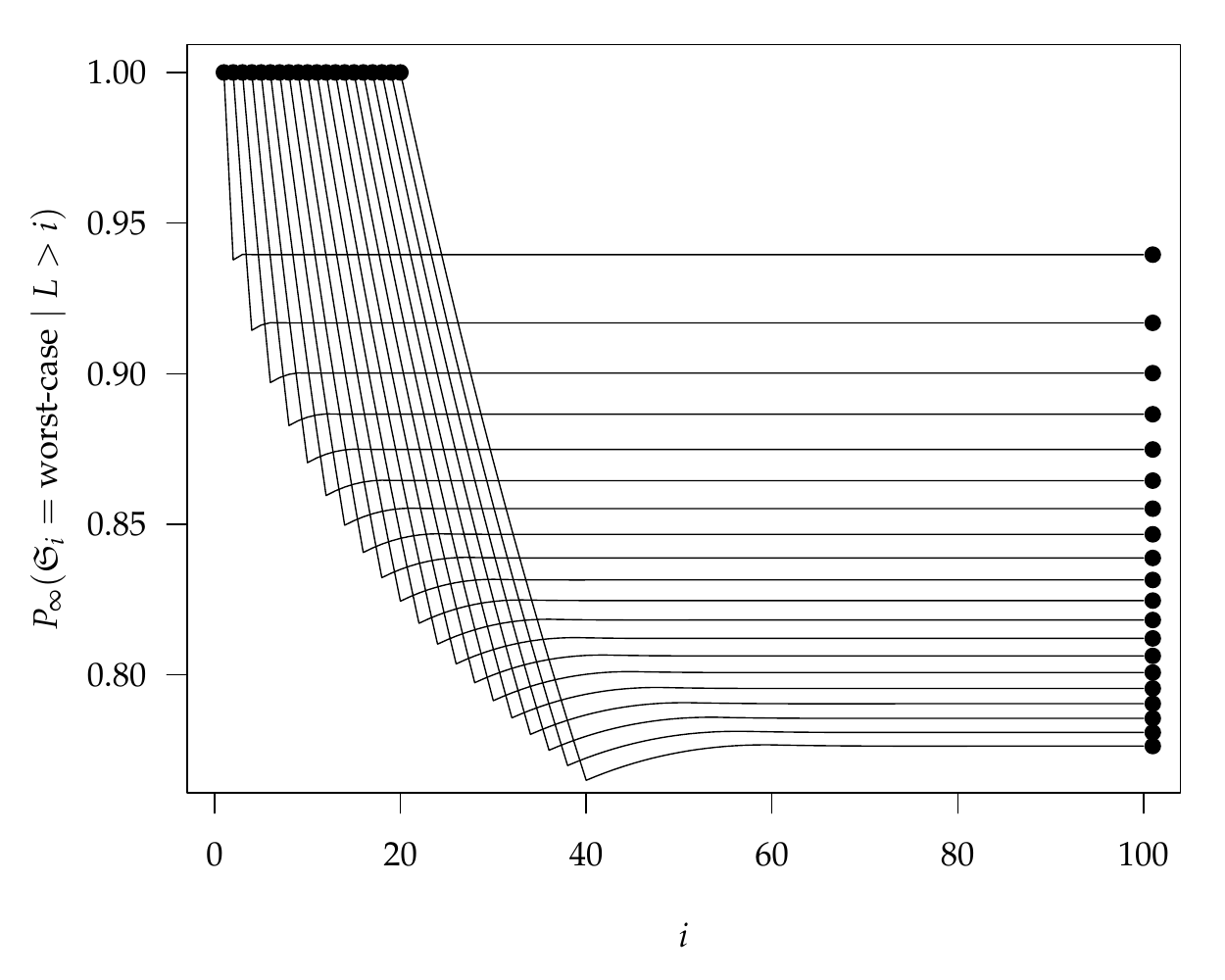} &	
  \includegraphics[width=.45\textwidth]{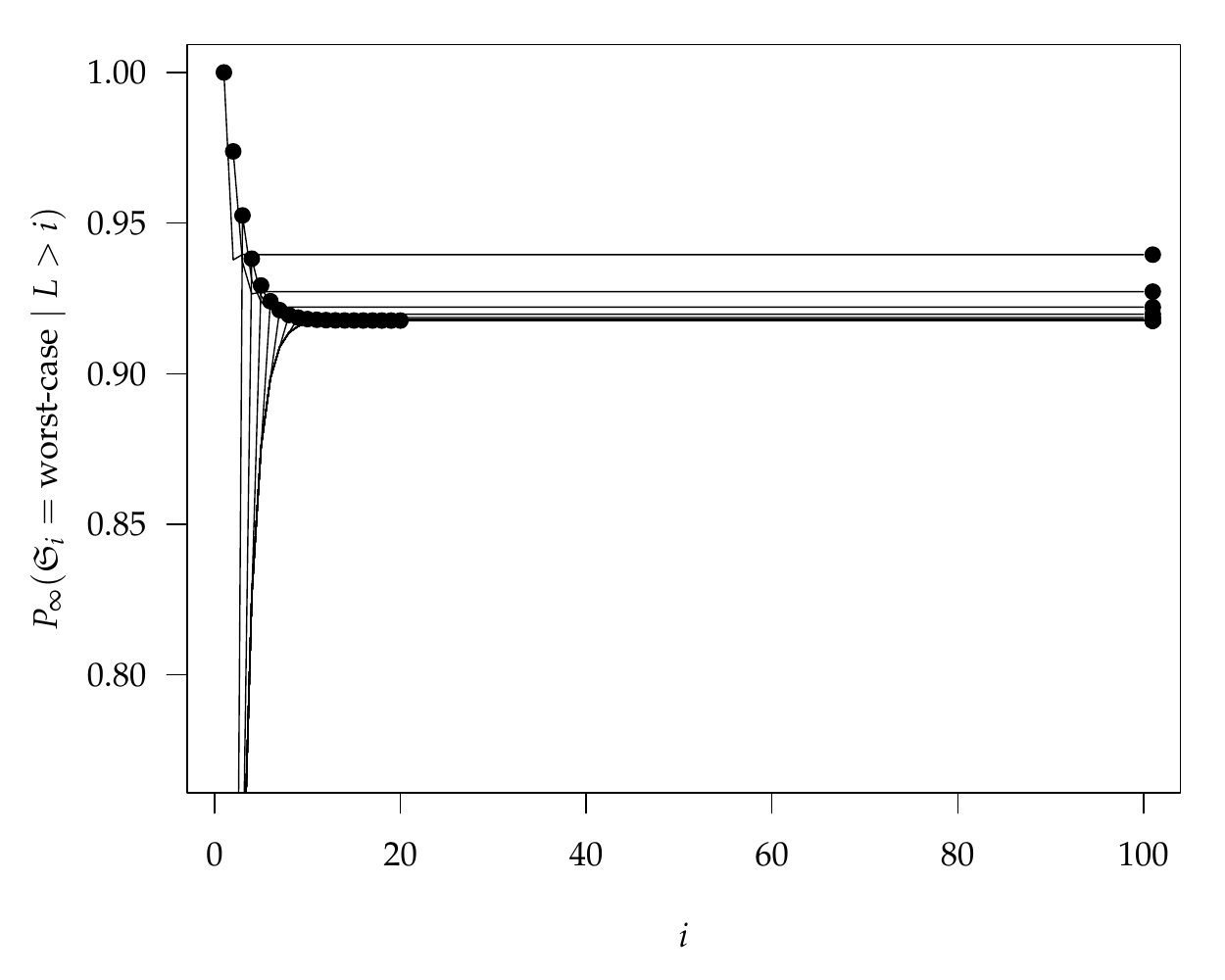}
\end{tabular}
\caption{Conditional (in-control) probability of being in the worst-case state,
$P_\infty(\mathfrak{S}_i = \text{worst-case}\mid L > i)$; in-control ARL 500, $H = 1, 2, \ldots, 20$,
the larger $H$ the lower the asymptotic level.} \label{fig:ppop}
\end{figure}

First, we observe that the detection performance gets better along $S_1$, \ldots, $S_4$.
Second, there is a pronounced difference between $S_4$ and the other synthetic-type charts.
For the latter, there is a clearly identifiable CED maximum at $\tau = H+1$.
Later we will learn about the root cause of this behavior (see Figure~\ref{fig:ppop}).
The larger $H$, the sharper is the increase from $\tau = 1$ to $\tau = H+1$.
In case of $S_4$ for all $H = 1, \ldots, 25$, stability of the $D_\tau$ is reached before $\tau = 10$.
However, the $S_4$ profiles are only a smoothed version of the other much more pronounced ones. Looking at the actual
numbers, we receive the same $\tau = H + 1$ as argument of the maximum.
Nonetheless, the $S_4$ version could be sufficiently well characterized by the zero-state and the steady-state ARL, whereas for the others the
inner maximum is important too, because it is considerably larger than the other two measures.
In Figure~\ref{fig:ced1}, we marked the profiles with the lowest zero-state and steady-state ARL, by
bold dashed and dash-dotted lines, respectively, and annotated the related $H$ value on the right-hand margin.
The $H$ for the minimum zero-state ARL ($H=21$, 13, 14) is substantially larger than for the steady-state one ($H=8$, 5, 5), except for $S_4$
(values are quite similar: $H = 19$ and 18).
For all synthetic charts with head-start, the zero-state ARL is markedly smaller than the steady-state ARL.
Therefore, judging these charts by only using zero-state ARL values is misleading. From all synthetic profiles we conclude
that the steady-state ARL is a much more representative measure than the more popular zero-state ARL, in particular for $S_4$.
Turning to the established competitor, we look at the EWMA profiles (two-dash and dotted line for $\lambda = 0.25$ and $= 0.1$, respectively).
These two profiles reside clearly below all synthetic-type chart counterparts.
Thereby, the $\lambda = 0.1$ EWMA is slightly better than the $\lambda = 0.25$ one (will change for larger $\delta$).
Eventually, the Shewhart chart ARL at $\delta = 1$ is with 54.58 too large to be seen in Figure~\ref{fig:ced1}.

We conclude that for $\delta = 1$, the ``old'' EWMA control chart exhibits the best performance.
Later we will see that the version with $\lambda = 0.25$ does a good job for all considered shifts.
For smaller shifts $\delta < 1$, the advantage of EWMA is even more pronounced.
Before looking at larger changes, we want to emphasize that $S_2$ and $S_3$ show nearly the same profiles,
with a slight advantage for the latter.

\begin{figure}[hbt]
\centering
\begin{tabular}{cc}
  \footnotesize $S_1$ & \footnotesize $S_2$ \\[-1ex]
  \includegraphics[width=.48\textwidth]{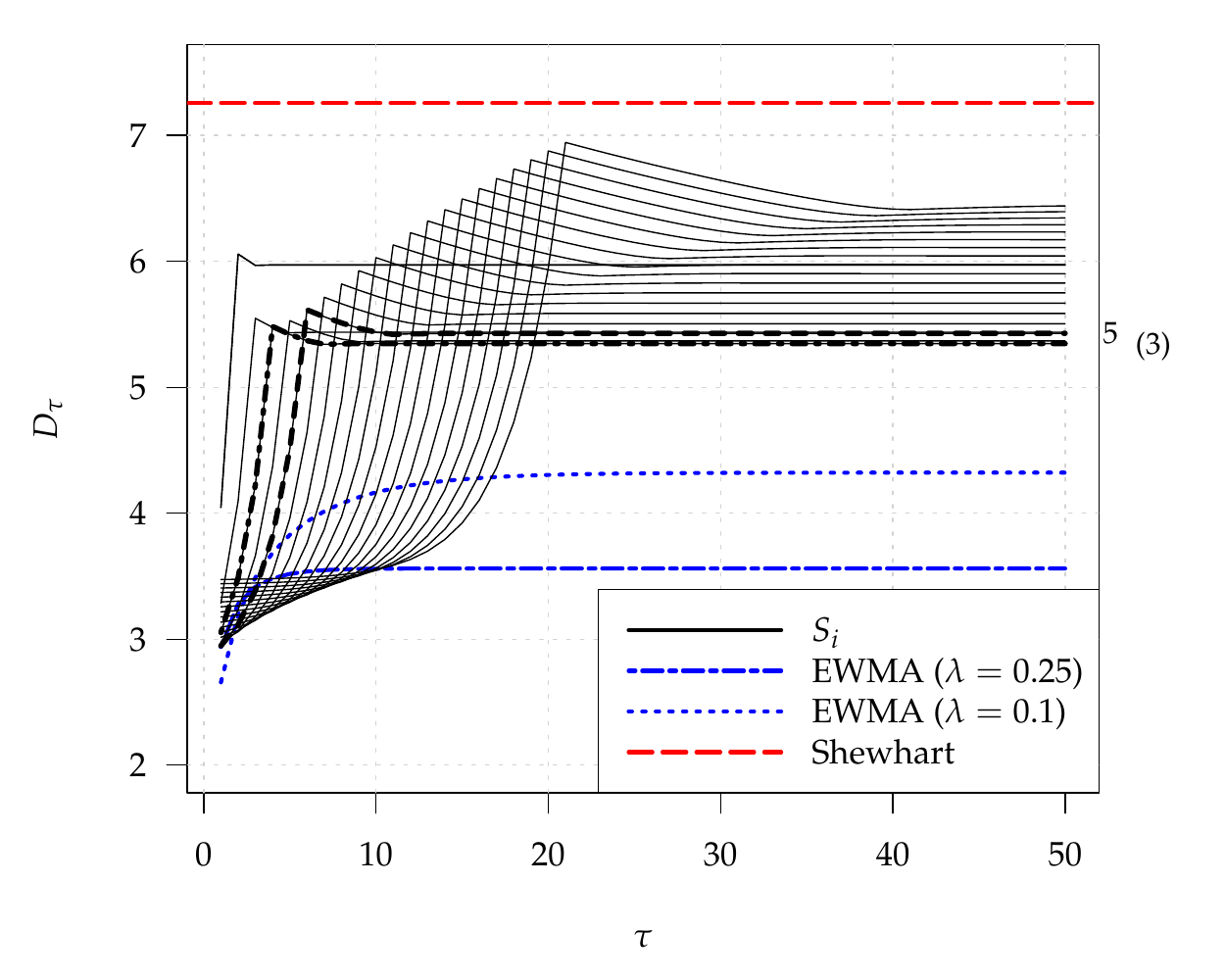} &	
  \includegraphics[width=.48\textwidth]{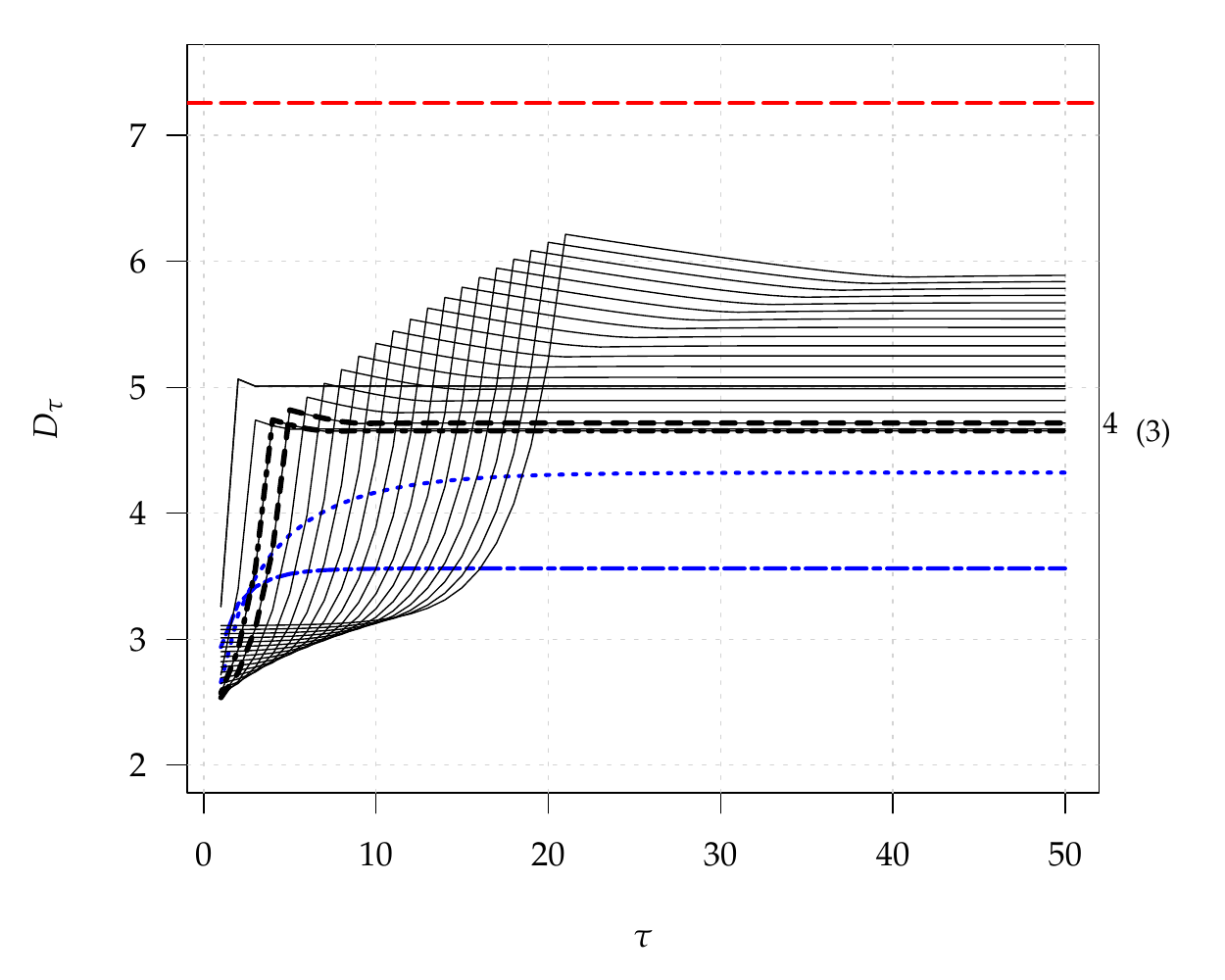} \\[1ex]
  \footnotesize $S_3$ & \footnotesize $S_4$ \\[-1ex]
  \includegraphics[width=.48\textwidth]{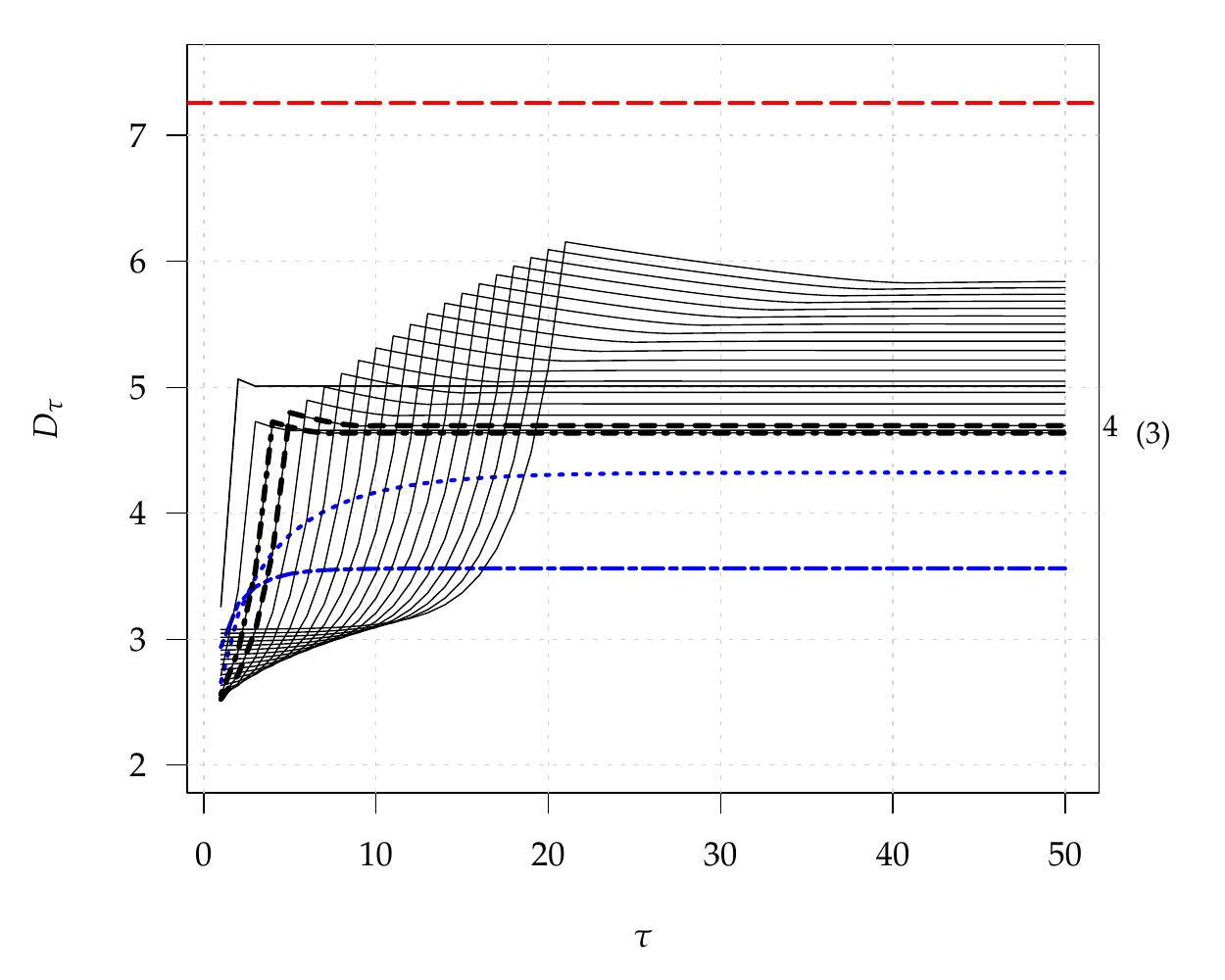} &	
  \includegraphics[width=.48\textwidth]{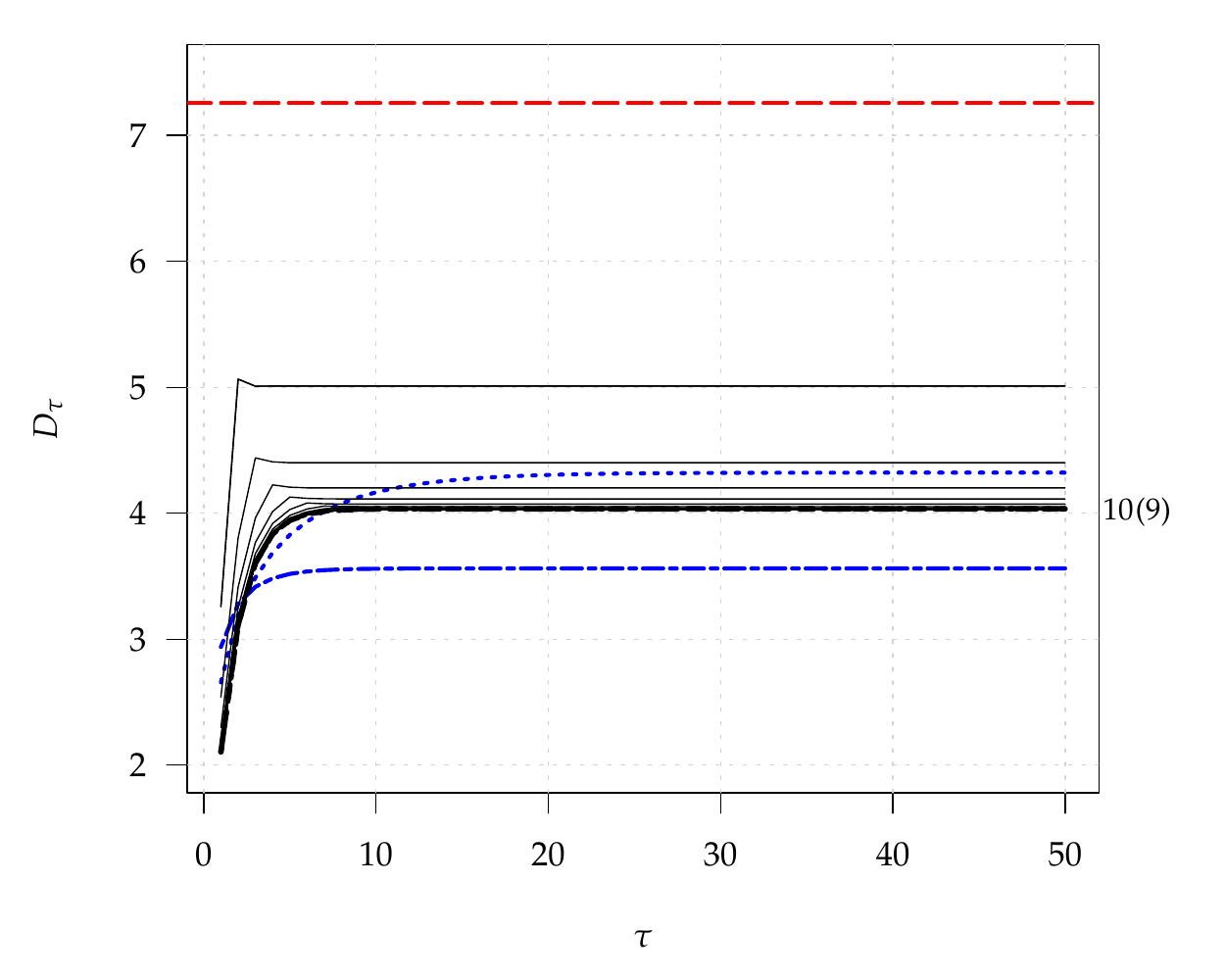}
\end{tabular}
\caption{$D_\tau$ profiles for four synthetic-type charts with head-start,
$H = 1, 2, \ldots, 25$, best scheme (zero-state and steady-state) bold (dashed and dash-dotted) lines,
shift $\delta = 2$, two EWMA charts; in-control ARL 500.} \label{fig:ced2}
\end{figure}

For the larger change $\delta = 2$ (see Figure~\ref{fig:ced2}), there are some clear overlappings between the synthetic and EWMA profiles.
However, for changes at $\tau > 20$ (remind here the in-control ARL 500), the EWMA chart with $\lambda = 0.25$
is again the clear winner (for $S_4$, $\tau > 3$ suffices).
Note that the synthetic-type chart ($S_1$, $S_2$, $S_3$) configurations which perform better than EWMA for $5\le \tau \le 20$,
exhibit heavily distorted performance for later changes, $\tau > 20$, which relegates them clearly from the competition.
The EWMA chart with the smaller $\lambda = 0.1$ can compete with $S_1$, $S_2$ and $S_3$, but not with most of the $S_4$ designs.
Thus, the actual competition is between the synthetic-type schemes and an EWMA chart with a mid-size $\lambda$.
Before discussing the profiles of the former more in detail, we want to note that all
charts behave better than the Shewhart chart (now its CED profile is visible).
Except for $S_4$, the differences between the profiles and within them are much more pronounced.
The optimal $H$ values are now smaller than for $\delta = 1$, that is we obtain
for the zero-state ARL $H=5$, 4, 4, 10 and for the steady-state ARL $H=3$, 3, 3, 9 for $S_1$,
\ldots, $S_4$, respectively. It is interesting that nearly the same $H$ makes the considered
ARL types minimal, for each chart type.
In sum we conclude that for $\delta = 2$, EWMA ($\lambda = 0.25$) is practically the best performing chart with
$S_4$ (and $H>2$) on the second place. For all four synthetic-type charts, choosing $H \in \{4, 5, 6, 7\}$
seems to be a good choice.

\begin{figure}[hbt]
\centering
\begin{tabular}{cc}
  \footnotesize $S_1$ & \footnotesize $S_2$ \\[-1ex]
  \includegraphics[width=.48\textwidth]{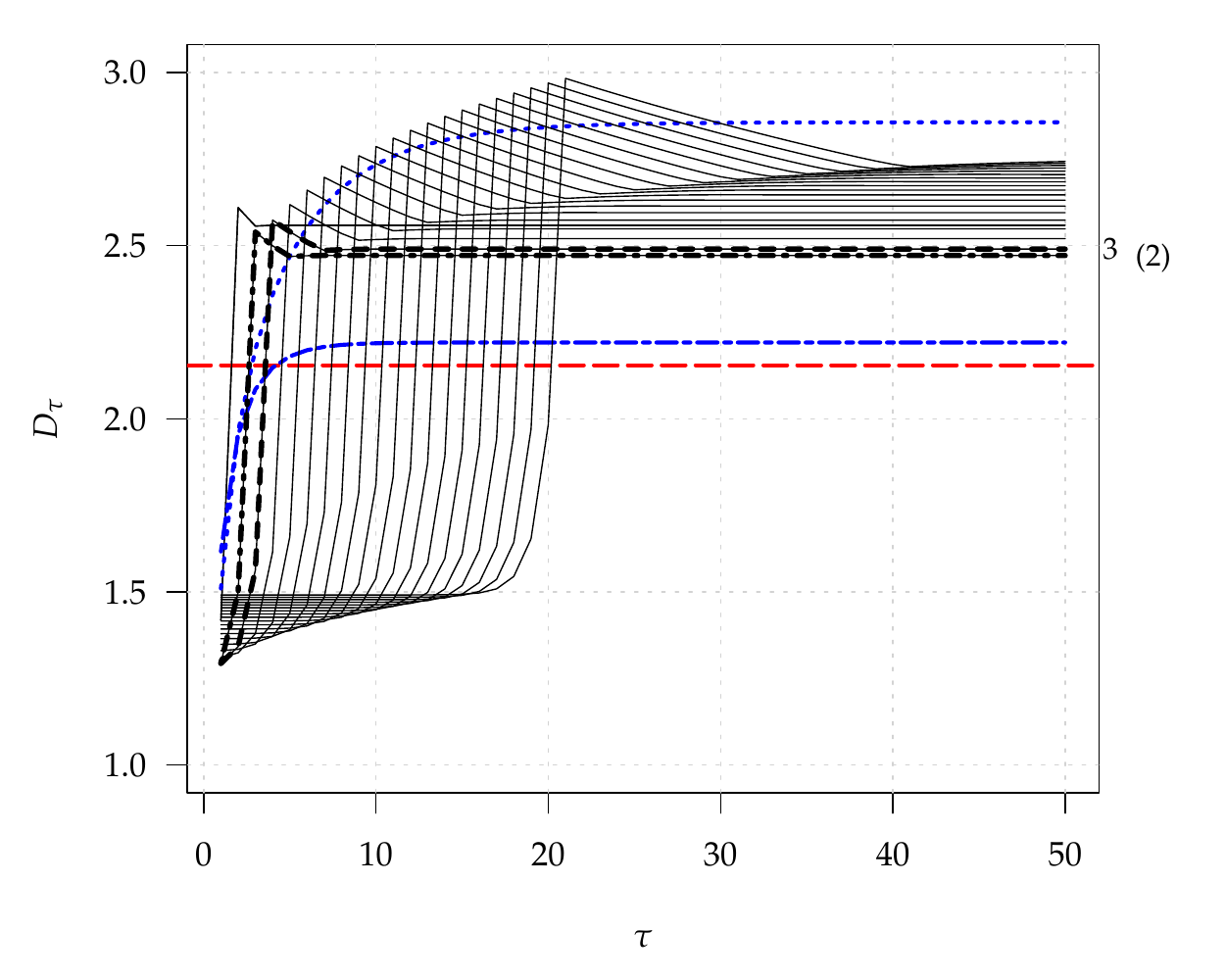} &	
  \includegraphics[width=.48\textwidth]{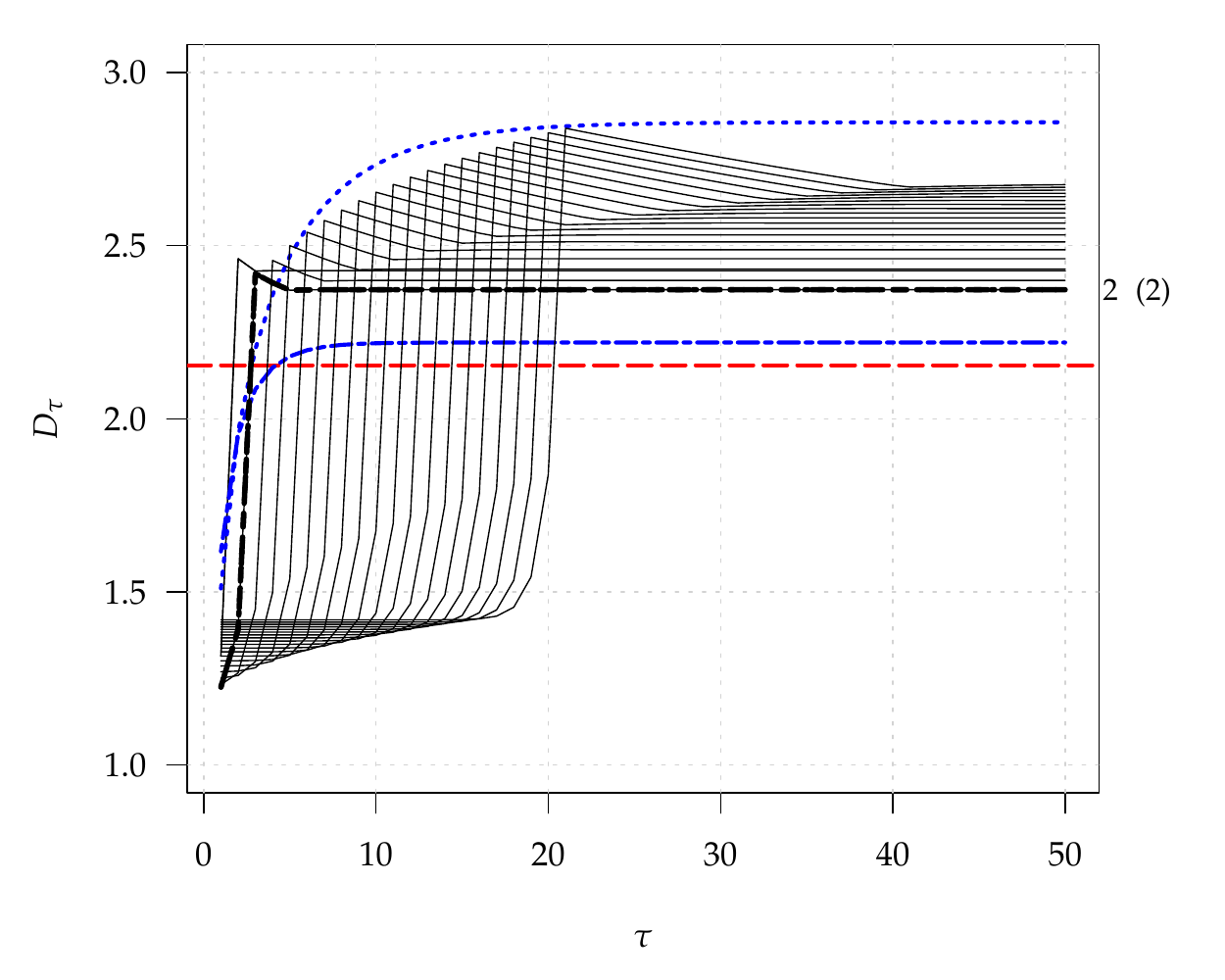} \\[1ex]
  \footnotesize $S_3$ & \footnotesize $S_4$ \\[-1ex]
  \includegraphics[width=.48\textwidth]{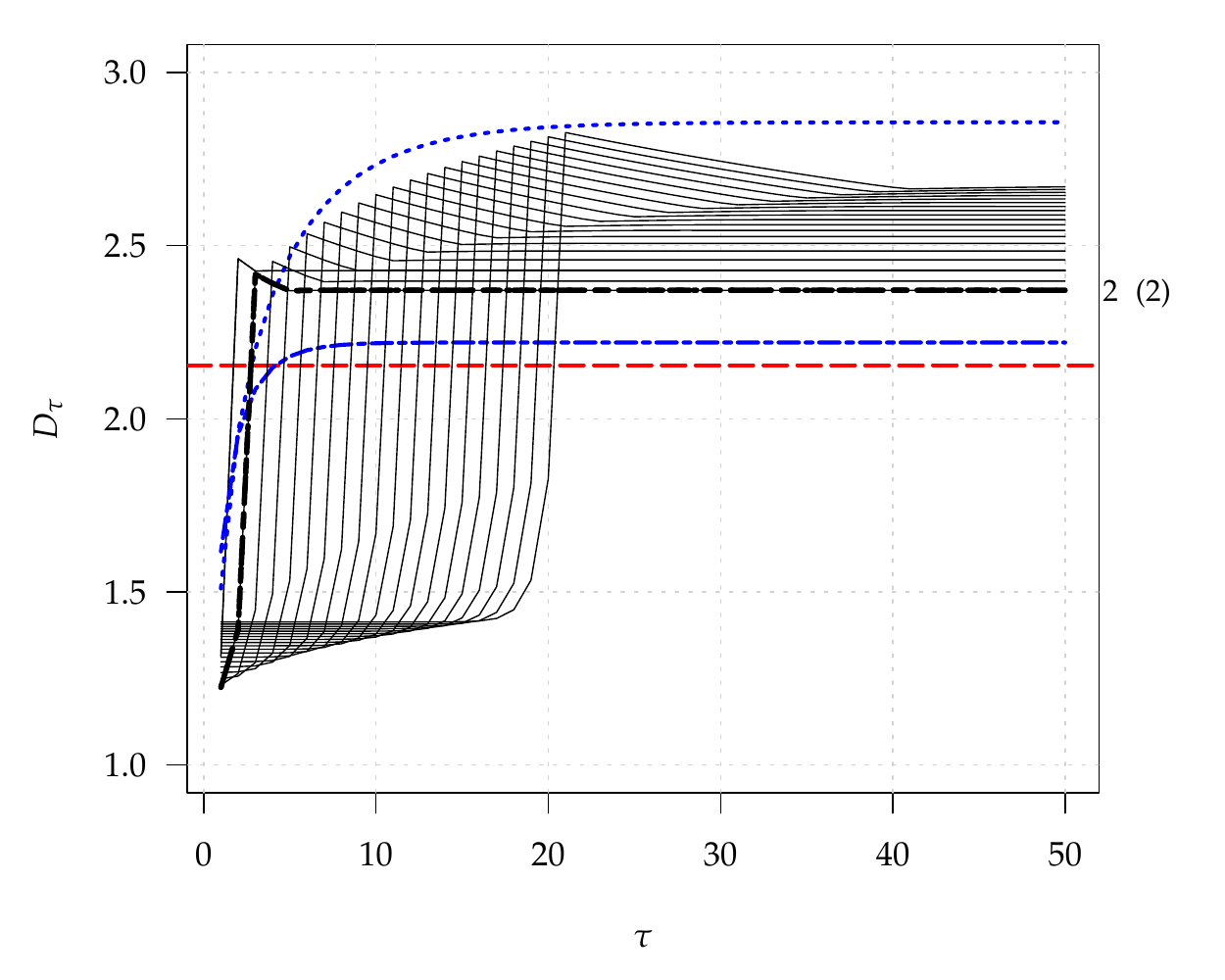} &	
  \includegraphics[width=.48\textwidth]{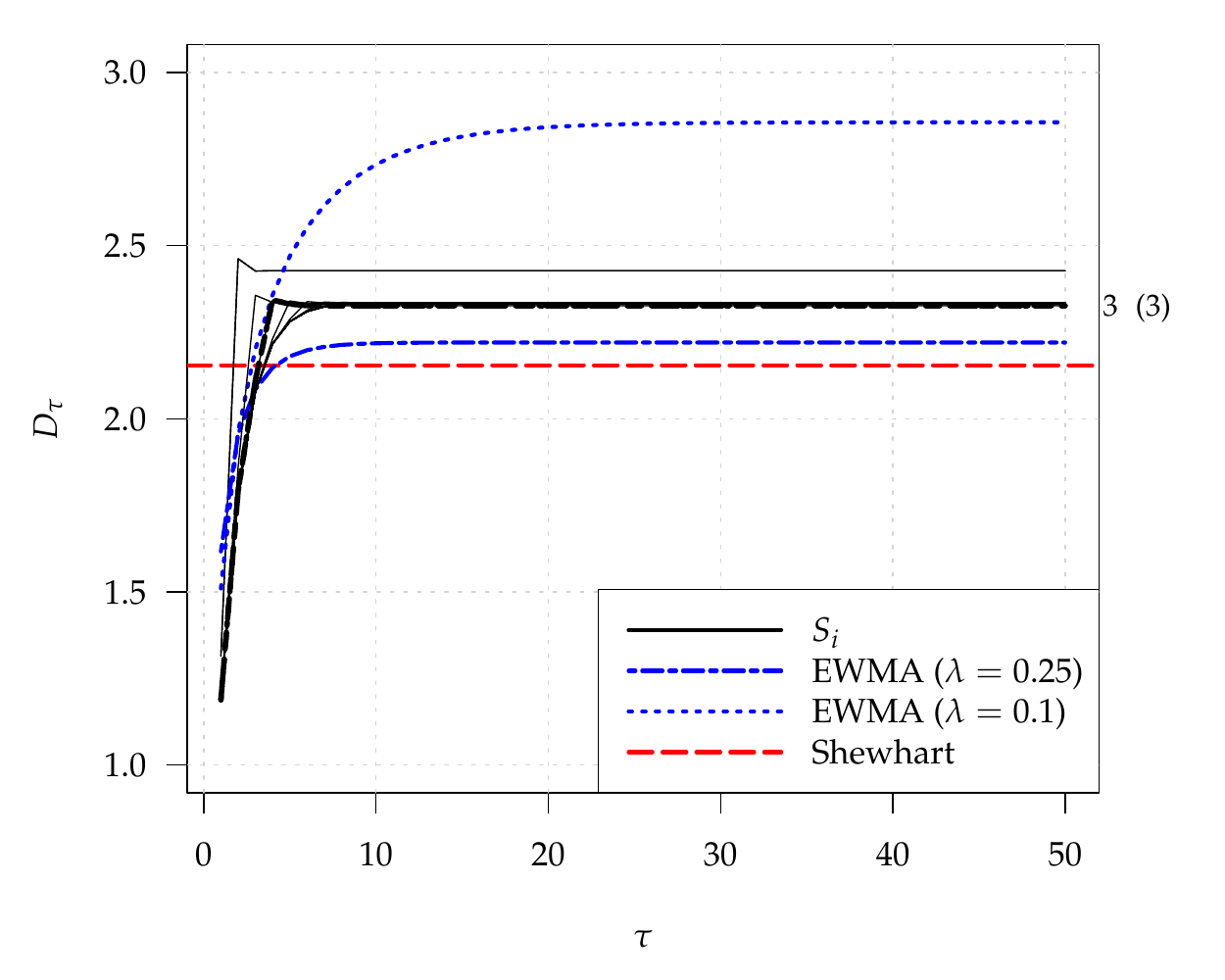}
\end{tabular}
\caption{$D_\tau$ profiles for four synthetic-type charts with head-start,
$H = 1, 2, \ldots, 25$, best scheme (zero-state and steady-state) bold (dashed and dash-dotted) lines,
shift $\delta = 3$, two EWMA charts; in-control ARL 500.} \label{fig:ced3}
\end{figure}

Next, we look at the change $\delta = 3$, where the Shewhart chart yields the smallest ARL (which resembles zero-state, steady-state and all CED
values). In Figure~\ref{fig:ced3} we see similar patterns as before in Figure~\ref{fig:ced2}. Starting with $S_1$, $S_2$ and $S_3$,
we observe the same (even much more) pronounced step shift of $D_\tau$
for $S_1$, $S_2$, $S_3$ at $\tau=H+1$. The best configuration, in terms of both ARL types, is either $H=2$ or $=3$.
The zero-state ARL (equal to $D_1$, of course) and the CED value $D_2$ are lower than for EWMA ($\lambda=0.25$), whereas
for $\tau > 2$, the latter chart exhibits the smallest $D_\tau$ including its limit, the steady-state ARL. Thus, again EWMA dominates over these three
synthetic-type charts clearly. The same has to be said about $S_4$ and EWMA ($\lambda=0.25$), except for the supplemental $D_3$,
where both feature similar values.
For all four synthetic-type charts with head-start we encounter, in case of the optimal setups with $H\in\{2,3\}$,
really low values for $E_1(L) = D_1$, $D_2$ and (only for $S_4$) $D_3$.
But for every change after $\tau = 3$, the EWMA ($\lambda = 0.25$) chart beats all other charts under study. Thus, synthetic-type charts
could be recommended only for the very special situation of early changes ($\tau \le 3$) with considerable magnitude ($\delta \ge 2$).
Else the classical EWMA chart with a mid-size $\lambda = 0.25$ is a better choice.
Taking the ARL of the Shewhart chart into account, we conjecture that even a combination of synthetic-type charts
and Shewhart charts \citep[called improved synthetic charts in][]{Raki:EtAl:2019} will not be much better, for changes $\delta < 2$.
Before we, however, provide our final judgment,
some more comparisons (for a larger set of $\delta$ values) focusing on zero- and steady-state ARL will be performed.
To make presentation more concise, we focus to type \#4 charts, i.\,e. AR and MSS or $R_4$ and $S_4$, respectively.
Thus we include as well the no head-start version (AR) proposed by \cite{Antz:Raki:2008a}.

To allow some overall judgment, we calculate ARL envelopes \citep{Drag:1994b} for $R_4$ and $S_4$. In detail, for each $\delta$ (on a rather fine grid) we pick
$H$ making the related out-of-control ARL minimal. The corresponding ARL values form the $R_4$ and $S_4$ envelope, respectively. In the here following Table~\ref{tab:Hopt}, we present some examples for these $H$.
\begin{table}[hbt]
\centering
\caption{Optimal values of $H$ for minimizing zero- and steady-state out-of-control ARL;
in-control zero-state ARL is set to 500.}\label{tab:Hopt}
\begin{tabular}{*{12}{c}} \toprule
  $\delta$ && 0.25 & 0.5 & 0.75 &  1 & 1.5 &  2 & 2.5 & 3 & 4 & 5 \\ \midrule
  && \multicolumn{10}{c}{zero-state} \\[0.5ex]
  $R_4$    &&   12 &  15 &   17 & 17 &  14 &  8 &   4 & 3 & 2 & 2 \\
  $S_4$    &&   12 &  15 &   18 & 19 &  15 & 10 &   6 & 3 & 2 & 2 \\ \midrule
    && \multicolumn{10}{c}{steady-state} \\[0.5ex]
  $R_4$    &&   12 &  15 &   17 & 17 &  14 &  9 &   5 & 3 & 2 & 4 \\
  $S_4$    &&   12 &  15 &   17 & 18 &  14 &  9 &   5 & 3 & 2 & 4 \\ \bottomrule
\end{tabular}
\end{table}
We obtained the results by searching over $H \in \{1,2,\ldots,200\}$. Because for small and mid-size $\delta$,
the ARL minimum is achieved for quite large $H$
while simultaneously the changes from $H = 5$ on are nearly negligible,
we replace the ``global'' $H$ by the smallest member of the above set, where the corresponding ARL is not larger by 0.1\% than the overall minimum.
We deployed the same approach for the values given in Figures~\ref{fig:ced1}, \ref{fig:ced2} and \ref{fig:ced3} for $S_4$.
It is not surprising that $R_4$ and $S_4$ exhibit nearly the same optimal $H$ values. Additionally,
aiming at small zero- and steady-state ARLs results in similar $H$ choices too.
In the appendix, we provide in Figure~\ref{fig:optimalH} two diagrams illustrating the dependence of the optimal $H$ from
$\delta$ in a more elaborated way. Here we want to emphasize that the actual choice of $H$ is not
really important, as long as it is not too small. Thus, some $5\le H\le 10$ does the job sufficiently
well. For the envelope, however, we use the best choice over $H \in \{1,2,\ldots,200\}$.

\begin{figure}[hbt]
\centering
\begin{tabular}{cc}
  \footnotesize zero-state & \footnotesize steady-state \\[-1ex]
  \includegraphics[width=.48\textwidth]{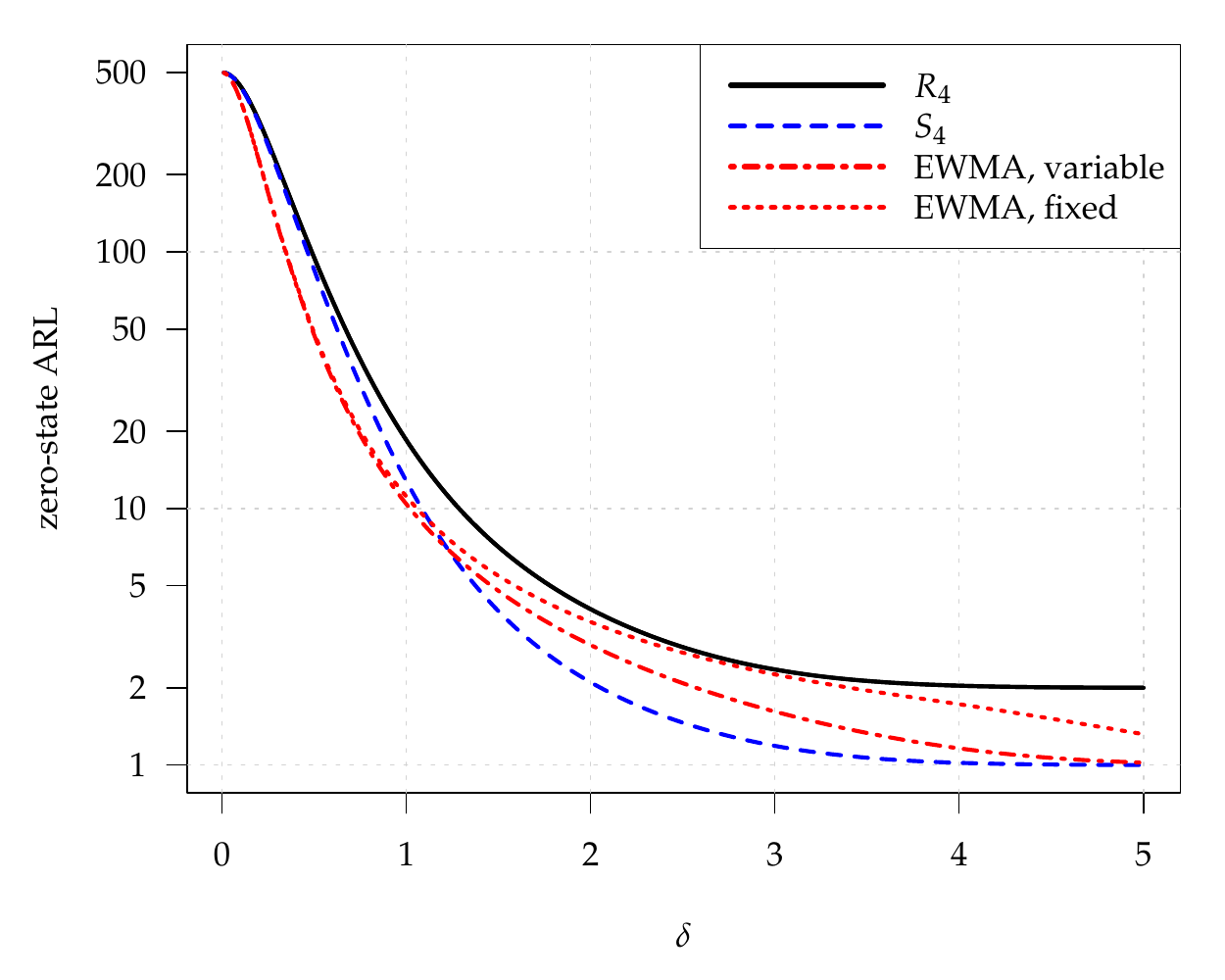} &	
  \includegraphics[width=.48\textwidth]{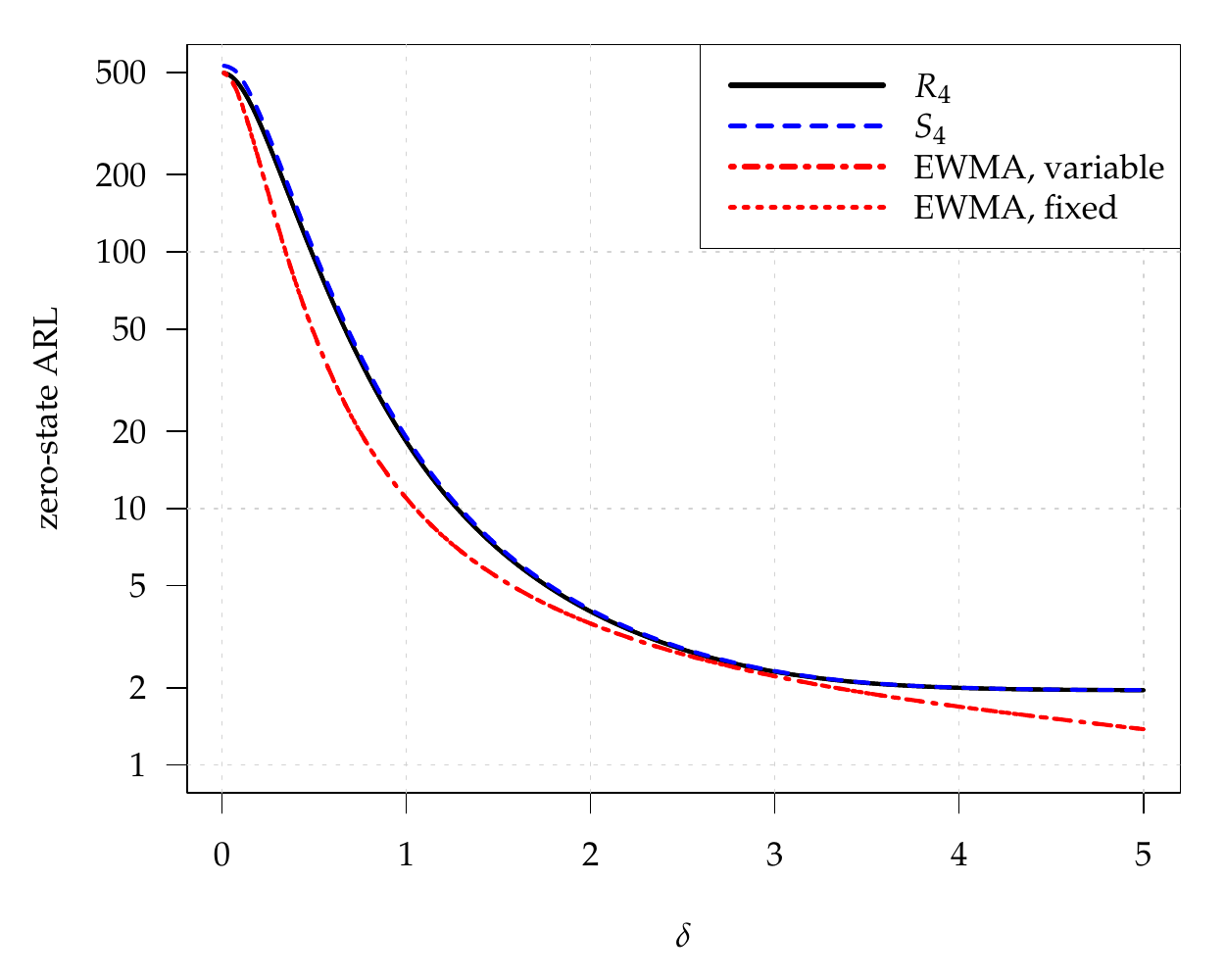}
\end{tabular}
\caption{ARL Envelopes (point-wise minimal, $ARL \to \min_{1\le H\le 200}$) of $R_4$ and $S_4$ (alias AR and MSS); in-control ARL 500;
EWMA (E) with $\lambda = 0.25$.} \label{fig:envelope}
\end{figure}

Turning now to Figure~\ref{fig:envelope} presenting the envelopes,
we want to note that
besides the already utilized EWMA chart with alarm rule \eqref{eq:ewmarule} we deploy as well an EWMA chart with constant limits, namely
with alarm condition $\tilde{c}_E \sqrt{\lambda/(2-\lambda)}$ relying on the asymptotic standard deviation of the EWMA statistics.
Thereby, the factor $\tilde{c}_E = 2.998$ is slightly smaller than $c_E = 3.000$ in \eqref{eq:ewmarule} for
the same in-control zero-state ARL ($A=500$).
The fixed limits EWMA as the antagonist of $R_4$ is more popular in SPM literature and related software packages, because
the ARL is more feasible (numerically). In Figure~\ref{fig:envelope}, we consider $0 < \delta \le 5$.
From the envelope diagram for the zero-state ARL we may conclude the popular statement that synthetic-type charts, here $S_4$ in particular,
perform well for change sizes $\delta > 1$. These statements, however, are only valid for the head-start versions $S_i$. From Figures~\ref{fig:ced1},
\ref{fig:ced2} and \ref{fig:ced3} we know that this advantage vanishes as soon the change does not take place during the first few (less than 10)
observations. That is, for most of the change point positions, the steady-state ARL is much more representative.
Looking at the corresponding ARL envelope on the right-hand side of Figure~\ref{fig:envelope} we conclude that EWMA with $\lambda = 0.25$
uniformly dominates the \textbf{point-wise} best $R_4$ and $S_4$ configurations. Besides, now the charts with head-start
($S_4$, EWMA with \eqref{eq:ewmarule} limits)
and without head-start ($R_4$, EWMA with fixed limits) behave alike. Interestingly, the steady-state ARL values for $2\le \delta \le 3$
do not differ considerably between EWMA and the \#4 charts.
But for smaller \textbf{and} larger values of $\delta$, EWMA performs much better than $R_4$/$S_4$.
While there is some remedy for the large values of $\delta$, nothing helps to improve the synthetic-type charts for changes smaller than $\delta \le 2$.
The dominating competitor is a standard EWMA chart with $\lambda = 0.25$, which could be even tuned to improve either the performance for smaller or larger
$\delta$. Eventually we want to remember that changes of size $\delta \ge 3$ constitute the realm of Shewhart control charts. In
sum, synthetic-type charts (here \#4) feature a decent detection performance for mid size changes, uniformly dominated by common EWMA control charts
and partially overshadowed by Shewhart charts.

\begin{figure}[hbt]
\centering
\begin{tabular}{cc}
  \footnotesize $R_4$, zero-state & \footnotesize $S_4$, zero-state \\[-1ex]
  \includegraphics[width=.48\textwidth]{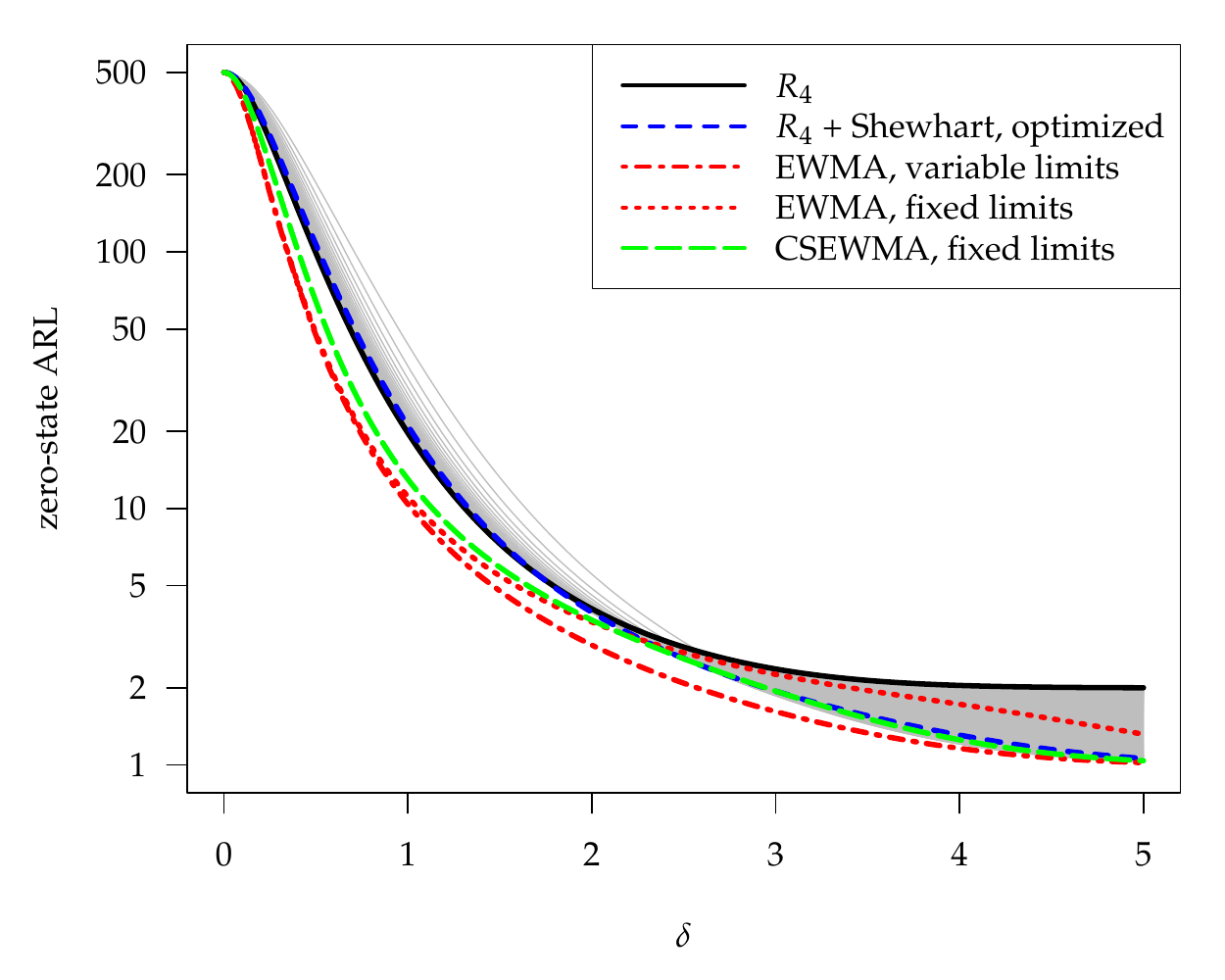} &	
  \includegraphics[width=.48\textwidth]{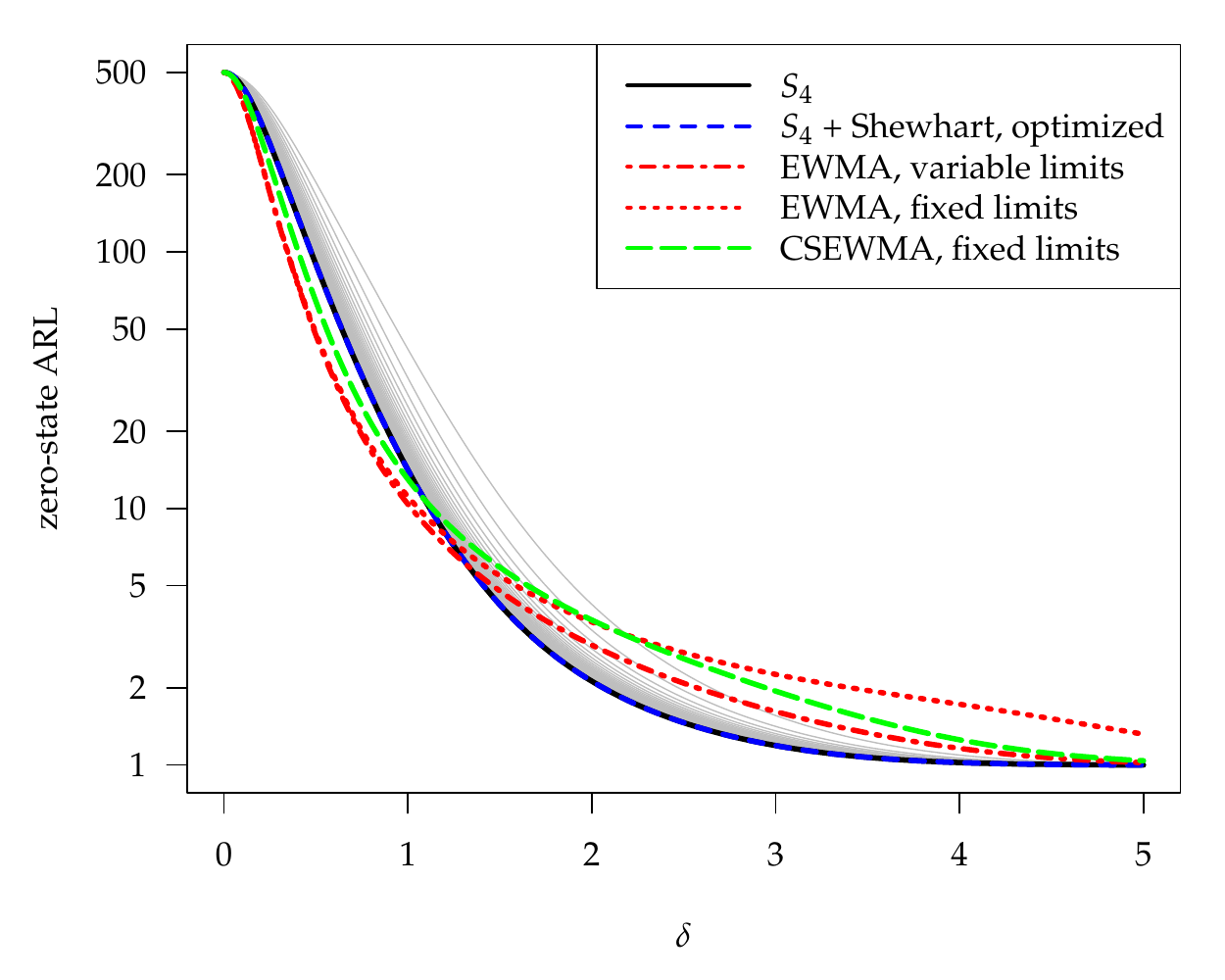} \\[1ex]
  \footnotesize $R_4$, steady-state & \footnotesize $S_4$, steady-state \\[-1ex]
  \includegraphics[width=.48\textwidth]{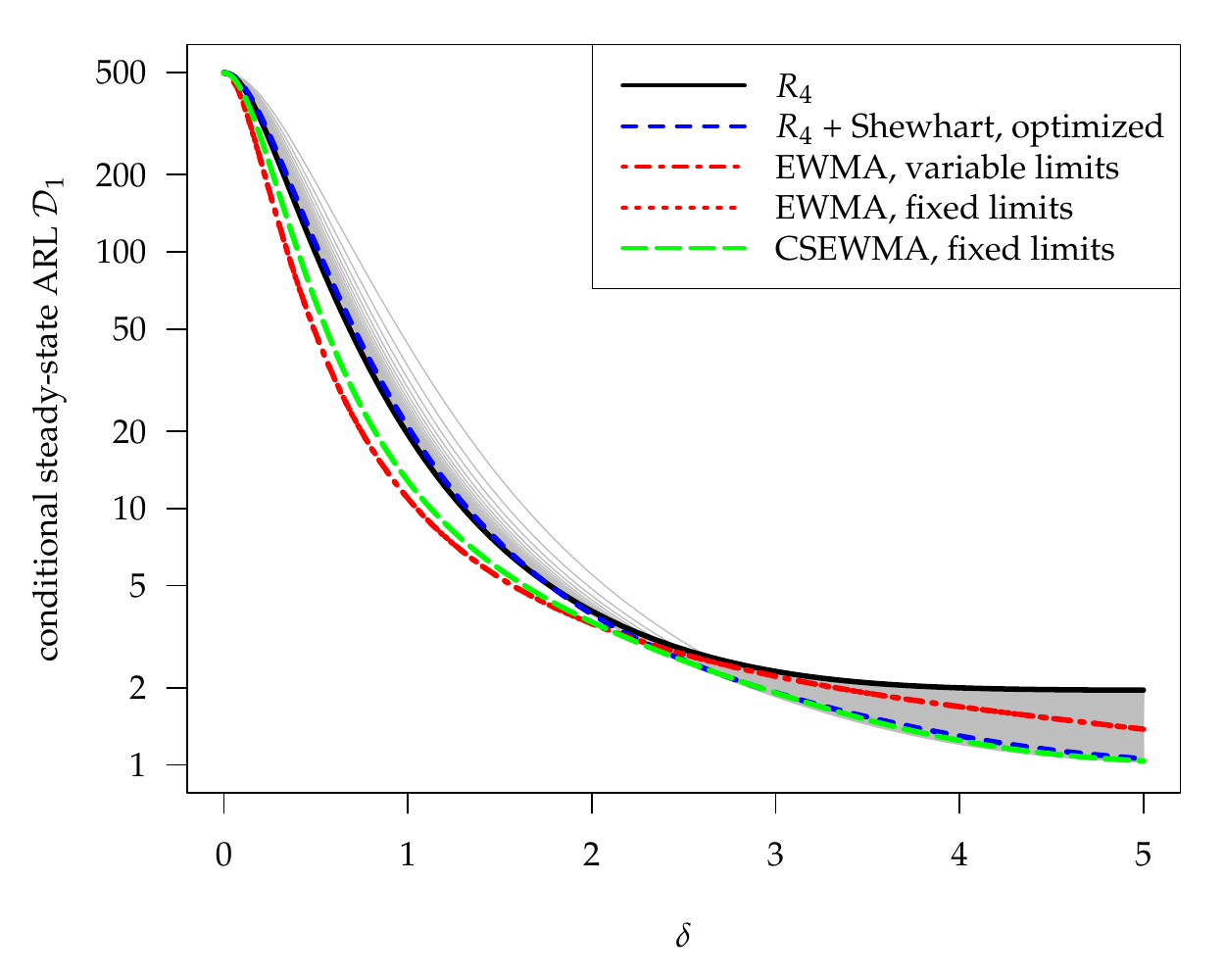} &	
  \includegraphics[width=.48\textwidth]{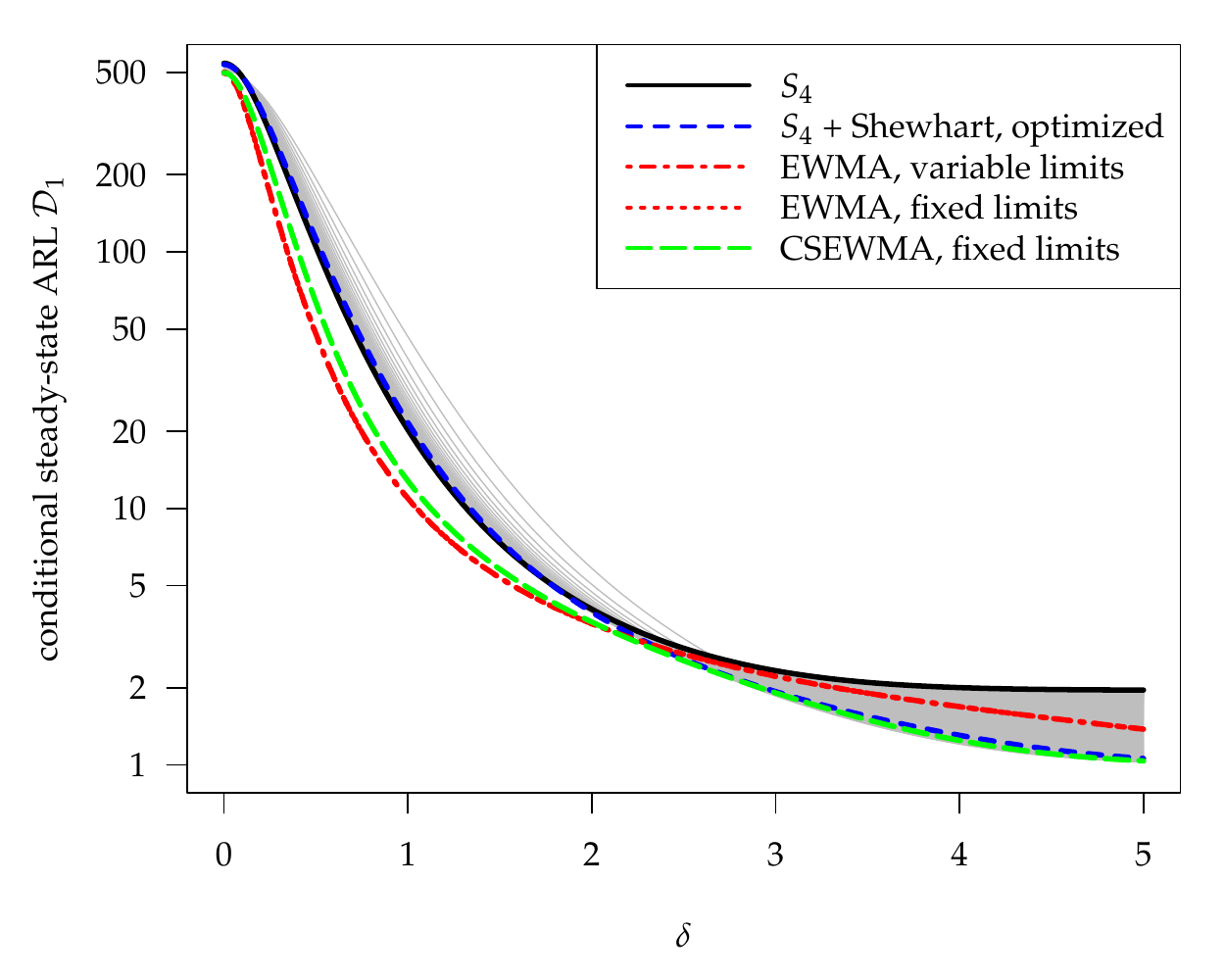}
\end{tabular}
\caption{ARL performance of single charts and combos; \#4 charts with $H=6$, EWMA with $\lambda = 0.25$; in-control ARL 500.} \label{fig:combos}
\end{figure}

Next, we want to deal with the combination of synthetic-type charts and Shewhart charts, which was proposed in \cite{Raki:EtAl:2019}, and earlier
in \cite{Wu:Ou:Cast:Khoo:2010} and \cite{Shon:Grah:2016a}. As we will later see, it improves the out-of-control steady-state ARL
results for $\delta > 3$, helping to close the gap between the right tails in Figure~\ref{fig:envelope}.
The adoption of the Markov chain models applied for all synthetic-type charts
for incorporating the Shewhart limit is straightforward \citep{Shon:Grah:2016a, Shon:Grah:2017a}.
It is more difficult for Shewhart-EWMA charts, but the Markov chain approximation described in \cite{Luca:Sacc:1990a}
works sufficiently well. We deploy here the more accurate algorithm introduced by \cite{Capi:Masa:2010a}.
In order to illustrate the potential impact of adding the Shewhart rule, we consider for $R_4$ and $S_4$ the case $H = 6$,
which is a reasonably general choice. Besides the above single EWMA charts with $\lambda = 0.25$ (exact and fixed limits),
we consider a Shewhart-EWMA combo with $\lambda = 0.25$, Shewhart limit $k_2 = 3.25$ and EWMA threshold $\tilde{c}_E = 3.2097$
(in-control ARL 500), where the EWMA component features constant limits (otherwise ARL calculation becomes more complicated).
For the two synthetic-type charts we look at many combinations of $(k_1, k_2)$, where $k_1$ replaces $k$ in \eqref{eq:kpqr} and
$k_2$ is again the Shewhart limit (of course, $k_2 > k_1$). We start with $k_2 = 3.1$ (the limit for a standalone Shewhart
chart with in-control ARL 500 is $k_2 = 3.09$) and increase it by $0.02$ steps (up to $7$). The $k_1$ of the inner synthetic rule is
determined (for $H=6$) to attain the in-control zero-state ARL 500 of the combo.
The resulting bundles of Shewhart-\#4 charts provide the grey areas in Figure~\ref{fig:combos}, where two
members are highlighted.
The black solid line marks the original pure \#4 chart, whereas the blue dashed lines
presents an optimal member. Optimal means here, that the measure $EQL = 1/\delta_\text{max} \sum_i \delta_i^2 ARL_i$
is minimized ($ARL_i$ is the out-of-control ARL for shift $\delta_i$). The utility function $EQL$ was used in \cite{Shon:Grah:2016a} in order to evaluate the detection
performance over a range of shifts with a single number. We set $\delta_\text{max}=5$ and $\delta_i = 0.01 i$.
The impact of small shifts (our $\delta_1 = 0.01$ is really small) to $EQL$ is rather minuscule because of the
weights $\delta_i^2$. The resulting Shewhart limits $k_2$ are 4.78 for $S_4$ in case of the zero-state ARL,
3.46 in case of the steady-state ARL, and 3.48 for $R_4$ in both cases. Only the first value, $k_2=4.78$ sticks out,
which does not really surprise because the zero-state performance of the head-start scheme $S_4$ is already sound.
The other numbers are quite similar. And for $R_4$ it is not important, what ARL measure is utilized.
We recognize the improvement potential for shifts $\delta \ge 3$ (the Shewhart realm). The best Shewhart-\#4 charts
exhibit profiles that are slightly lifted for changes $\delta\le 2$ and substantially lowered for $\delta \ge 3$.
For the Shewhart-$R_4$ combo we observe quite similar patterns for the zero-state and the steady-state ARL.
It is different for Shewhart-$S_4$, where many members of the aforementioned Shewhart-$S_4$ family yield agreeably small
ARL values for $\delta > 1$. We notice as well that the standalone EWMA chart with exact limits, see \eqref{eq:ewmarule},
exhibits the best uniform performance within the EWMA charts. But the other two entertain constant limits,
which results in higher zero-state ARL values by construction.
However, the more interesting comparison is the one for
the steady-state ARL. And here we conclude that all three EWMA designs behave better for changes $\delta < 2$, again.
For changes $2\le\delta\le3$, all considered charts perform similarly. And for large changes, $\delta > 3$,
the combo schemes (Shewhart-\#4 and Shewhart-EWMA) are the best ones and display roughly the same performance.
Then the two standalone EWMA charts follow, as we observed already in Figure~\ref{fig:envelope}.
The worst chart types are the standalone \#4 charts ($R_4$/$S_4$).
In summary, the merge of Shewhart with synthetic-type charts helps to close the $\delta > 3$ ARL gap.
However, the Shewhart-EWMA combo shows much better performance for changes $\delta \le 2$, whereas for larger changes
it behaves like the optimal Shewhart-\#4. Thus, a clear recommendation could be given:
Use either single EWMA or Shewhart-EWMA combo charts.

Finally, we want to stress the expedience of choosing the so-called wildcard head-start in contrast to the standard setup, which
was chosen without further ado for the $R_i$ charts, namely DR, KL, MC1 and AR by \cite{Derm:Ross:1997}, \cite{Klei:2000a}, \cite{Mach:Cost:2014b}
and \cite{Antz:Raki:2008a}, respectively. While the majority of the runs rules chart literature picked this initial state, which resembles
the worst-case (maximum out-of-control ARL), \cite{Wu:Sped:2000a} started a movement to chose the best-case state.
In the here following Figure~\ref{fig:ppop} we illustrate, how quickly the control charts ``return'' to the worst-case after starting
in the best-case.
For an in-control ARL of 500 we plot the probability that after $i$ observations the synthetic chart arrives in the worst-case state.
With $\mathfrak{S}_i$ we denote the state at observation $i$. From Table~\ref{tab:classStates} we know the number of possible states
(for the simple \#1 chart, we observe 0 and $H$ as best- and worst-case state following the notation in \eqref{eq:kpqr},
that is, picking the states from the set $\{0,1,\ldots,H\}$).
To improve presentation, we started plotting at this $i$, where the said probability is positive. Interestingly, for $S_1$, $S_2$ and $S_3$,
we obtain $P(\mathfrak{S}_i = \text{worst-case}\mid L > i) = 0$ for $i < H$ and $P(\mathfrak{S}_H = \text{worst-case}\mid L > H) = 1$.
If there is no (false) alarm during the first $H$ observations, then we reach the worst case state with probability 1 at the $H$th
observation. Thus, the behavior of the head-start and the common design differs substantially only during the first $H$ observations.
Then the head-start type chart arrives in the worst-case state with (conditional) probability one. The common chart started
in the worst-case with probability one, but returns to it at index $H$ with a (conditional) probability, which is quite large, but smaller than one.
The bullets at the end of all profiles in Figure~\ref{fig:ppop} mark the conditional steady-state probability of the worst-case.
For all four chart types and all considered $H=1,2,\ldots,20$ the convergence to the latter values is quick.
Eventually, we put a bullet too at $P(\mathfrak{S}_H = \text{worst-case}\mid L > H)$ for all four chart types.

The $S_4$ chart differs slightly from the other ones. First, only for $H=1$ we observe $P(\mathfrak{S}_H = \text{worst-case}\mid L > H) = 1$.
For larger $H$, we neither get long series of zero probabilities (from $i=2$ on the probability is positive) nor
the probability one at $i = H$.
But more importantly, the dominating probability value is about 92\%. Thus the best design among all considered synthetic-type charts
with head-start, namely $S_4$, exhibits two faces: (i) It shows an excellent zero-state ARL profile, cf. to Figure~\ref{fig:envelope}. (ii)
But these low values are highly untypical facets of $S_4$, because it operates with a probability of more than
90\% in worst-case mode. Thus, a thorough and legitimate judgment of the $S_4$ chart would rely
on the ARL numbers we know for the no head-start version, i.\,e. for $R_4$. Another way of avoiding misjudgment is, of course,
considering the steady-state ARL. Finally we should mention that for $S_1$, $S_2$ and $S_3$ a similar statement could be given, because
the deplorable probability of being in the worst-case is not much smaller, it is for all considered configurations larger than 75\%.

\section{Conclusions} \label{sec:Concl}

Of course, synthetic-type charts ($R_1$, ..., $S_4$) are easy to build and to analyze. In particular, for the analysis one
can easily apply exact Markov chain models. 
For the simplest ones, $R_1$ and $S_1$, there are even explicit solutions for all considered measures.
EWMA control charts, however, are similarly easy to use. Their ARL
analysis needs more computational power, which is not a problem nowadays. Detection performance-wise, a clear recommendation
can be given: Apply EWMA, because it exhibits the best detection performance for small changes $\delta \le 1.5$ (in terms
of standard deviation) in our study, whereas for larger changes all the considered schemes differ not much. Without an added Shewhart rule,
synthetic-type charts perform worse than EWMA even for large changes ($\delta>3$). Adding this Shewhart rule improves
the large change detection behavior a lot, for both synthetic-type (here we focus to $R_4$ and $S_4$, the most recent phenotypes)
and EWMA control charts. Finally we want to urgently emphasize that for a sound analysis of a control chart device
dealing with the steady-state ARL is adamant. Naturally, a worst-case ARL analysis would be appropriate too.
The $R_i$ charts are designed through the lens of their worst-case ARL. In Appendix~\ref{app:R4vsCUSUM}
a rough comparison between $R_4$ and CUSUM \citep[cumulative sum charts introduced in][]{Page:1954c}
control charts (the worst-case ``experts'') is provided. Again, the older charts (CUSUM)
yield better ARL results. In summary, synthetic-type charts are somewhat easier to setup than the classical charts
such as EWMA and CUSUM, but the older ones exhibit the better detection performance.

\bibliographystyle{apalike} 
\bibliography{synthetic_another}

\appendix

\section{Appendix}

\subsection{Explicit formulae for steady-state ARL for $S_1$ ($R_1$) and its limit for $\delta \to 0$} \label{app:steadyARL}

Because in \cite{Shon:Grah:2016a} the steady-state ARL was deployed to determine $k$, we look 
at the most simple case, namely $S_1$ (and implicitly $R_1$) more thoroughly, augmenting somehow  \cite{Shon:Grah:2017a, Shon:Grah:2019}.
We consider all $\psi_i$, $i = 1, 2, 3, 4$ (see Section~\ref{sec:ststARL}).
The actual shift $\delta$ is added as subscript, for example $\bm{\psi}_{1;0}$ denotes the in-control case $\delta = 0$.
\begin{align*}
  \intertext{Conditional steady-state ARL, cf. to \cite{Knot:2016a}:}	
  \mathcal{D}_1 = \bm{\psi}_{1;0}^\prime\,\bm{\ell}_\delta & =
	\left( \frac{\varrho_0}{p_0} + \frac{1-\left(\frac{q_0}{\varrho_0}\right)^H}{1 - \frac{q_0}{\varrho_0}} \right) \frac{s_0}{p_\delta}
	+ \left( \frac{\varrho_0}{p_0} + q_\delta^H \frac{1 - \left(\frac{q_0}{q_\delta \varrho_0}\right)^H}{1 - \frac{q_0}{q_\delta\varrho_0}}\right)
	\frac{s_0}{r_\delta} \quad \xrightarrow[\delta\to 0]{} \quad \frac{1}{1-\varrho_0} \,.
  \intertext{Cyclical steady-state ARL, re-start at state $0$, cf. to \cite{Wu:Ou:Cast:Khoo:2010, Knot:2016a}:}
	\mathcal{D}_2 = \bm{\psi}_{2;0}^\prime\,\bm{\ell}_\delta & =
	\frac{1 + q_0 p_\delta \frac{q_0^H - q_\delta^H}{q_0-q_\delta}}{r_\delta} \quad \xrightarrow[\delta\to 0]{} \quad \frac{1 + H p_0 q_0^H}{r_0}
	= \frac{1}{r_0} + H \frac{q_0^H}{1-q_0^H} \,.
  \intertext{Cyclical steady-state ARL, re-start at state $H$, cf. to \cite{Shon:Grah:2019}:}
	\mathcal{D}_3 = \bm{\psi}_{3;0}^\prime\,\bm{\ell}_\delta & =
	\frac{1-q_\delta^H}{r_\delta} +  \frac{1 + p_0 q_\delta \frac{q_0^H - q_\delta^H}{q_0-q_\delta}}{r_\delta (2-q_0^H)}
	\quad \xrightarrow[\delta\to 0]{} \quad \frac{1-q_0^H}{r_0} + \frac{1 + H p_0 q_0^H}{r_0(2-q_0^H)} \,.
  \intertext{Wrong conditional steady-state ARL, cf. to \cite{Shon:Grah:2017a, Shon:Grah:2019}:}	
    \mathcal{D}_4 = \bm{\psi}_{4;0}^\prime\,\bm{\ell}_\delta & =
    \frac{1-q_\delta^H}{r_\delta} + \frac{1 + p_0 q_\delta \frac{1-q_\delta^H}{1-q_\delta}}{r_\delta(1+Hp_0)}
    \quad \xrightarrow[\delta\to 0]{} \quad \frac{1-q_0^H}{r_0} + \frac{1+q_0(1-q_0^H)}{r_0 (1+Hp_0)} \,.
\end{align*}
Next we apply the above formulas for a $S_1$ chart with $H=3$ and $k=2.2238$ (in-control
ARL 500). Besides the four different steady-state ARL results we show as well the zero-state ARL of an $R_1$
(alias true synthetic chart without head-start) in the following Table~\ref{tab:icARLvar}.
\begin{table}[hbt]
\centering
\caption{In-control ARL for $S_1$ and $R_1$, zero-state $\ell$ and steady-state $\mathcal{D}$, $H=3$, $k=2.2238$.}\label{tab:icARLvar}
\begin{tabular}{ccccccc} \toprule
  $\ell_{S_1}$ & $\ell_{R_1}$ && $\mathcal{D}_1$ & $\mathcal{D}_2$ & $\mathcal{D}_3$ & $\mathcal{D}_4$ \\ \midrule
  500 & 538.224 && 536.378 & 536.242 & 536.383 & 536.354 \\ \bottomrule
\end{tabular}
\end{table}
The four $\mathcal{D}_i$ values are nearly the same. Thus, using one of the correct or even the wrong ($\mathcal{D}_4$) formulas
makes not a big difference. Interestingly, the zero-state ARL $\ell_{R_1}$ chart is really close to these numbers as well.
Thus, for charts without head-start it is sufficient to look at the zero-state ARL (the mentioned behavior carries over
to the out-of-control case).

\begin{figure}[hbt]
\centering
\renewcommand{\tabcolsep}{-1mm}
\begin{tabular}{cc}
  \footnotesize $\delta \in (0,4)$ & \footnotesize $\delta \in [0,0.1)$ \\[-1ex]
  \includegraphics[width=.45\textwidth]{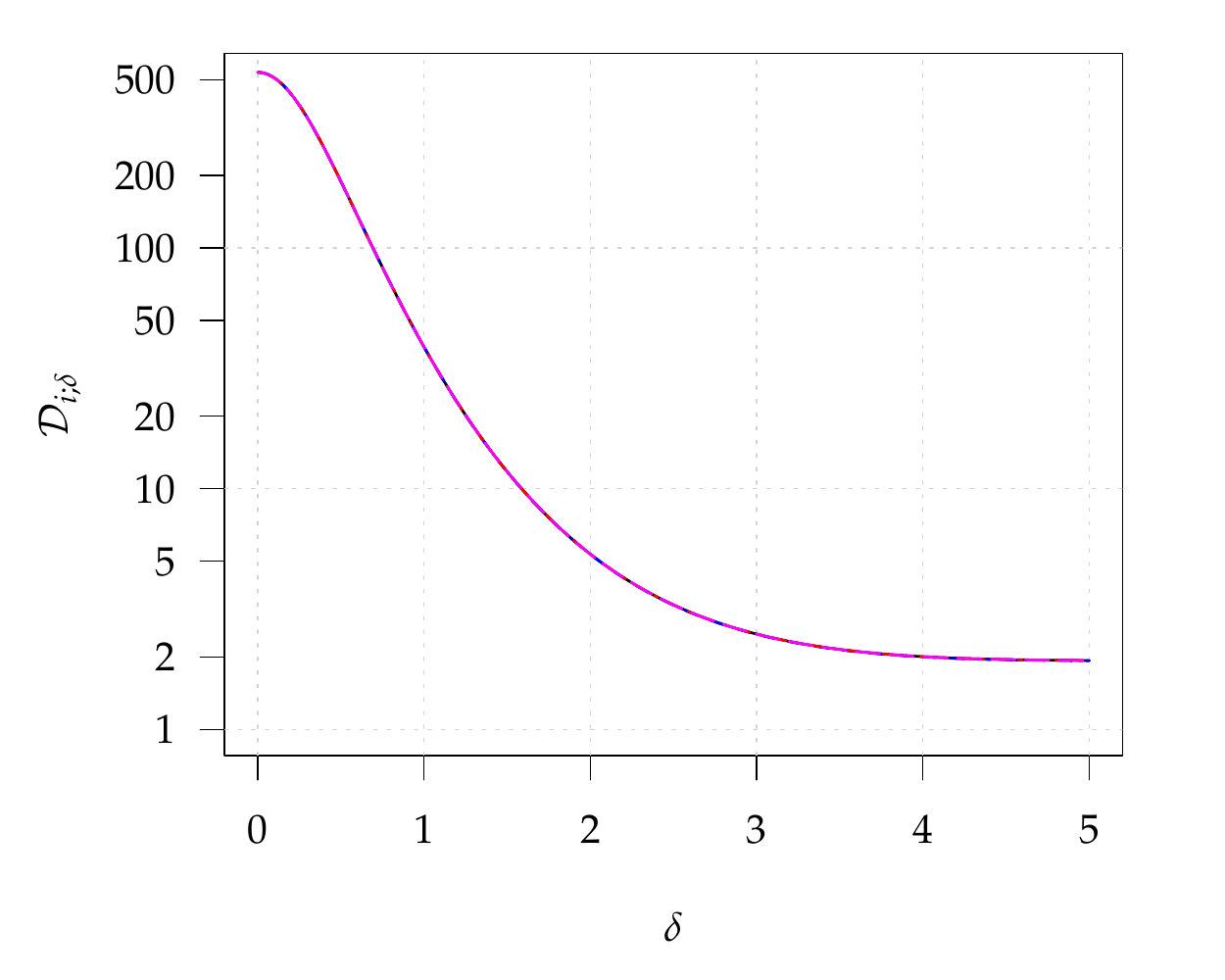} &	
  \includegraphics[width=.45\textwidth]{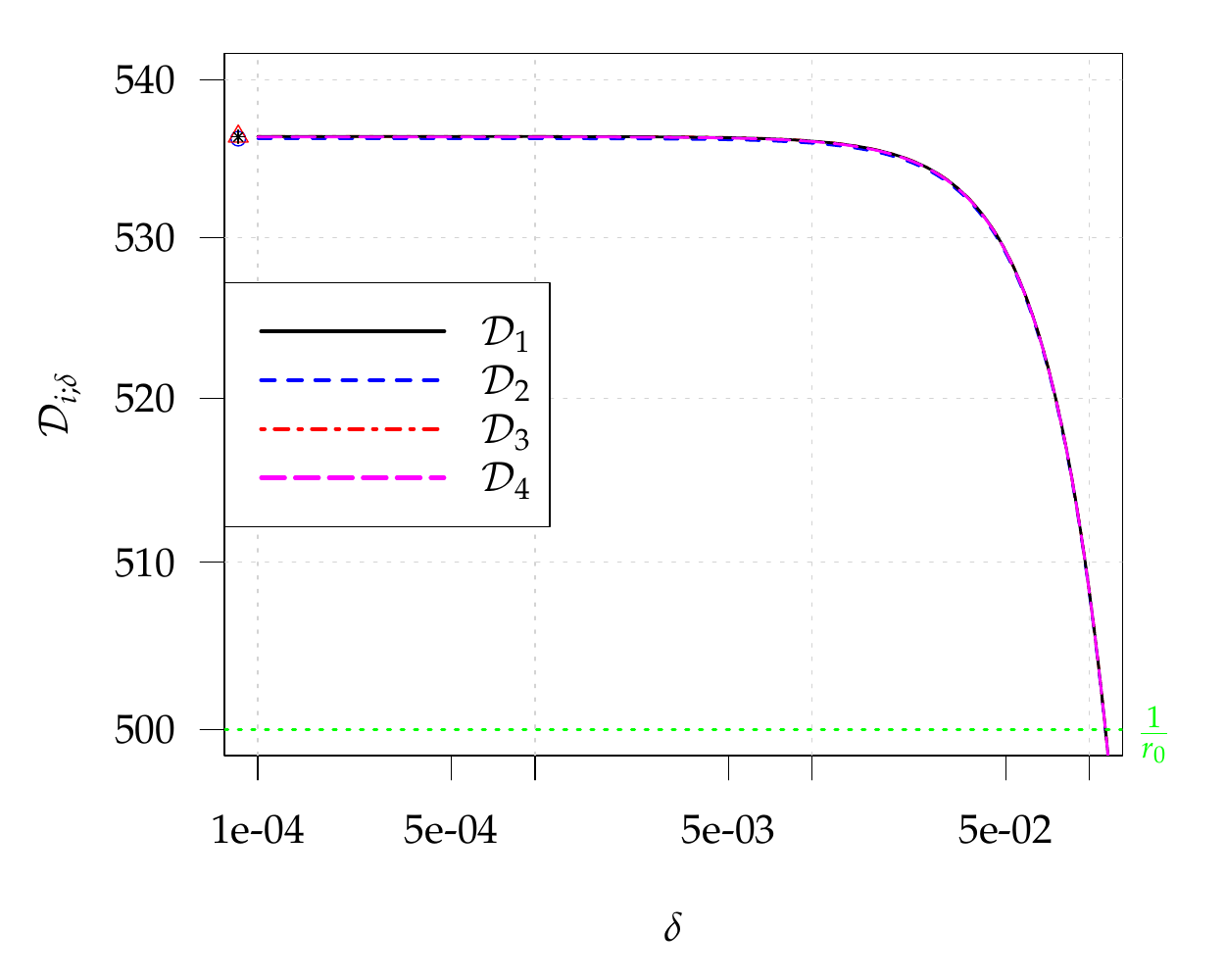}
\end{tabular}
\caption{Various types of steady-state ARL, $\mathcal{D}_i$, for a typical range of changes $(0,5)$ and for $\delta \to 0$,
true synthetic chart ($S_1$) with $H = 3$ and $k = 2.2238$ (in-control ARL 500).}\label{fig:DstarA}
\end{figure}

To judge this behavior for the out-of-control case, we plotted in Figure~\ref{fig:DstarA} some $\mathcal{D}_i$ profiles.
Not surprisingly, all these profiles coincide.
From these numbers we conclude that the wrongly chosen steady-state vector recipe in \cite{Shon:Grah:2017a, Shon:Grah:2019}
does not induce visible consequences.

Eventually we want to mention that calibrating (setting $k$ for synthetic-type charts) control charts to achieve a certain in-control steady-state ARL,
as it was done in \cite{Shon:Grah:2016a}, refers to starting the chart from its steady-state distribution.
This is certainly not a common task in SPM practice.

\subsection{Minimizing out-of-control ARL by tuning $H$} \label{app:optH}

\begin{figure}[hbt]
\centering
\renewcommand{\tabcolsep}{-1mm}
\begin{tabular}{cc}
  \footnotesize zero-state & \footnotesize steady-state \\[-1ex]
  \includegraphics[width=.45\textwidth]{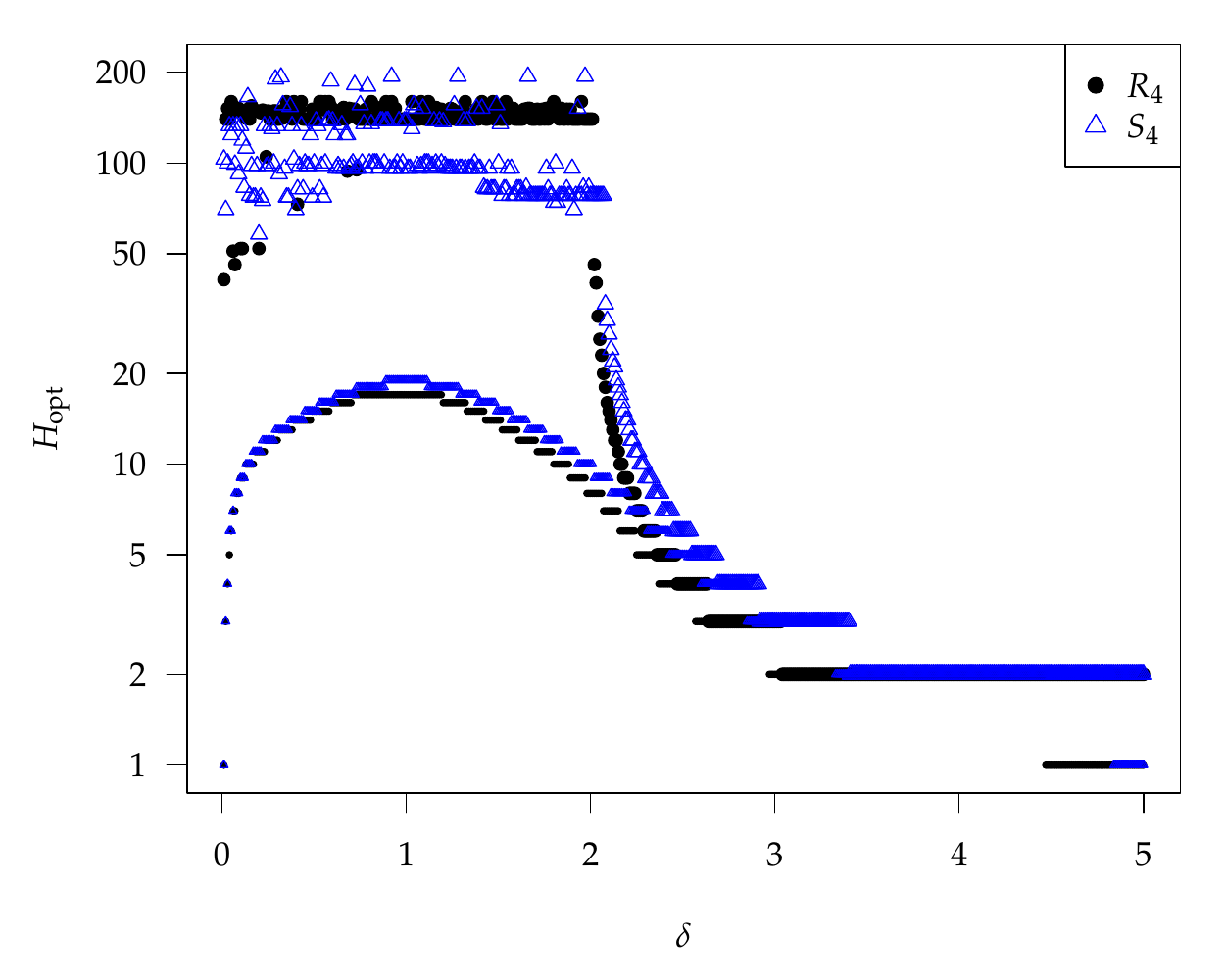} &	
  \includegraphics[width=.45\textwidth]{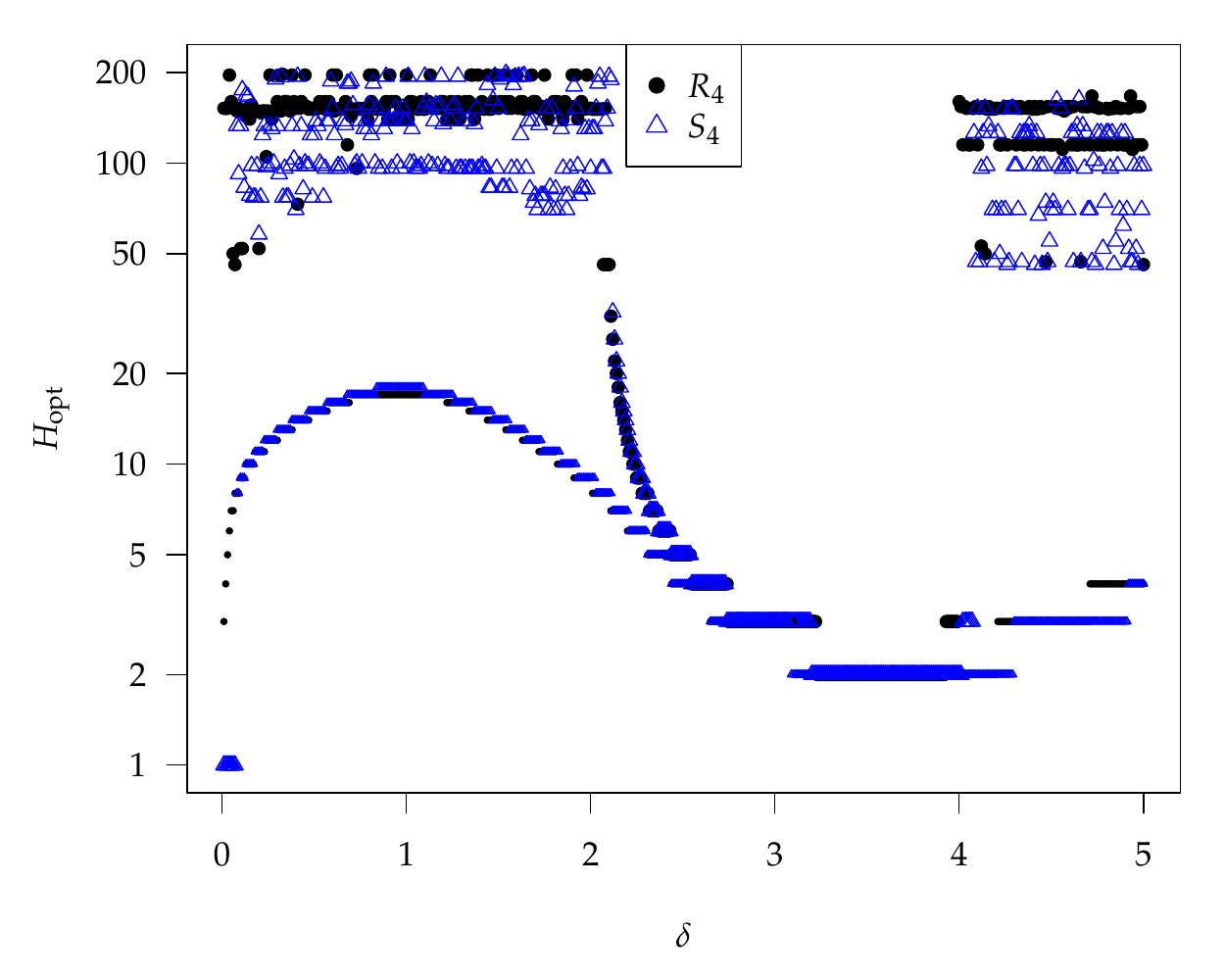}
\end{tabular}
\caption{Optimal $H$ values.} \label{fig:optimalH}
\end{figure}

In addition to the numbers given in Table~\ref{tab:Hopt} (Section~\ref{sec:cedStudy}) we show
here the complete output of our optimization procedure. For both \#4 charts ($R_4$ and $S_4$)
we tried $H \in \{1,2,\ldots,200\}$ and picked the $H$ value that either minimizes
the zero- or the steady-state ARL. In addition, we searched for small $H$ values that
yield ARL values not larger by 0.1\% than the overall minimum. In Figure~\ref{fig:optimalH}
all these $H$ values are plotted. The in-control ARL is set to 500.
We observe quite similar patterns for both ARl types. The most pronounced difference
could be seen for $\delta > 4$. Fortunately, tuning synthetic-type charts for so large
changes is quite uncommon.

\subsection{Worst-case ARL competition} \label{app:R4vsCUSUM}

Here, we compare the zero-state ARL of $R_4$ ($H = 8$ -- optimal for $\delta = 2$)
and of two-sided CUSUM control charts. For the latter
we choose $k = 1$ ($k$ denotes here the reference value of a CUSUM control chart)
to achieve good performance for mid-size changes ($\delta = 2$ and its neighborhood).
\begin{figure}[hbt]
\centering
\includegraphics[width=.5\textwidth]{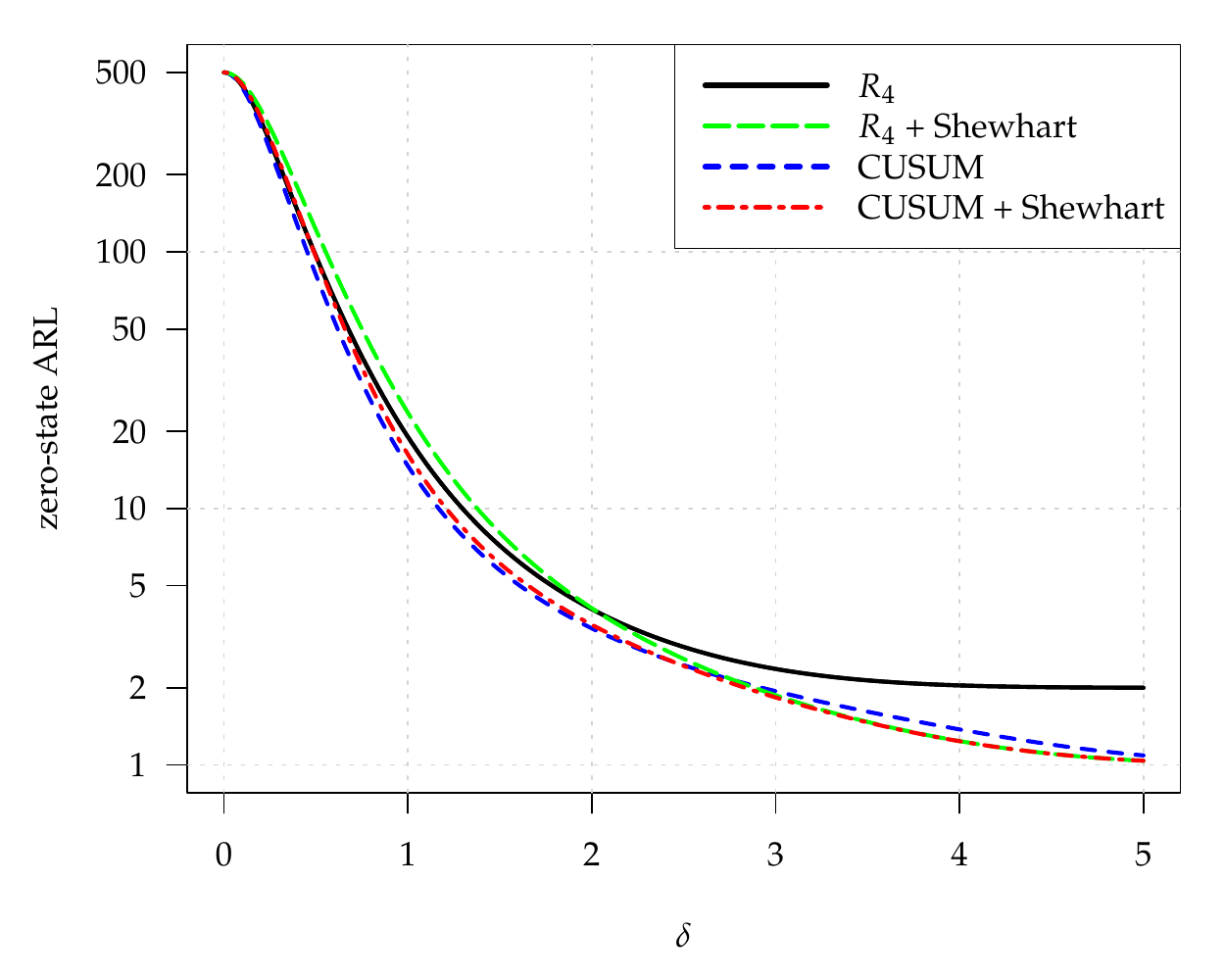}
\caption{Zero-state ARL comparison between $R_4$ ($H=8$) and CUSUM control charts
($k = 1$); in-control ARL 500; for both a combo version (+ Shewhart with alarm threshold $k_2 = 3.25$)
is added.} \label{fig:R4vsCUSUM}
\end{figure}
We add as well a combo of $R_4$ and Shewhart ($k_2 = 3.25$) to deal with the weak right-hand tail of $R_4$.
The $k=1$ CUSUM ($h=2.665$) is uniformly better than $R_4$.
Compared to $R_4$, the Shewhart-$R_4$ combo exhibits a better detection performance for
$\delta \ge 3$. Finally, the Shewhart-CUSUM combo ($k = 1$, $k_2 = 3.25$ and $h = 2.947$)
shows lower out-of-control ARL results for $\delta < 3$ and more or less the same values for $\delta \ge 3$
like the Shewhart-$R_4$ combo. Thus, the CUSUM schemes win both worst-case ARL competitions.
Eventually we want to note that the ARL values of the Shewhart-CUSUM combo are determined with the algorithms
given in \cite{Knot:2018a}. For the standard CUSUM the function \texttt{xcusum.arl()} from
the \textsf{R} package \texttt{spc} is utilized.

\subsection{Further CED profiles} \label{app:furtherCED}

In addition to the cases $\delta \in \{1,2,3\}$ we plot here some CED profiles for further changes,
namely $\delta \in \{0.5,1.5\}$.

\begin{figure}[hbt]
\centering
\begin{tabular}{cc}
  \footnotesize $S_1$ & \footnotesize $S_2$ \\[-1ex]
  \includegraphics[width=.5\textwidth]{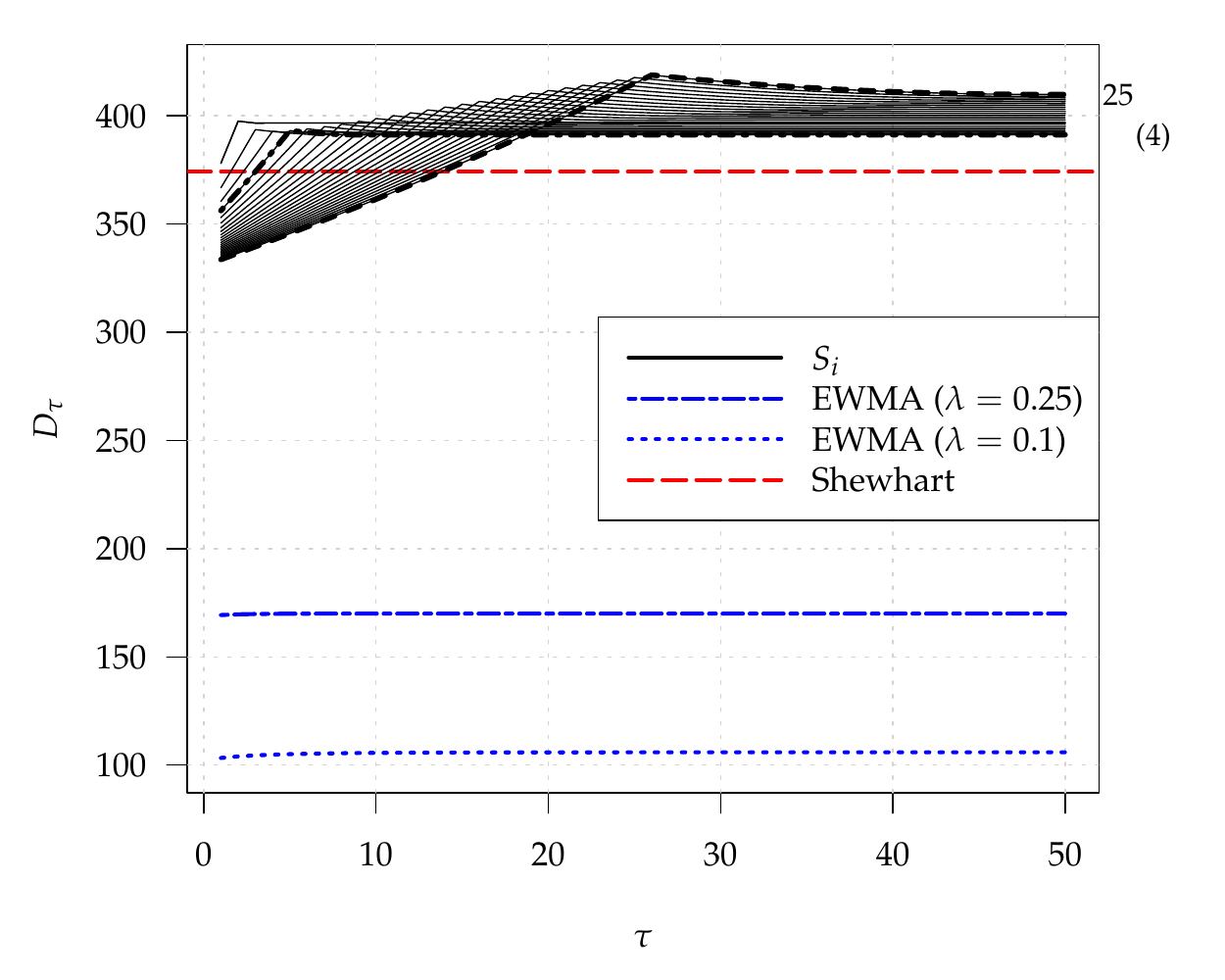} &	
  \includegraphics[width=.5\textwidth]{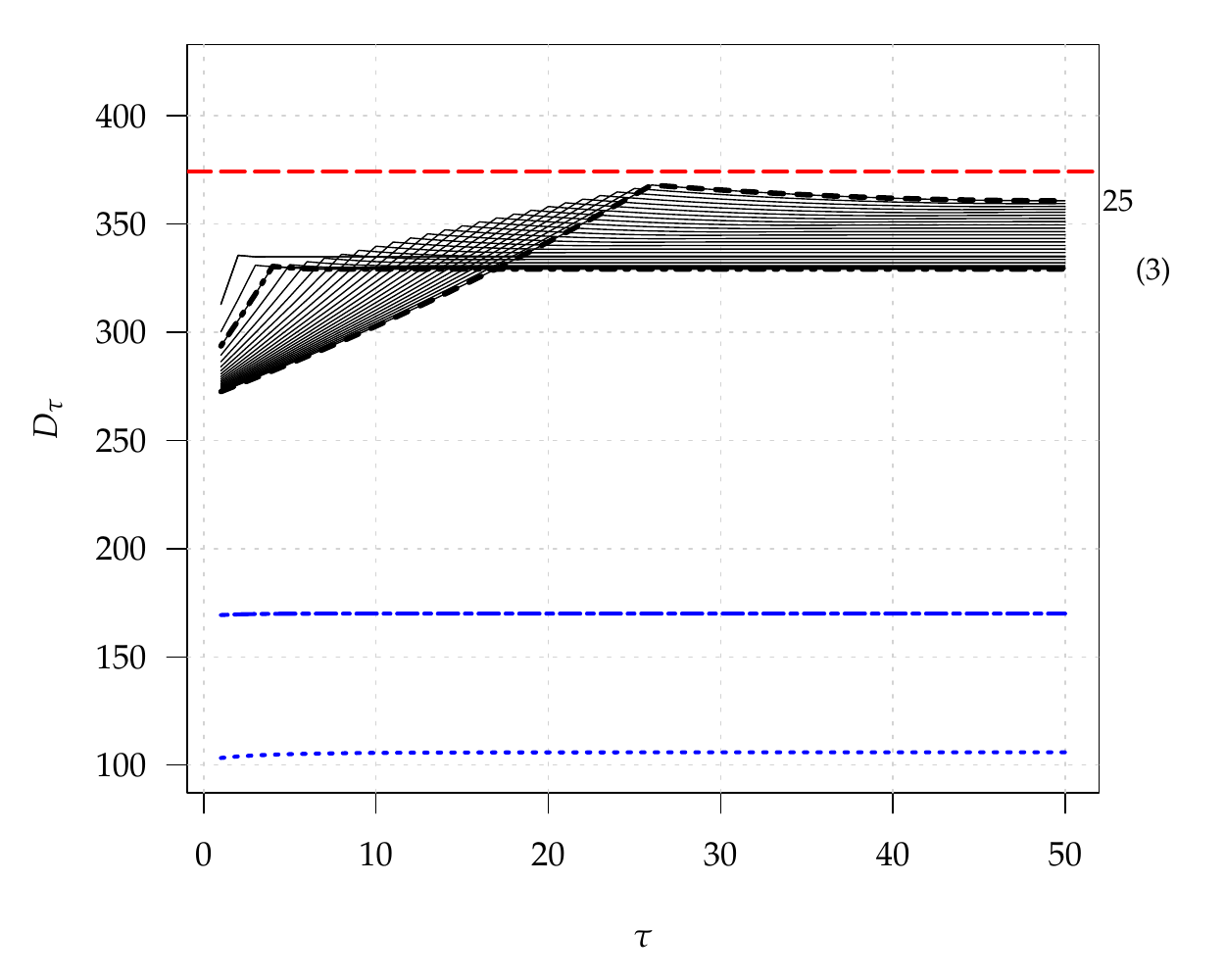} \\[1ex]
  \footnotesize $S_3$ & \footnotesize $S_4$ \\[-1ex]
  \includegraphics[width=.5\textwidth]{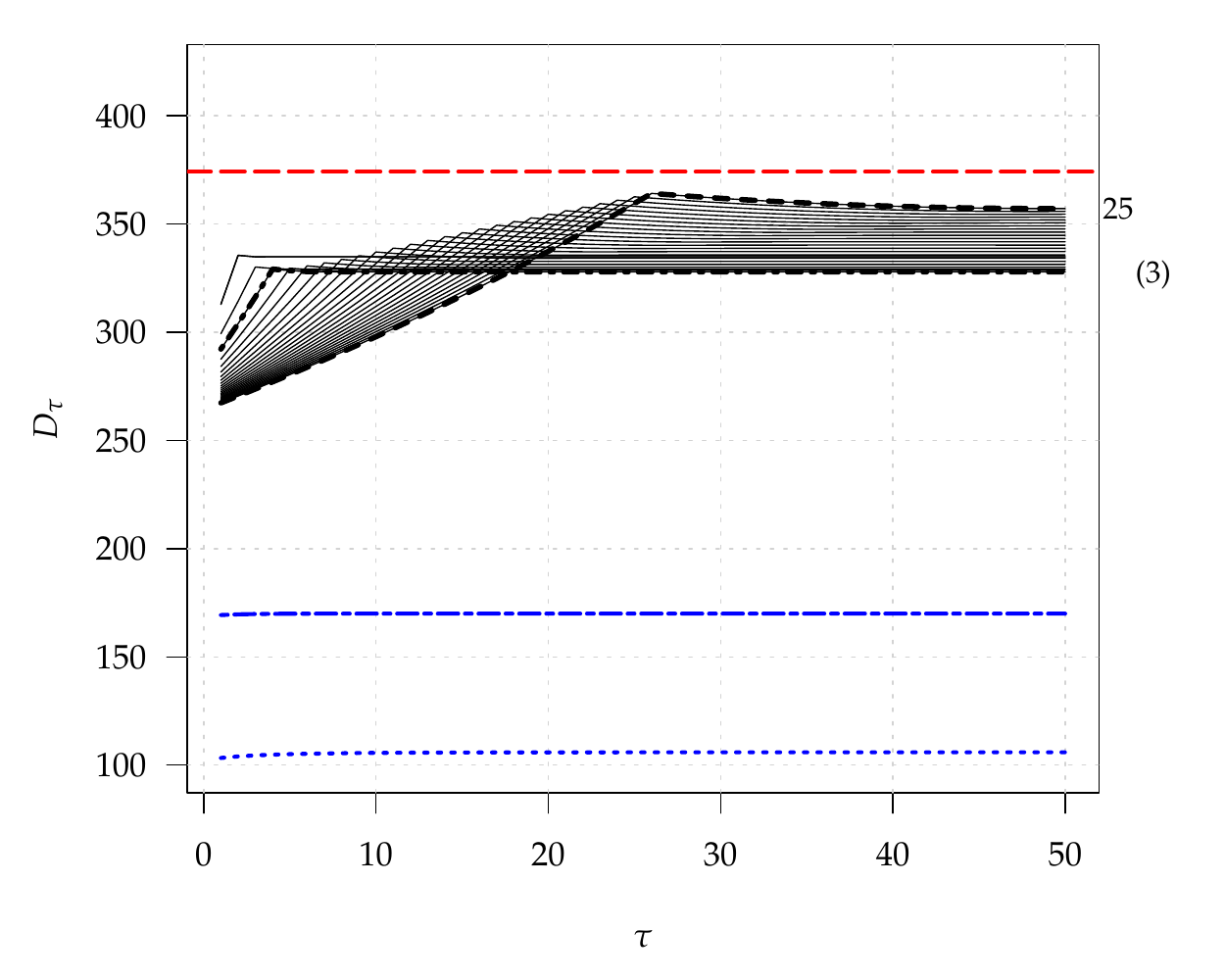} &	
  \includegraphics[width=.5\textwidth]{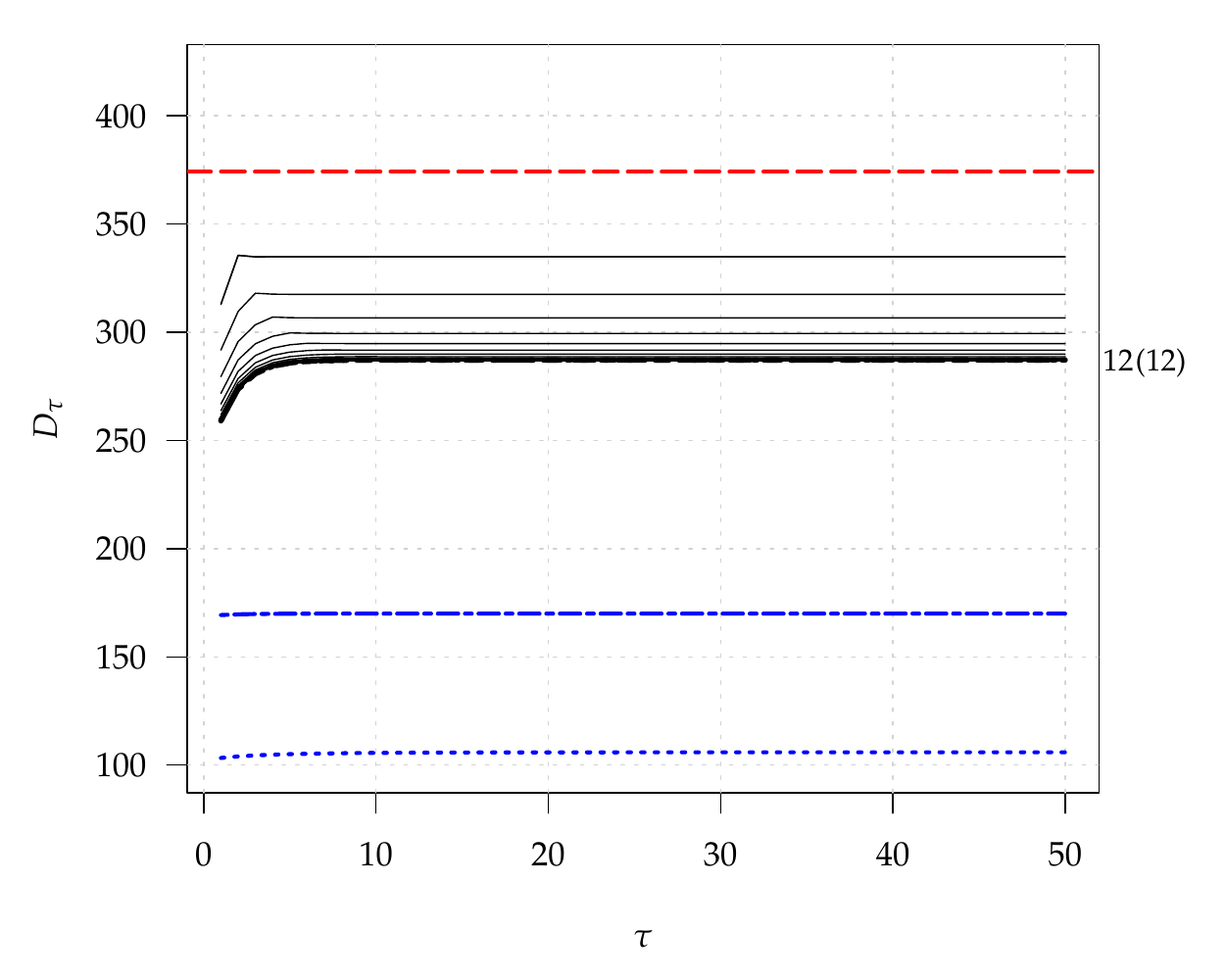}
\end{tabular}
\caption{$D_\tau$ profiles for four synthetic-type charts with head-start,
$H = 1, 2, \ldots, 25$, best scheme (zero-state and steady-state) bold (dashed and dash-dotted) lines,
shift $\delta = 0.25$, two EWMA charts; in-control ARL 500.} \label{fig:ced025}
\end{figure}

\begin{figure}[hbt]
\centering
\begin{tabular}{cc}
  \footnotesize $S_1$ & \footnotesize $S_2$ \\[-1ex]
  \includegraphics[width=.5\textwidth]{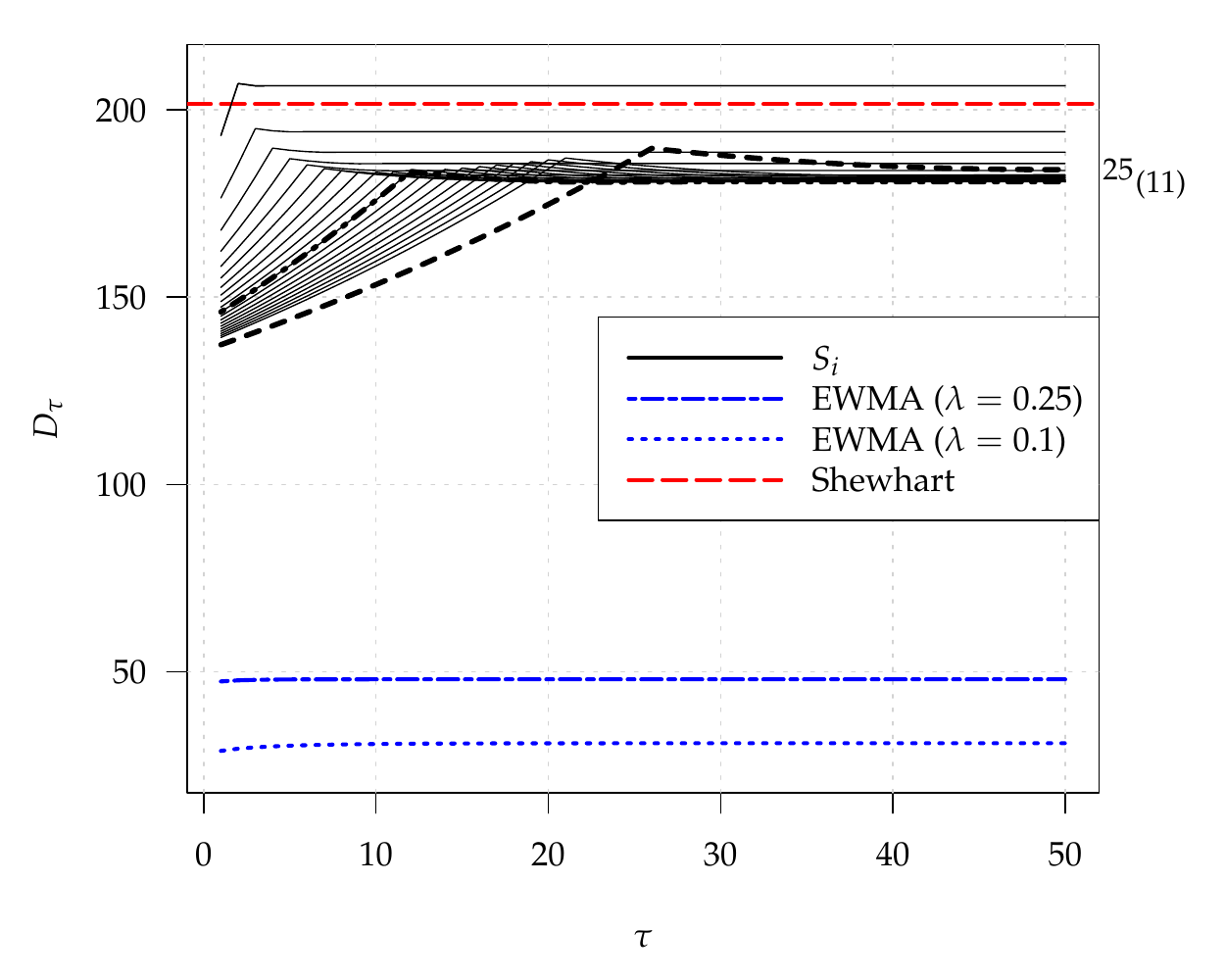} &	
  \includegraphics[width=.5\textwidth]{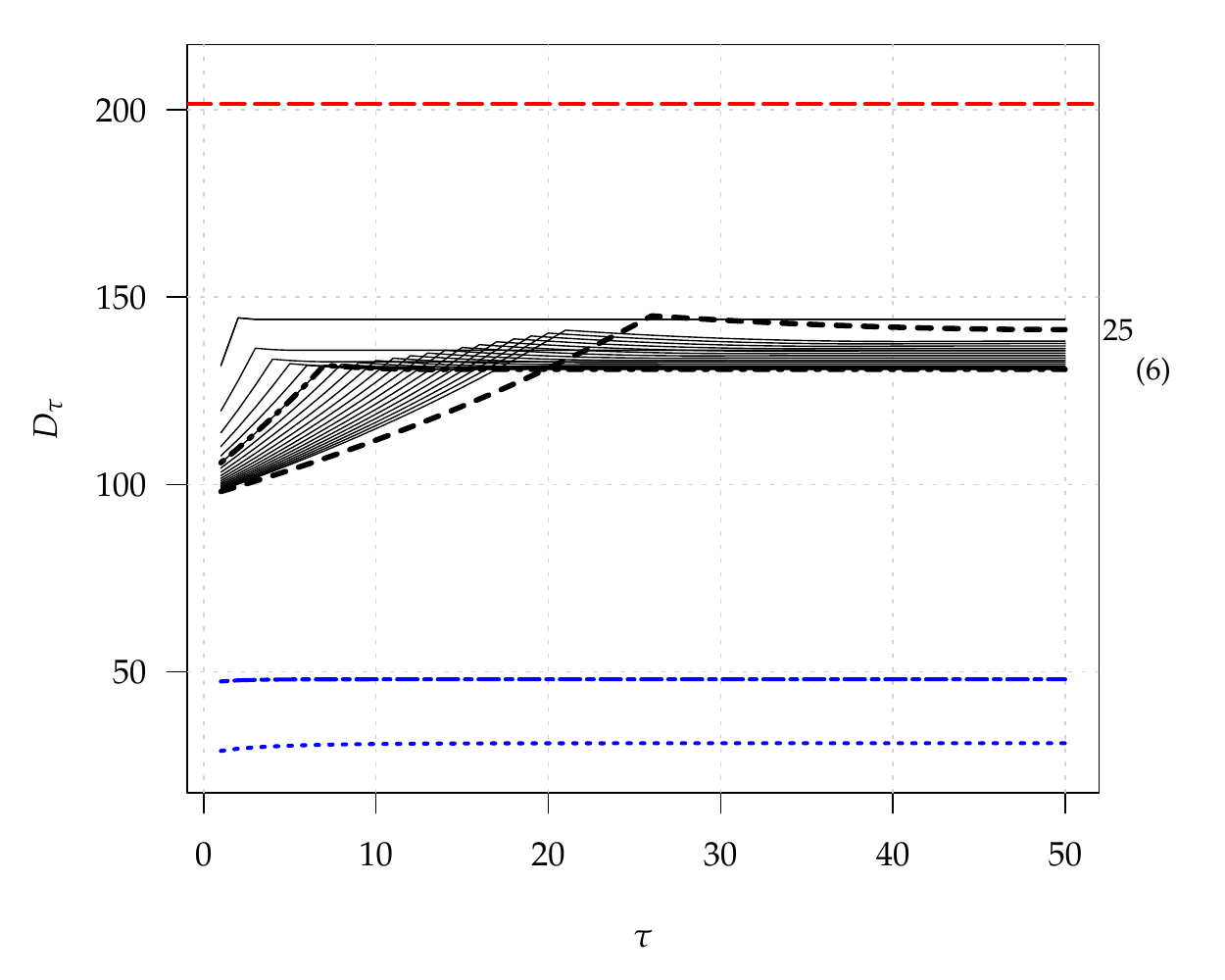} \\[1ex]
  \footnotesize $S_3$ & \footnotesize $S_4$ \\[-1ex]
  \includegraphics[width=.5\textwidth]{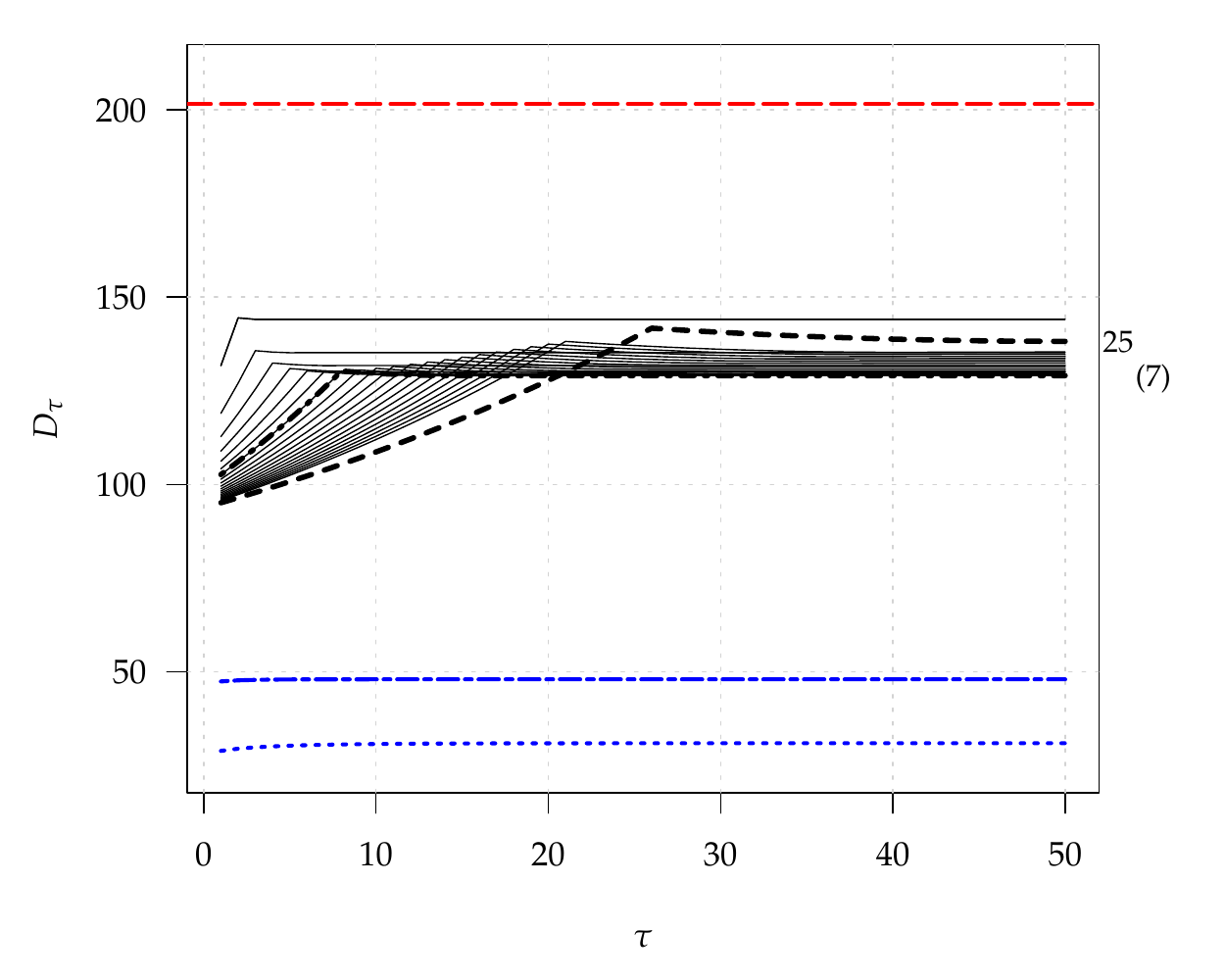} &	
  \includegraphics[width=.5\textwidth]{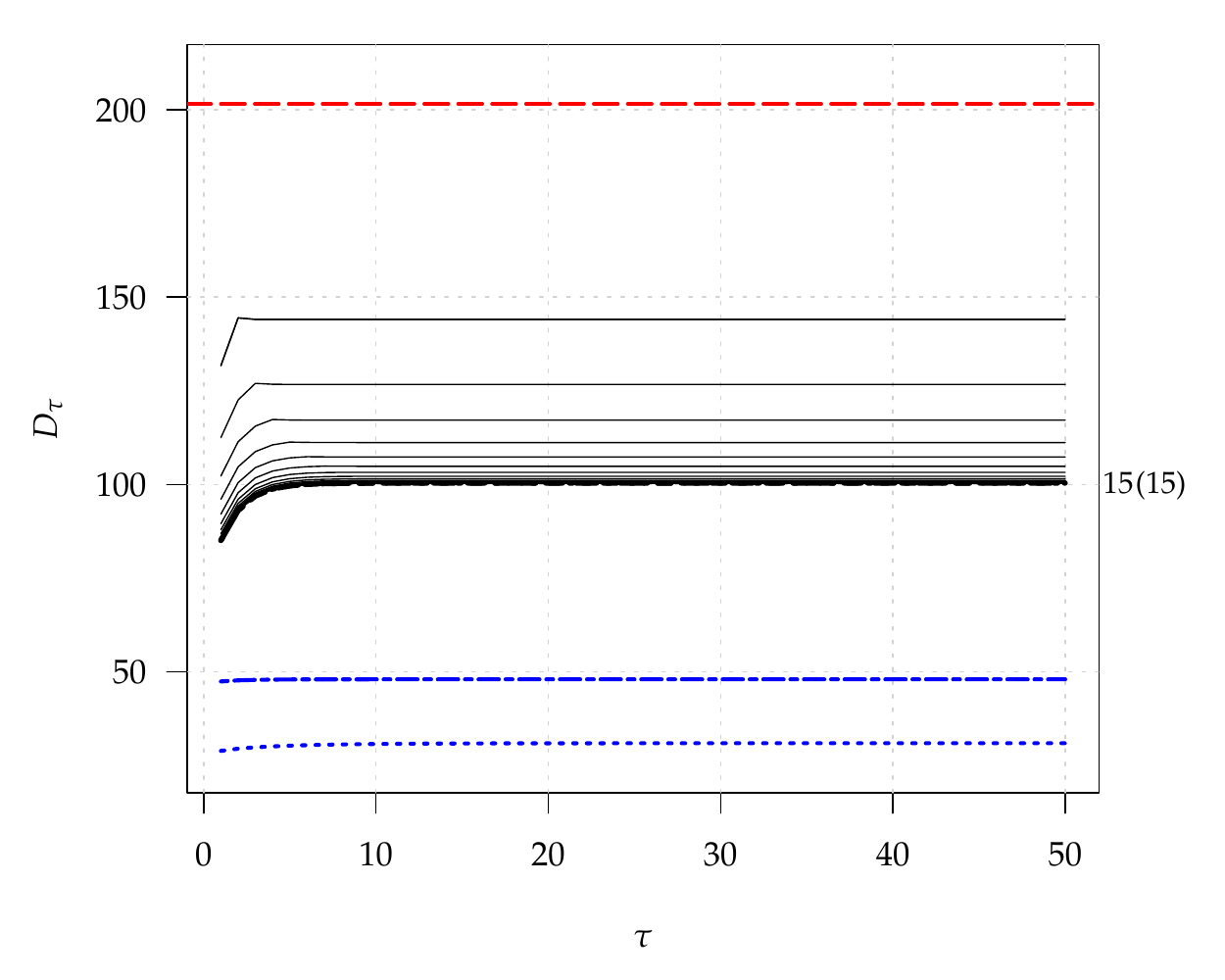}
\end{tabular}
\caption{$D_\tau$ profiles for four synthetic-type charts with head-start,
$H = 1, 2, \ldots, 25$, best scheme (zero-state and steady-state) bold (dashed and dash-dotted) lines,
shift $\delta = 0.5$, two EWMA charts; in-control ARL 500.} \label{fig:ced05}
\end{figure}

\begin{figure}[hbt]
\centering
\begin{tabular}{cc}
  \footnotesize $S_1$ & \footnotesize $S_2$ \\[-1ex]
  \includegraphics[width=.5\textwidth]{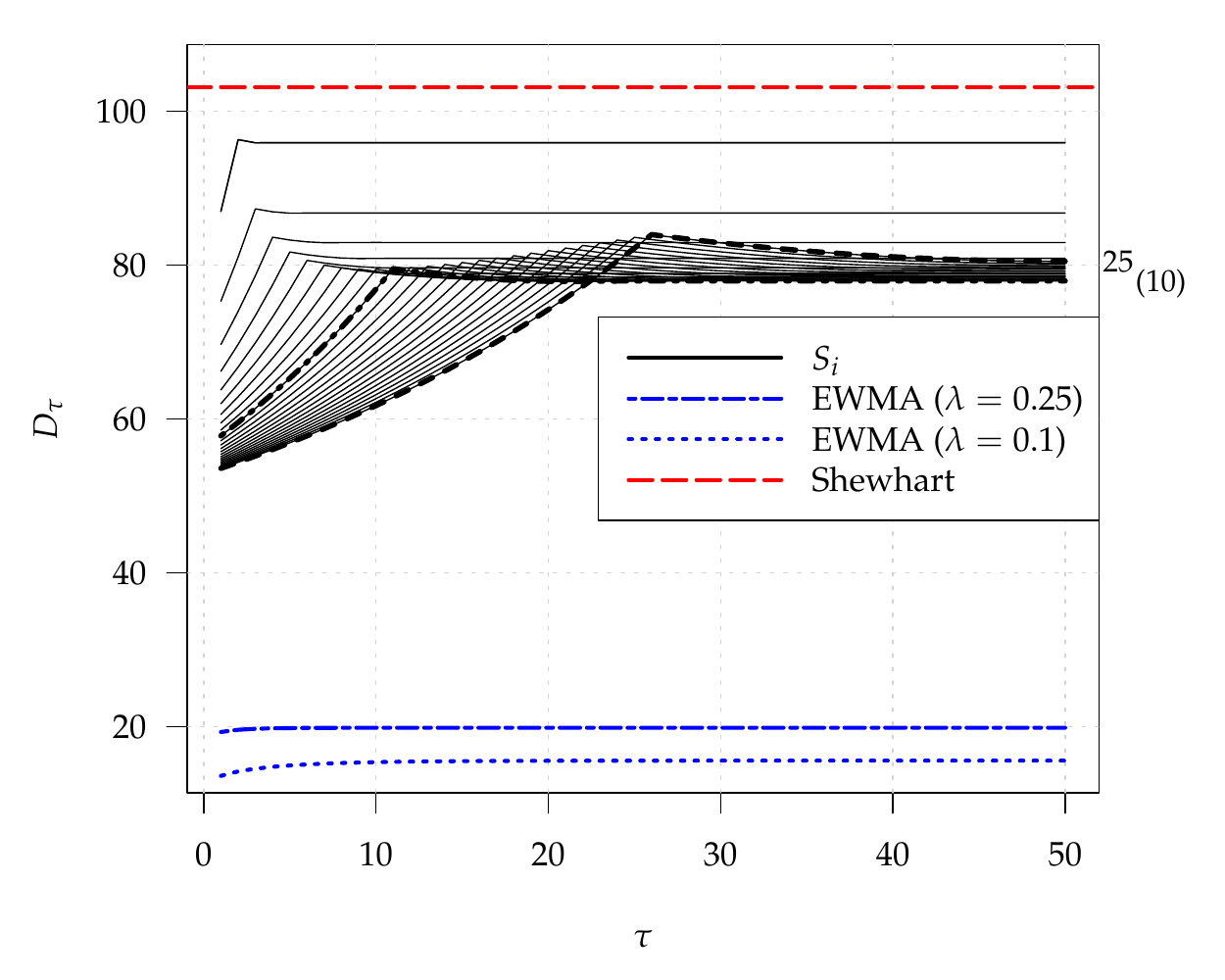} &	
  \includegraphics[width=.5\textwidth]{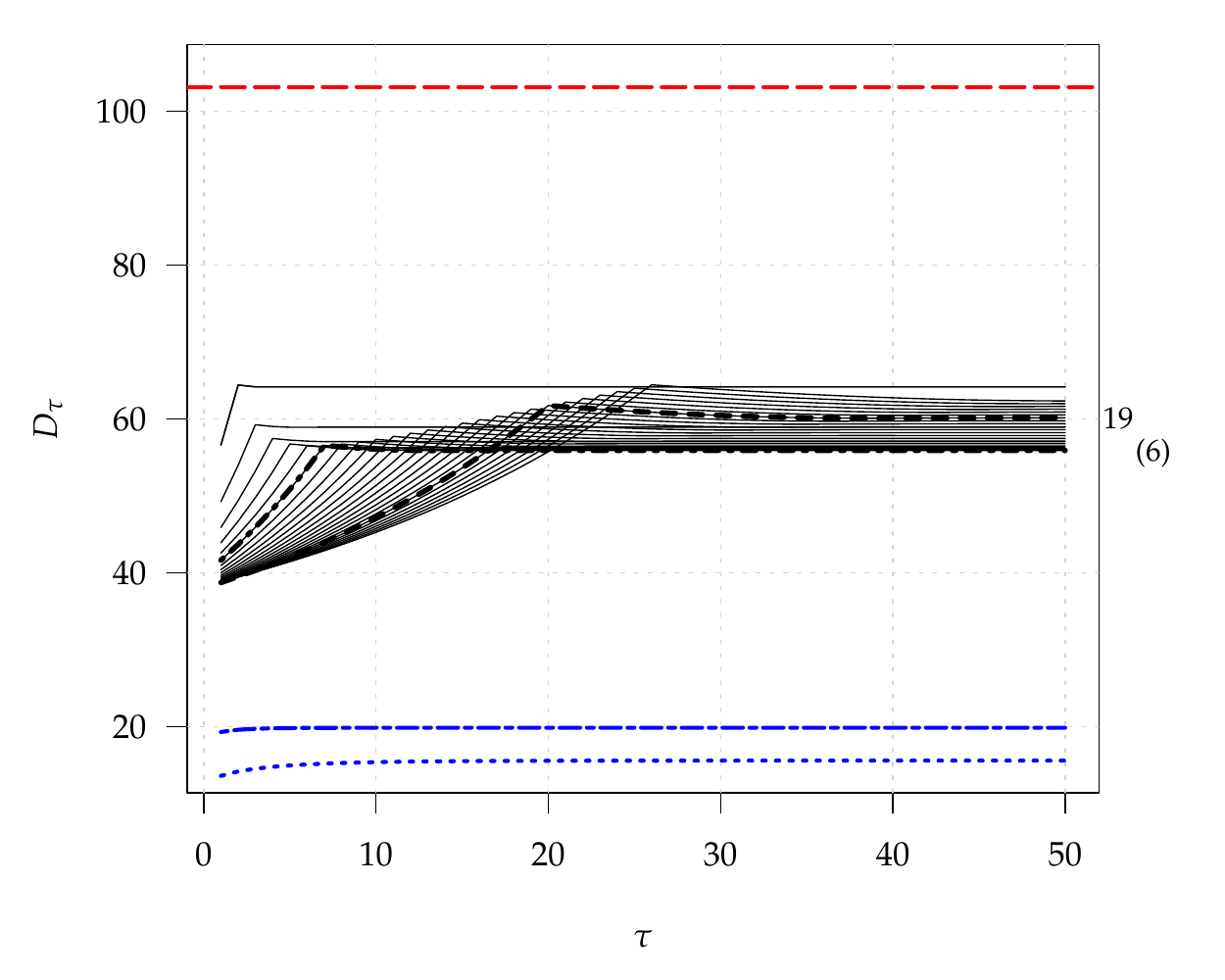} \\[1ex]
  \footnotesize $S_3$ & \footnotesize $S_4$ \\[-1ex]
  \includegraphics[width=.5\textwidth]{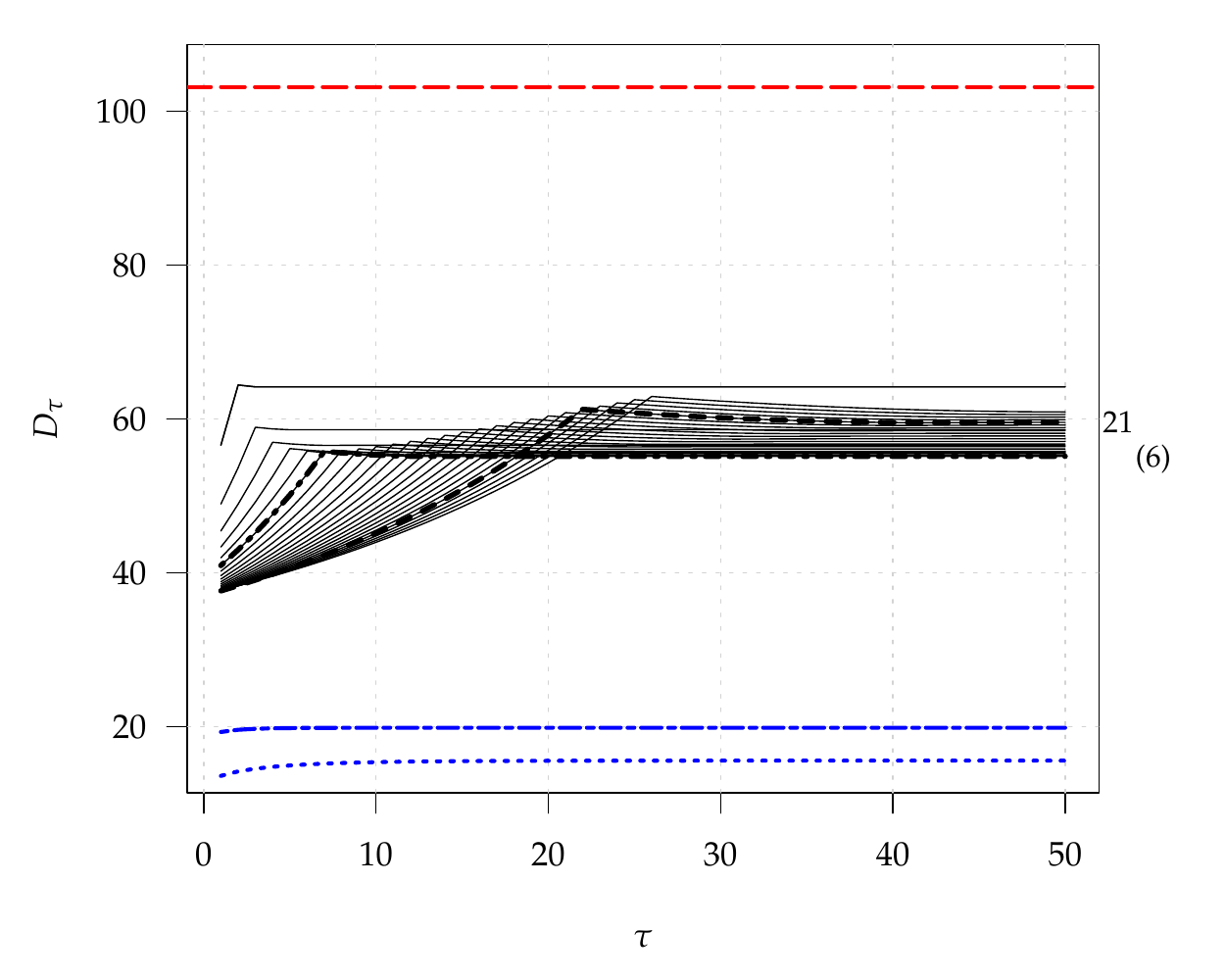} &	
  \includegraphics[width=.5\textwidth]{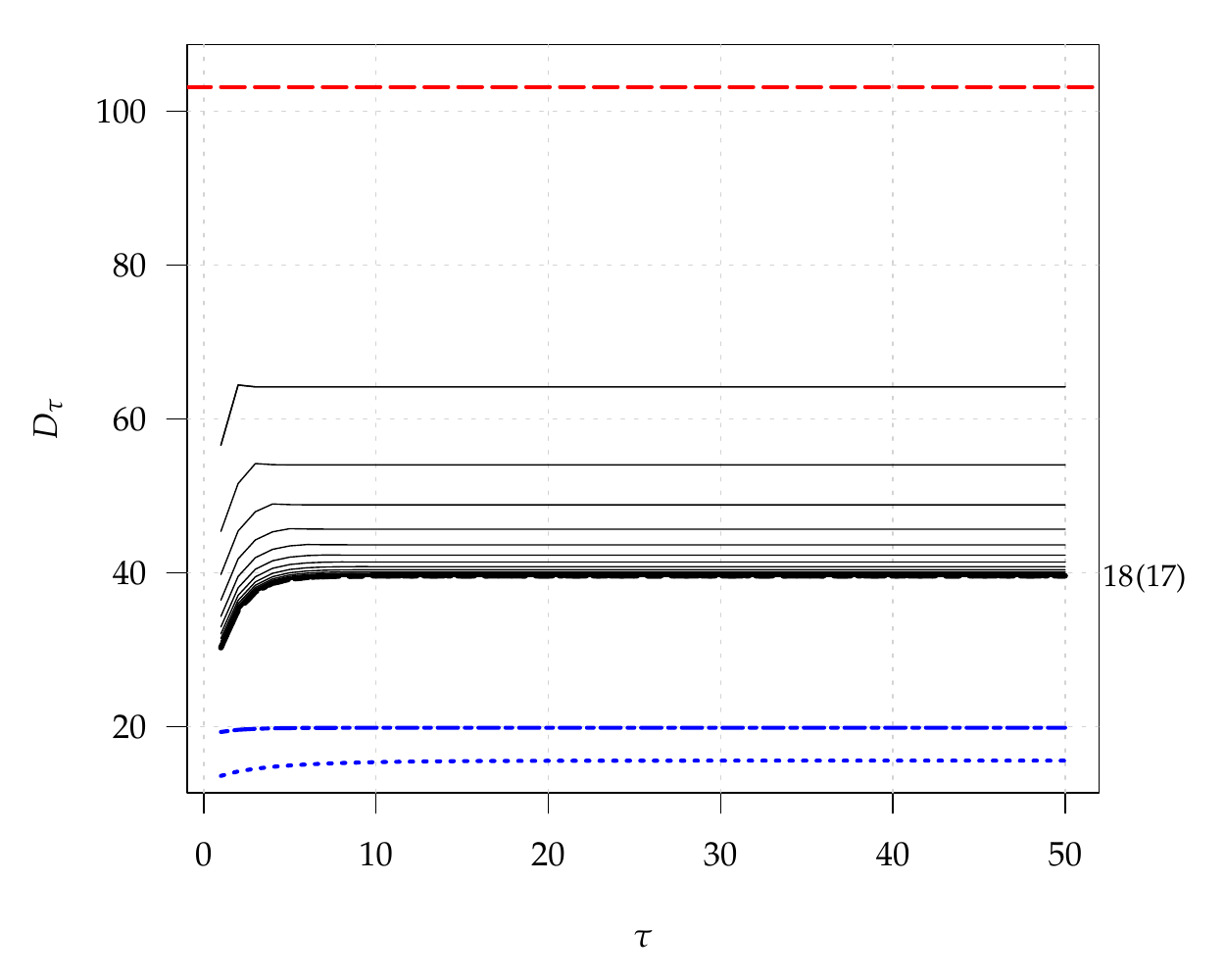}
\end{tabular}
\caption{$D_\tau$ profiles for four synthetic-type charts with head-start,
$H = 1, 2, \ldots, 25$, best scheme (zero-state and steady-state) bold (dashed and dash-dotted) lines,
shift $\delta = 0.75$, two EWMA charts; in-control ARL 500.} \label{fig:ced075}
\end{figure}

\begin{figure}[hbt]
\centering
\begin{tabular}{cc}
  \footnotesize $S_1$ & \footnotesize $S_2$ \\[-1ex]
  \includegraphics[width=.5\textwidth]{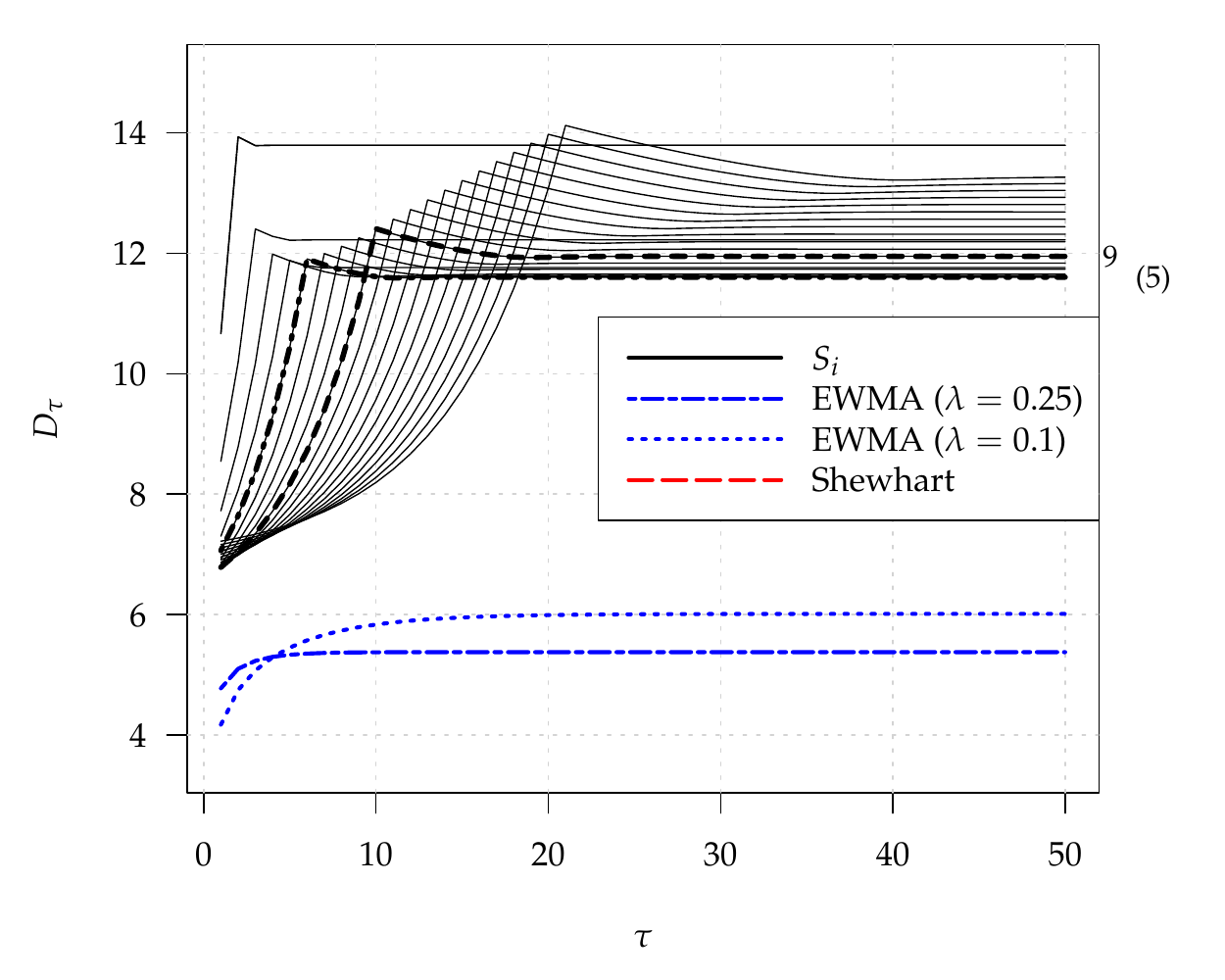} &	
  \includegraphics[width=.5\textwidth]{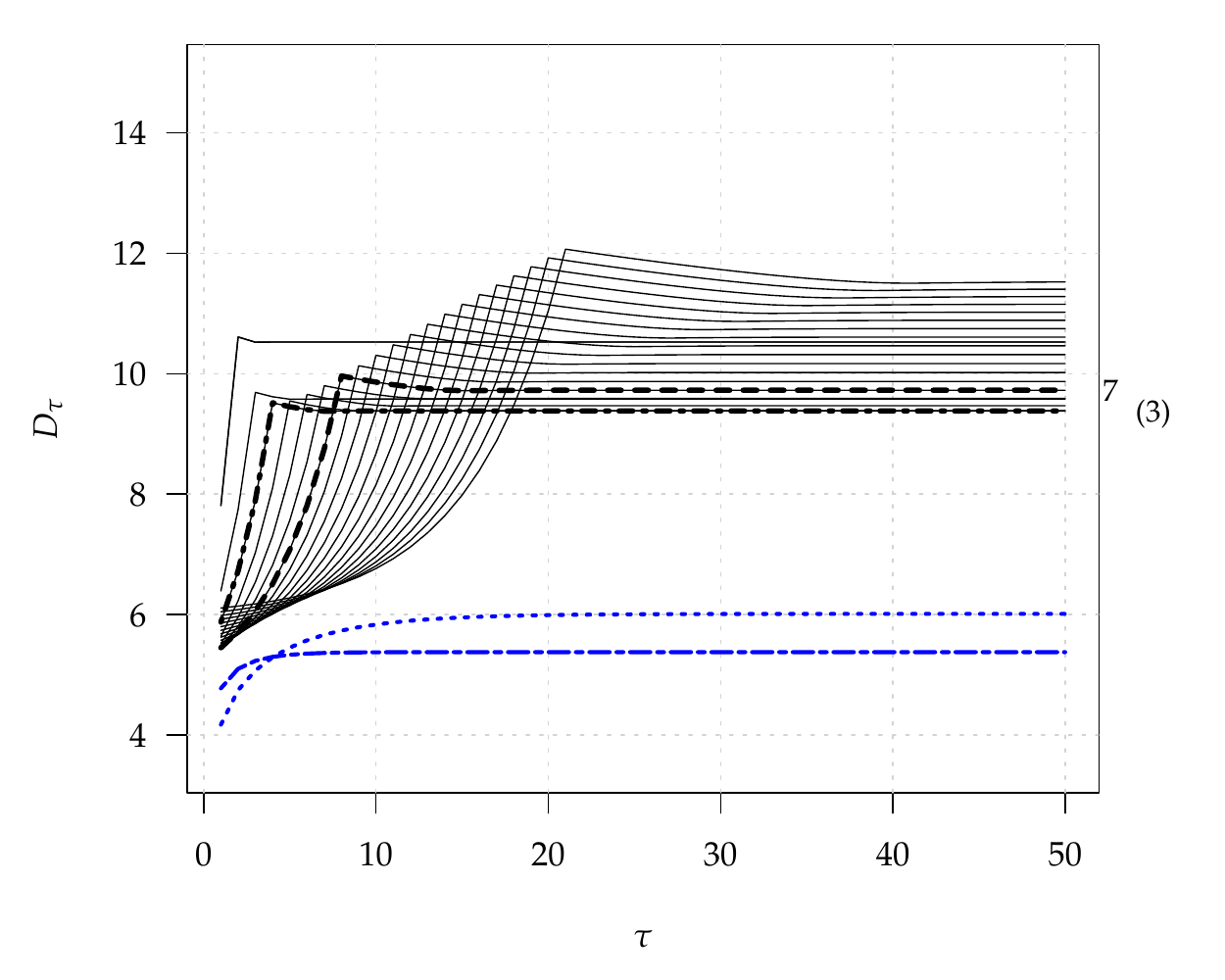} \\[1ex]
  \footnotesize $S_3$ & \footnotesize $S_4$ \\[-1ex]
  \includegraphics[width=.5\textwidth]{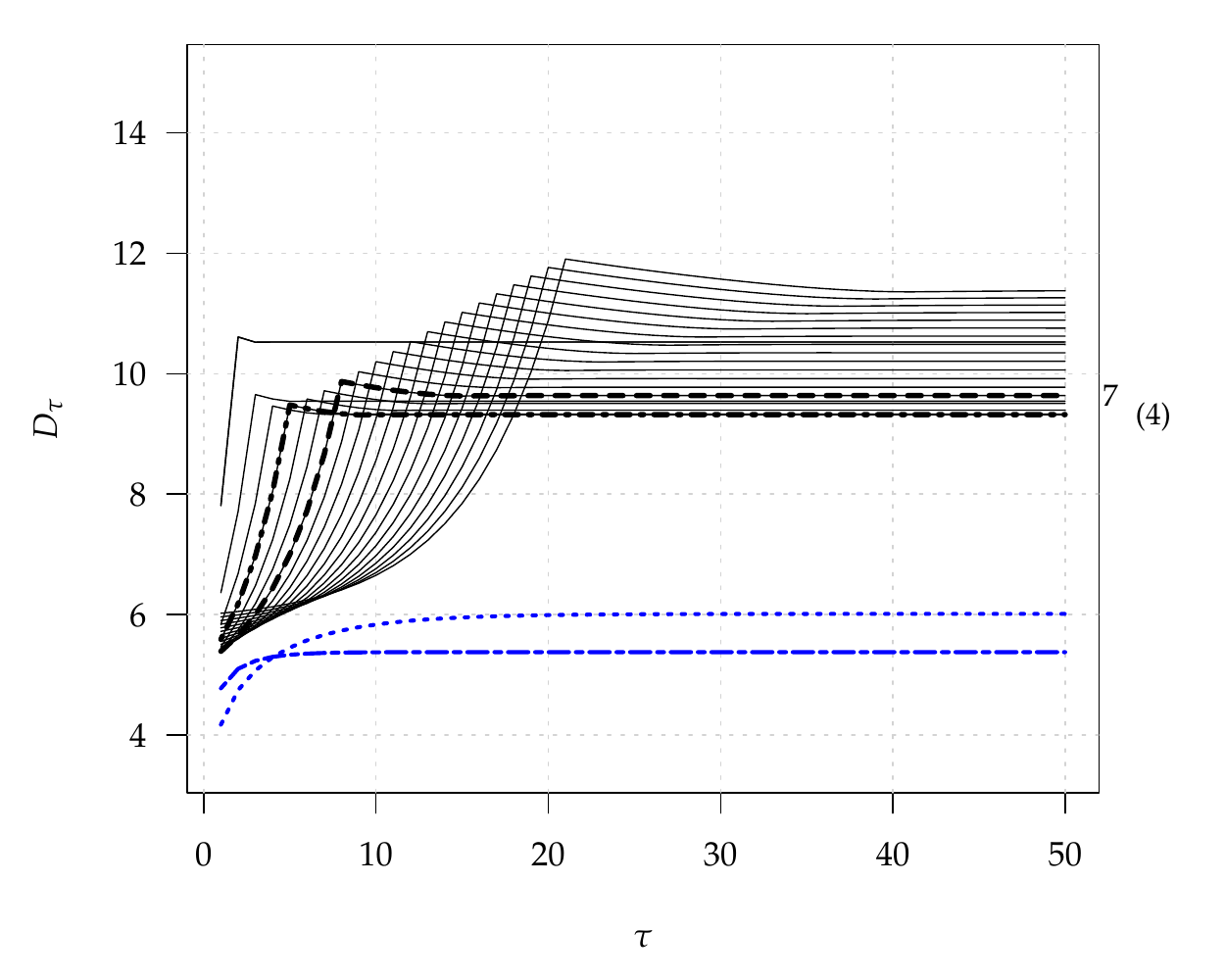} &	
  \includegraphics[width=.5\textwidth]{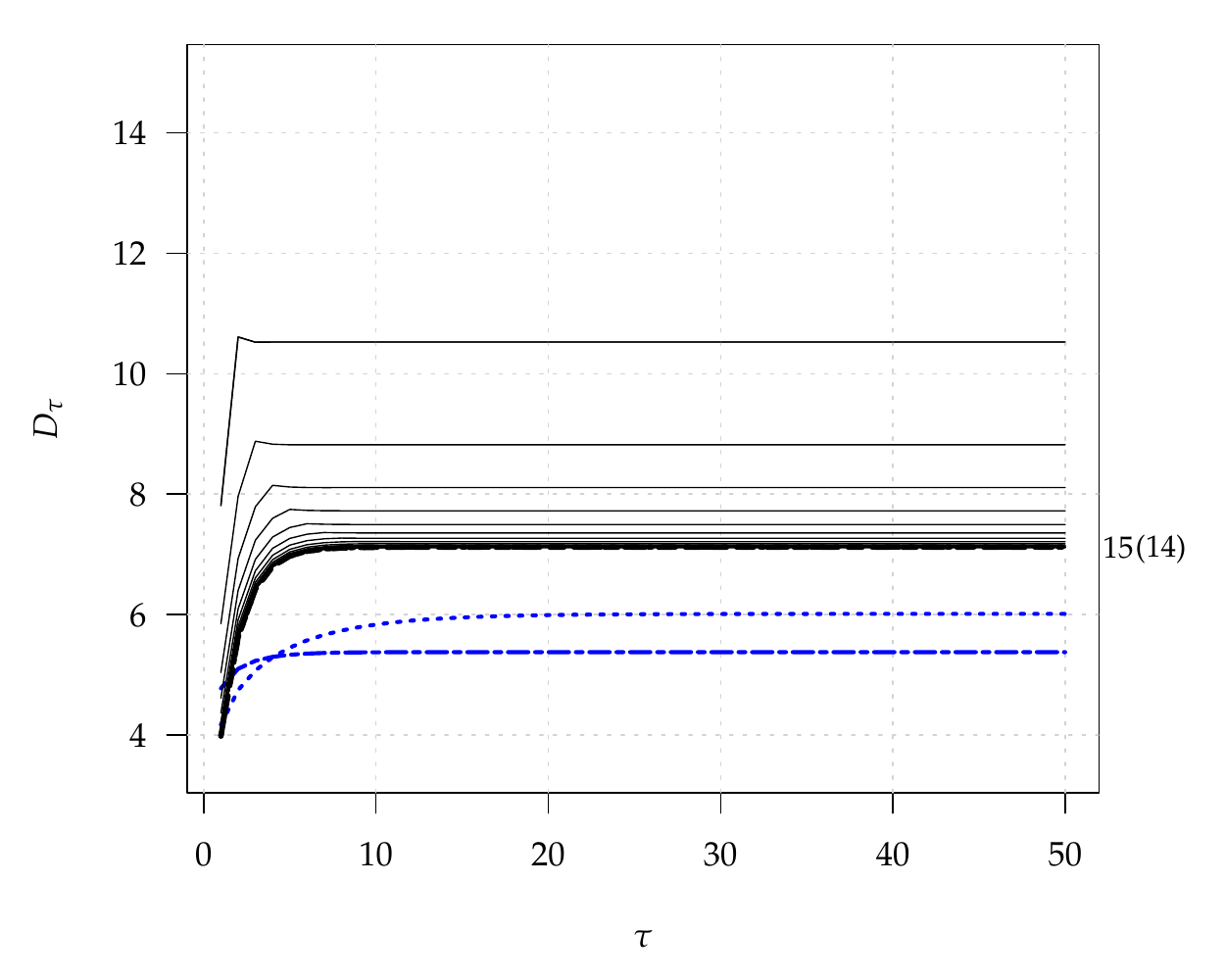}
\end{tabular}
\caption{$D_\tau$ profiles for four synthetic-type charts with head-start,
$H = 1, 2, \ldots, 25$, best scheme (zero-state and steady-state) bold (dashed and dash-dotted) lines,
shift $\delta = 1.5$, two EWMA charts; in-control ARL 500.} \label{fig:ced15}
\end{figure}

\begin{figure}[hbt]
\centering
\begin{tabular}{cc}
  \footnotesize $S_1$ & \footnotesize $S_2$ \\[-1ex]
  \includegraphics[width=.5\textwidth]{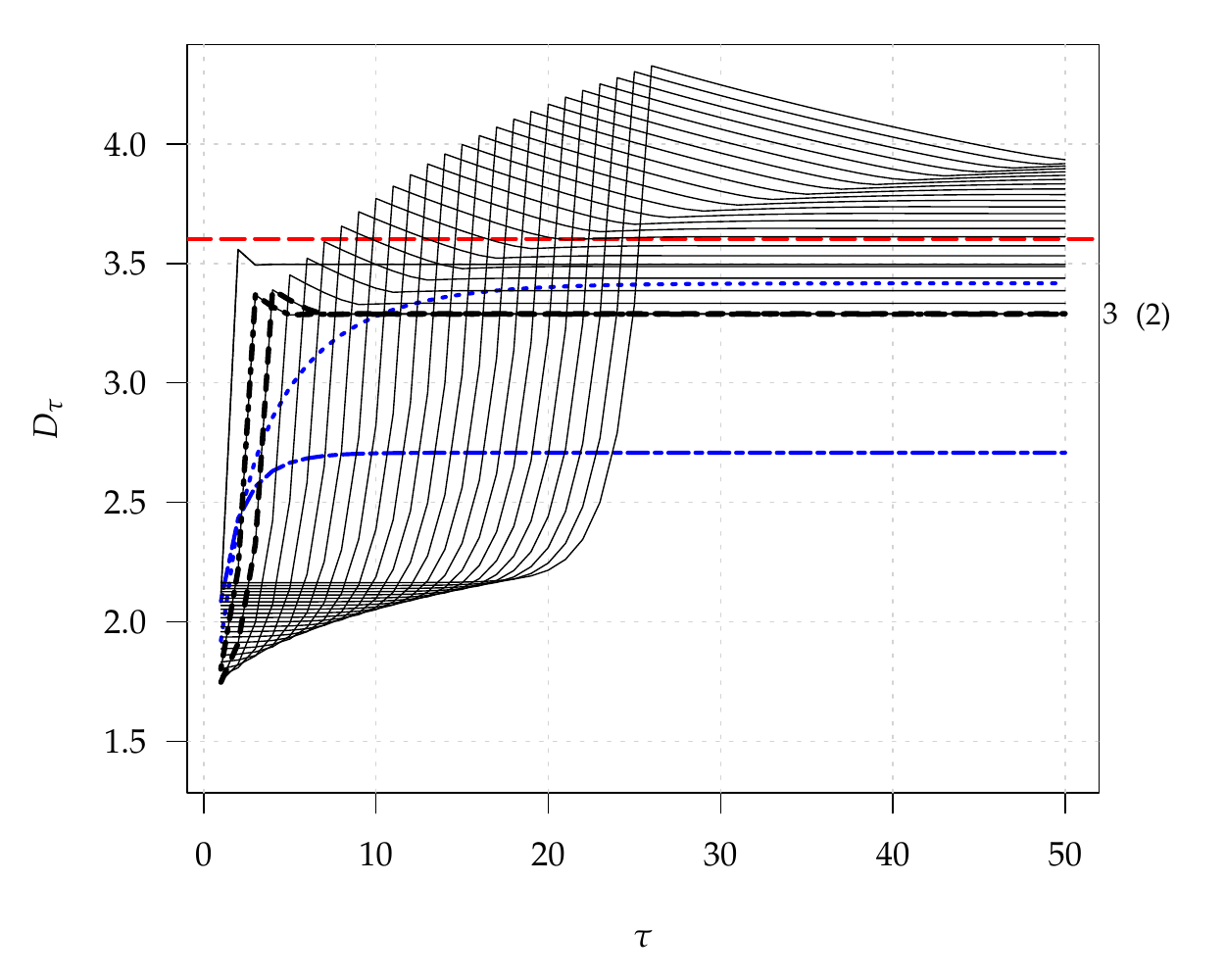} &	
  \includegraphics[width=.5\textwidth]{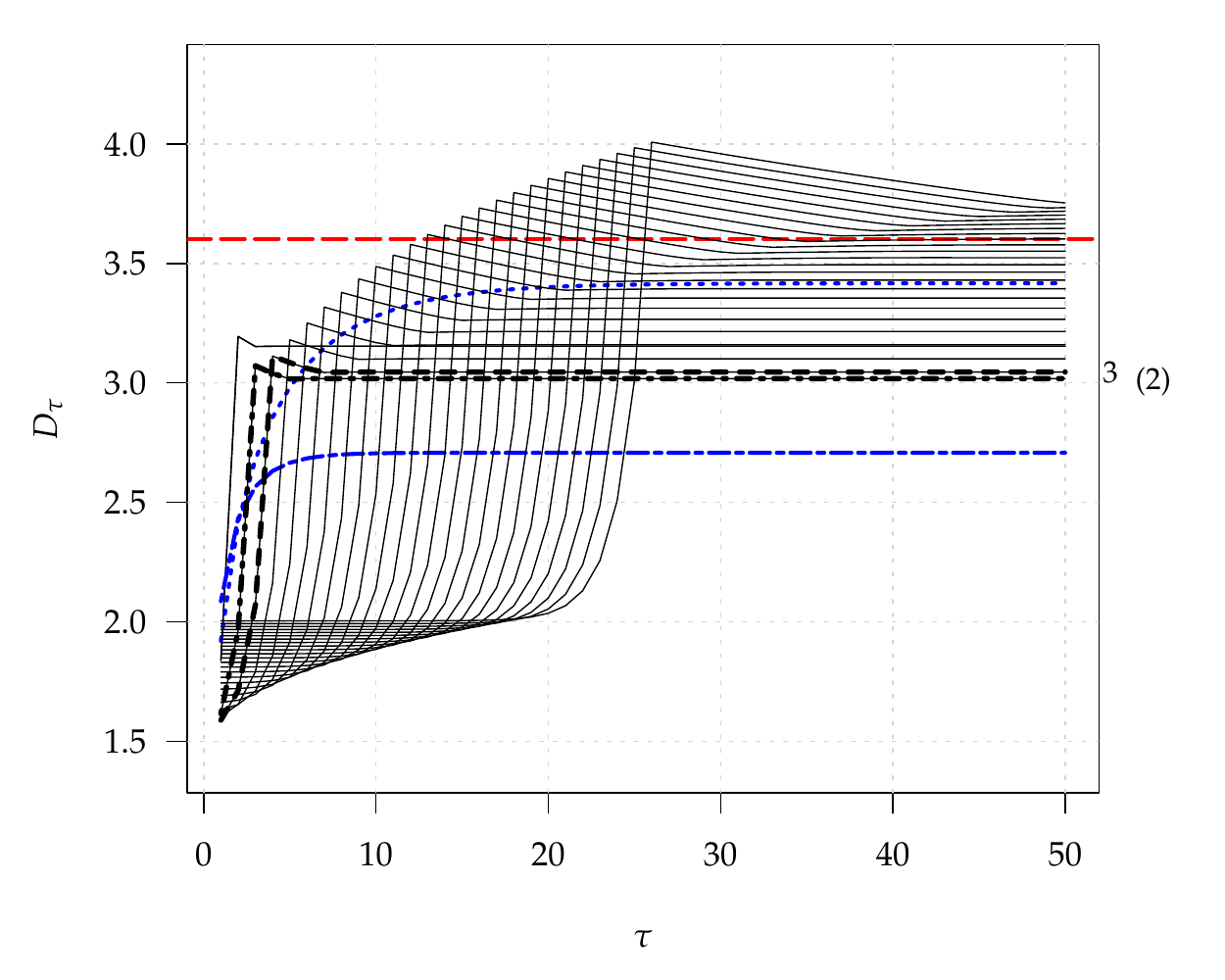} \\[1ex]
  \footnotesize $S_3$ & \footnotesize $S_4$ \\[-1ex]
  \includegraphics[width=.5\textwidth]{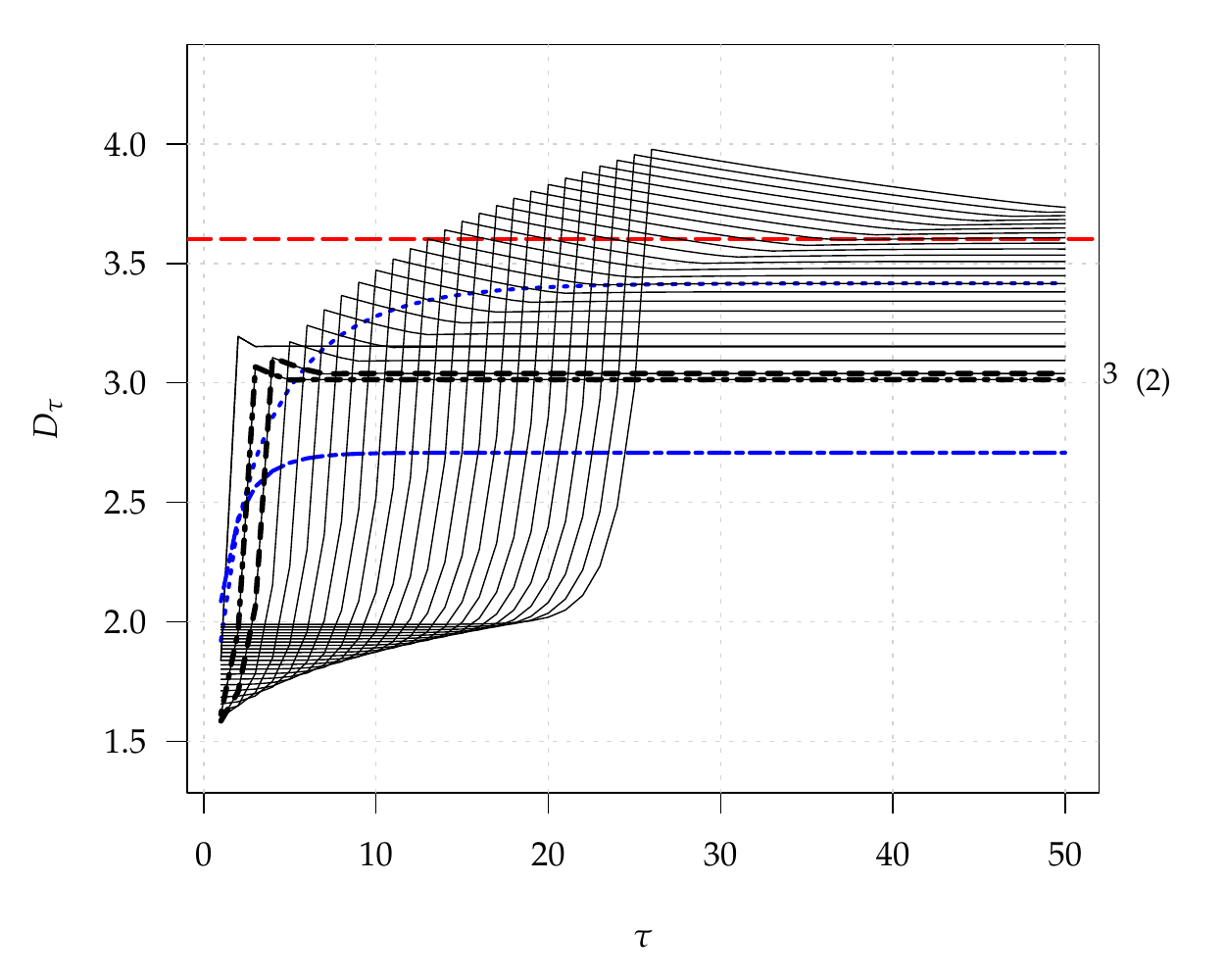} &	
  \includegraphics[width=.5\textwidth]{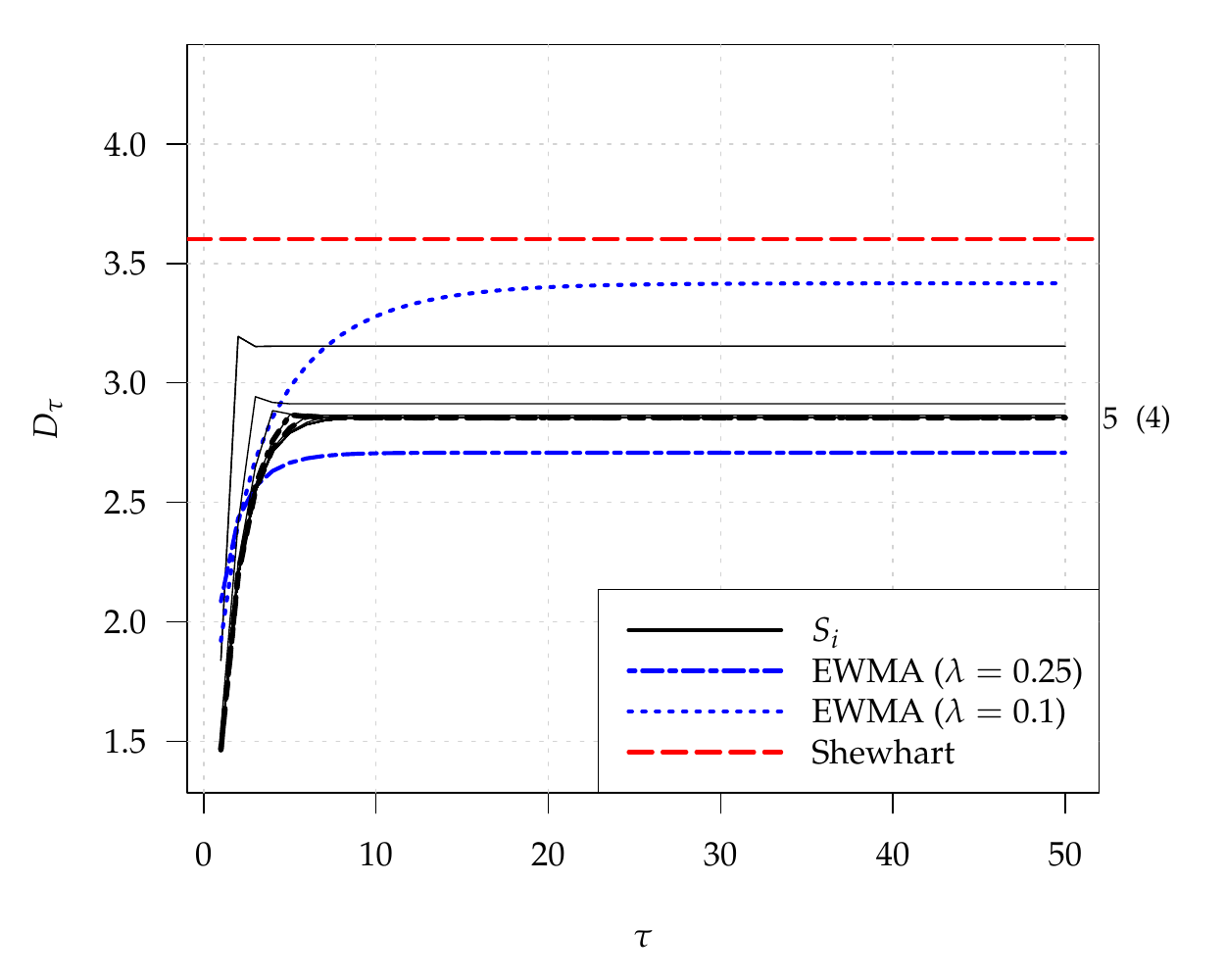}
\end{tabular}
\caption{$D_\tau$ profiles for four synthetic-type charts with head-start,
$H = 1, 2, \ldots, 25$, best scheme (zero-state and steady-state) bold (dashed and dash-dotted) lines,
shift $\delta = 2.5$, two EWMA charts; in-control ARL 500.} \label{fig:ced25}
\end{figure}

\begin{figure}[hbt]
\centering
\begin{tabular}{cc}
  \footnotesize $S_1$ & \footnotesize $S_2$ \\[-1ex]
  \includegraphics[width=.5\textwidth]{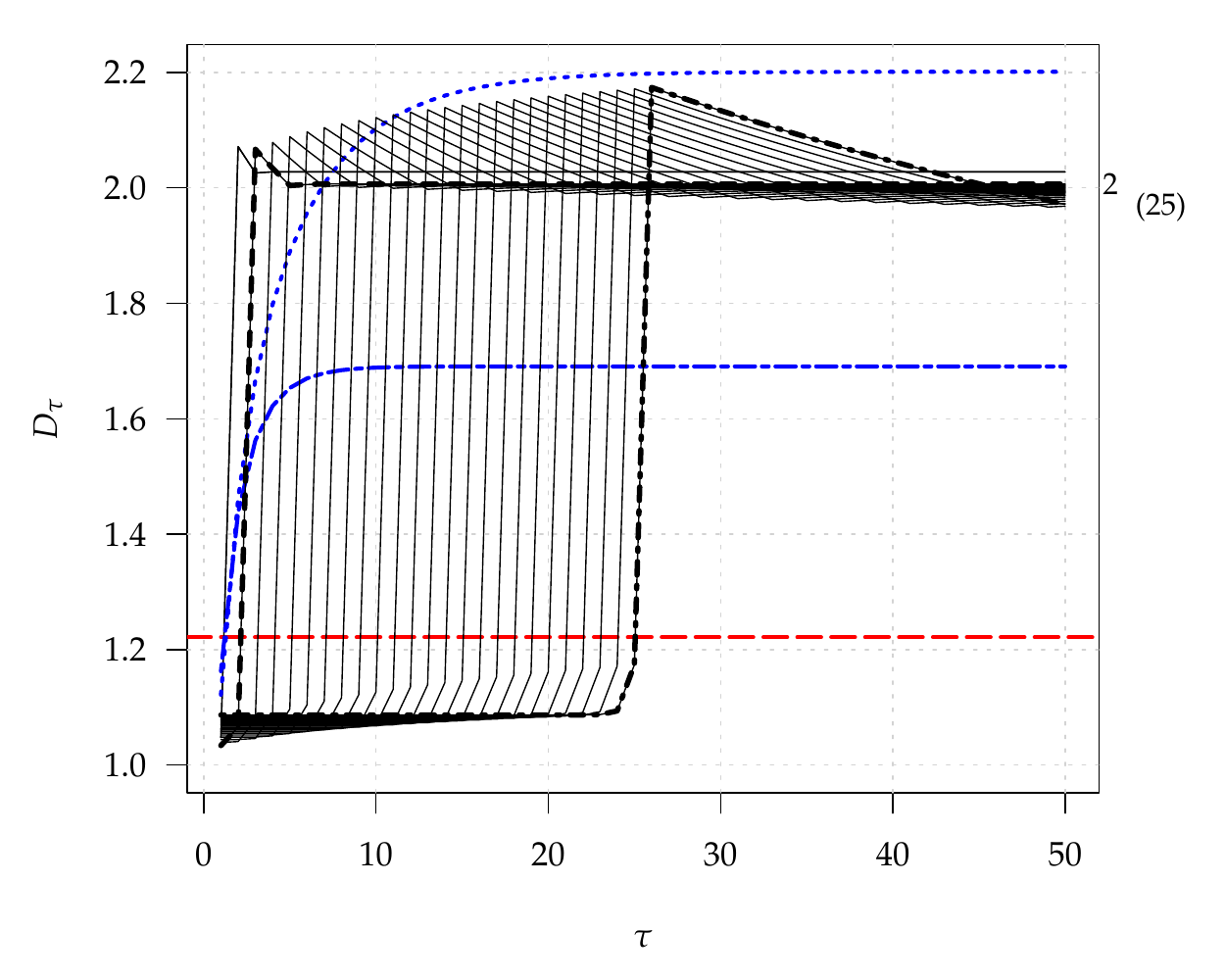} &	
  \includegraphics[width=.5\textwidth]{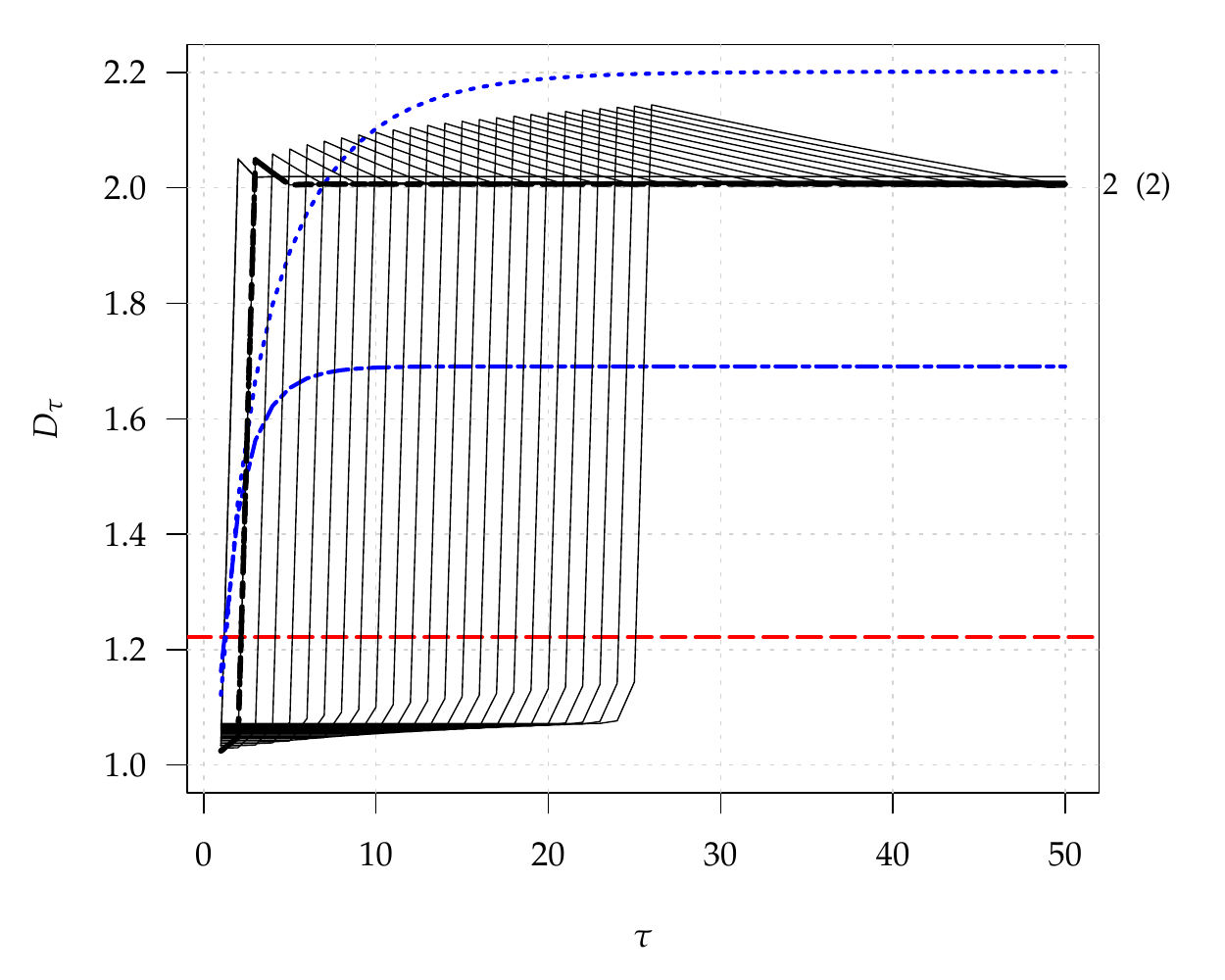} \\[1ex]
  \footnotesize $S_3$ & \footnotesize $S_4$ \\[-1ex]
  \includegraphics[width=.5\textwidth]{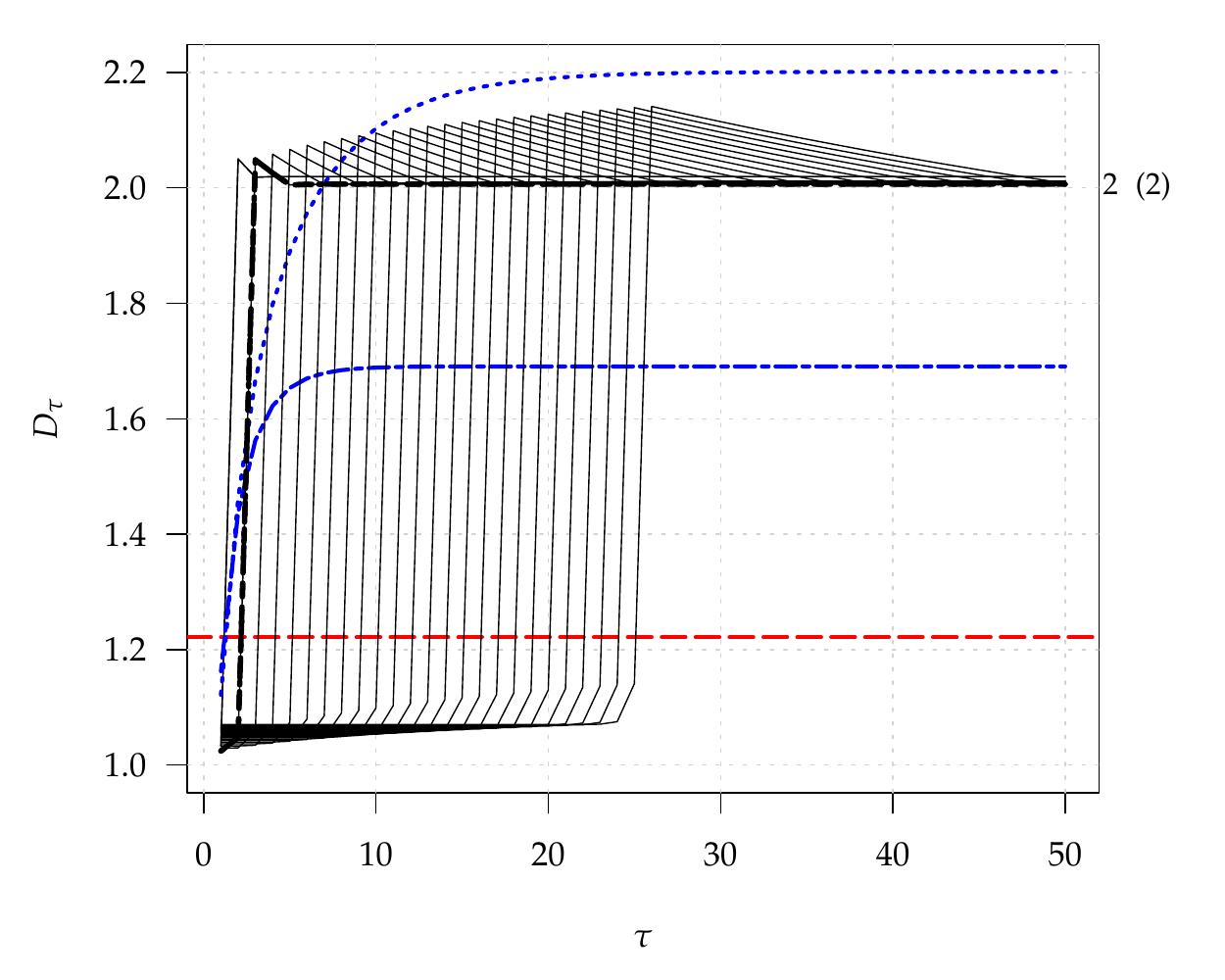} &	
  \includegraphics[width=.5\textwidth]{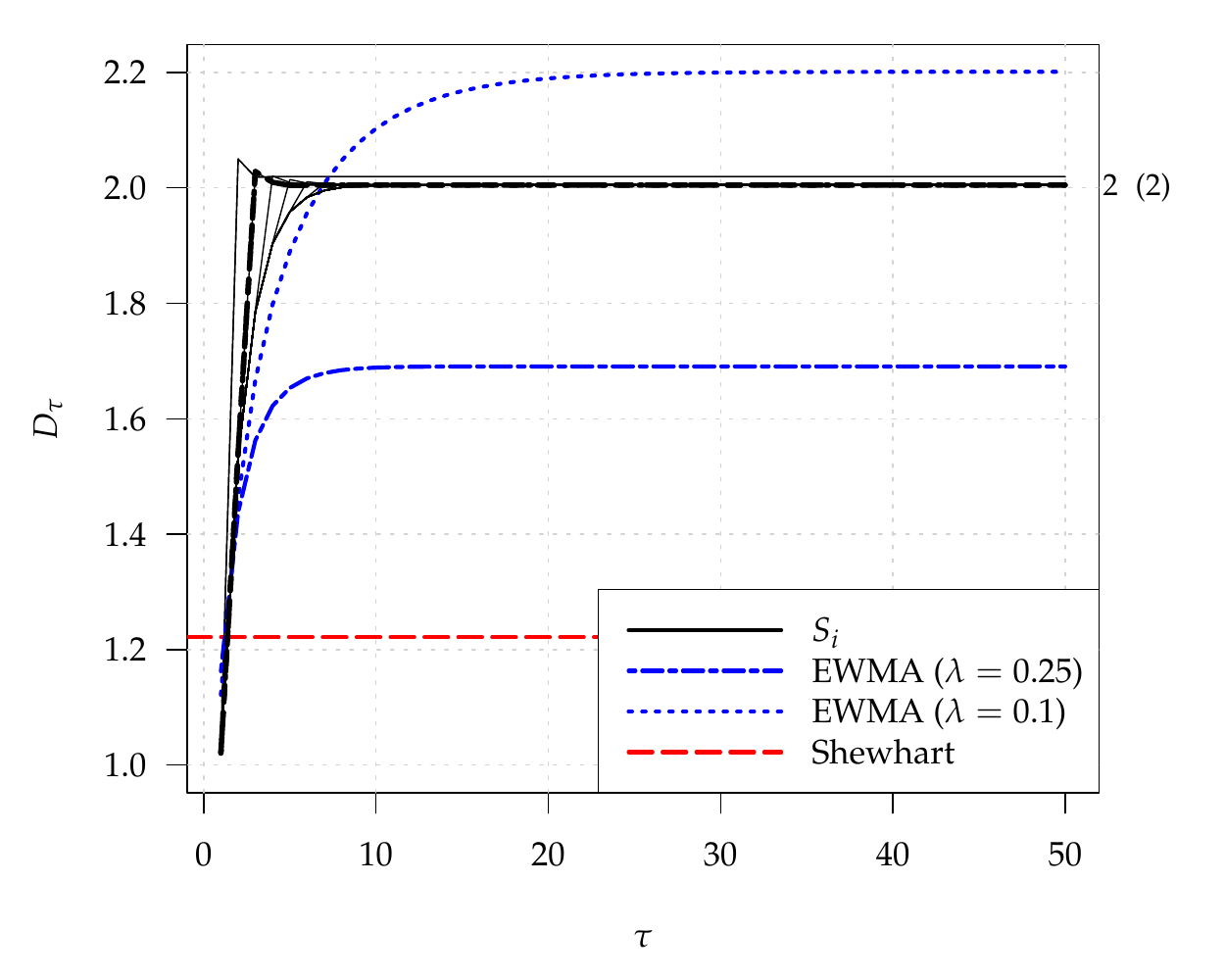}
\end{tabular}
\caption{$D_\tau$ profiles for four synthetic-type charts with head-start,
$H = 1, 2, \ldots, 25$, best scheme (zero-state and steady-state) bold (dashed and dash-dotted) lines,
shift $\delta = 4$, two EWMA charts; in-control ARL 500.} \label{fig:ced4}
\end{figure}

\begin{figure}[hbt]
\centering
\begin{tabular}{cc}
  \footnotesize $S_1$ & \footnotesize $S_2$ \\[-1ex]
  \includegraphics[width=.5\textwidth]{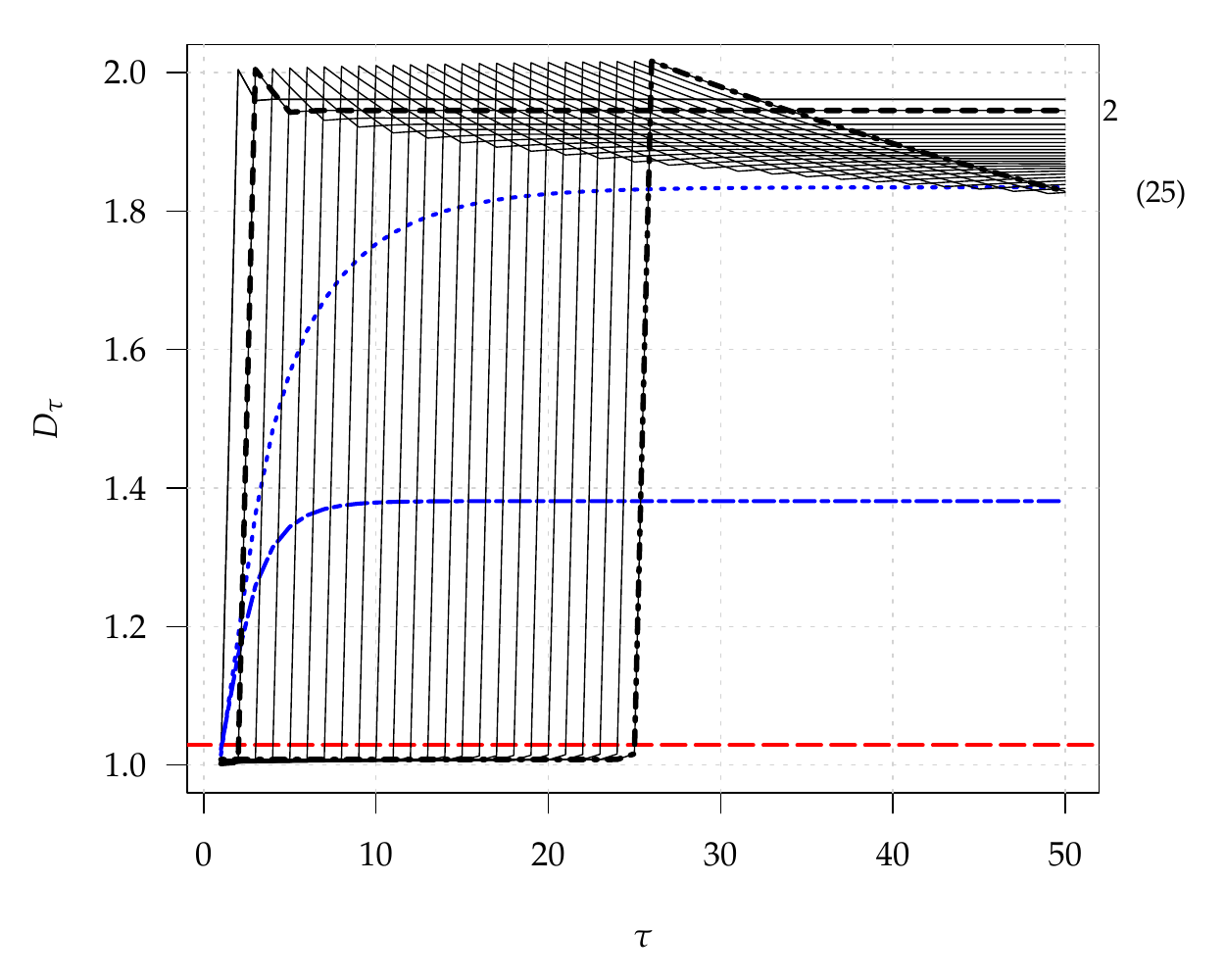} &	
  \includegraphics[width=.5\textwidth]{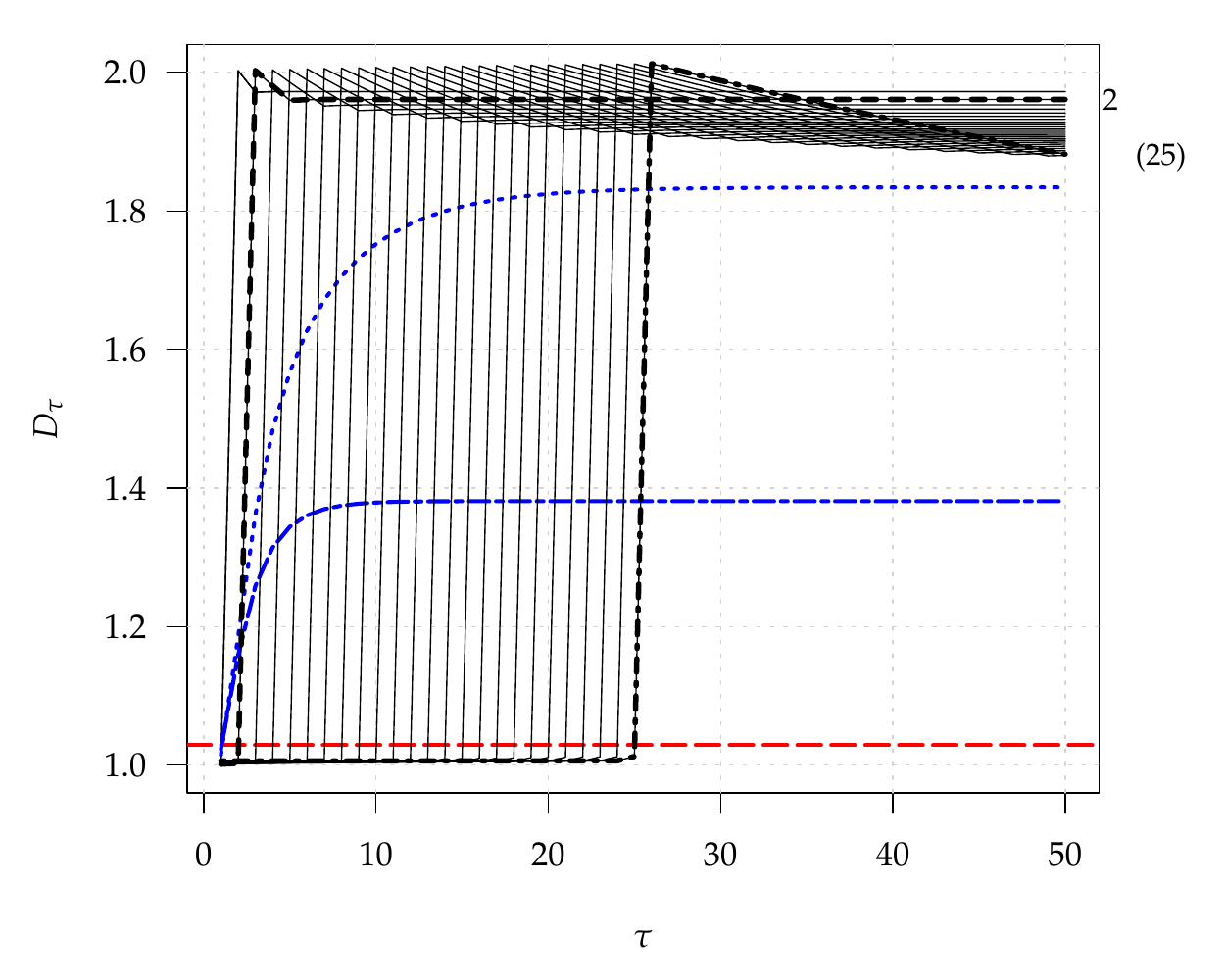} \\[1ex]
  \footnotesize $S_3$ & \footnotesize $S_4$ \\[-1ex]
  \includegraphics[width=.5\textwidth]{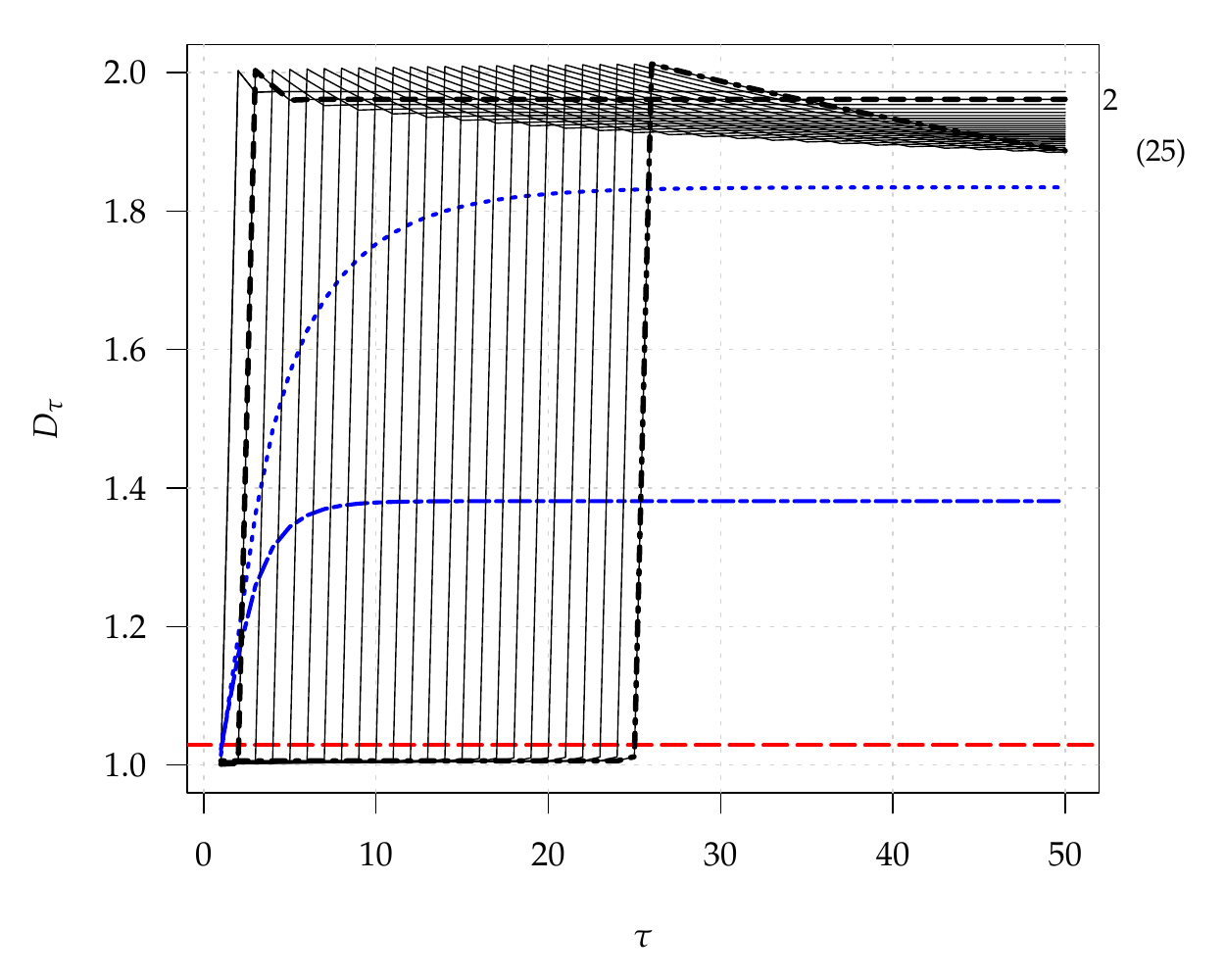} &	
  \includegraphics[width=.5\textwidth]{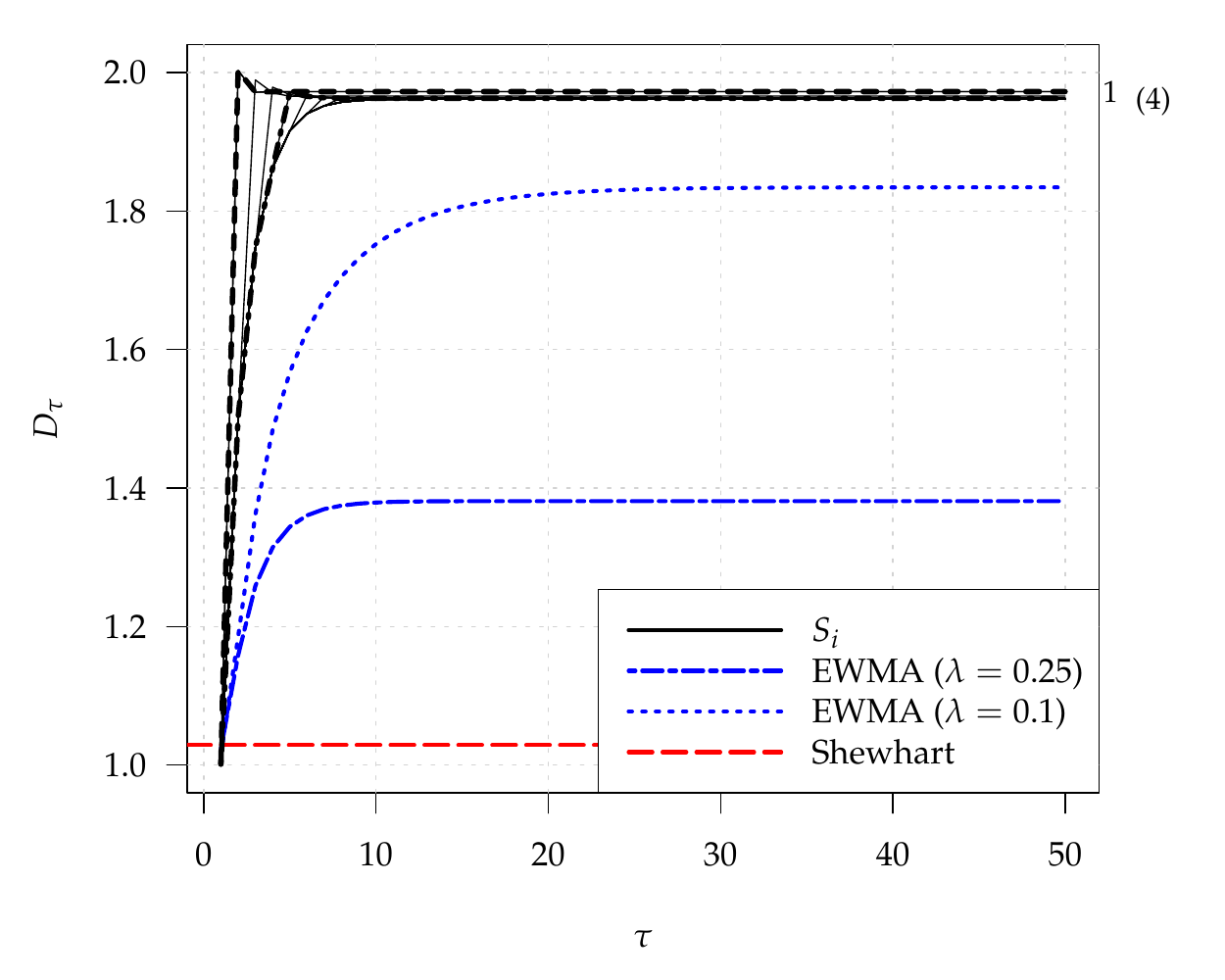}
\end{tabular}
\caption{$D_\tau$ profiles for four synthetic-type charts with head-start,
$H = 1, 2, \ldots, 25$, best scheme (zero-state and steady-state) bold (dashed and dash-dotted) lines,
shift $\delta = 5$, two EWMA charts; in-control ARL 500.} \label{fig:ced5}
\end{figure}

\end{document}